\documentclass[preprint]{aastex61}

\newcommand\aastex{AAS\TeX}

\received{\today}
\submitjournal{AAS}

\shorttitle{\aastex\ ALMA 0$\farcs$02-resolution dense molecular line 
observations of NGC 1068}
\shortauthors{Imanishi et al.}

\begin{document}

\title{ALMA 0$\farcs$02-resolution observations reveal
HCN-abundance-enhanced counter-rotating and outflowing dense molecular 
gas at the NGC 1068 nucleus}

\correspondingauthor{Masatoshi Imanishi}
\email{masa.imanishi@nao.ac.jp}

\author[0000-0001-6186-8792]{Masatoshi Imanishi}
\affil{National Astronomical Observatory of Japan, National Institutes 
of Natural Sciences (NINS), 2-21-1 Osawa, Mitaka, Tokyo 181-8588, Japan}
\affil{Department of Astronomy, School of Science, The Graduate University 
for Advanced Studies, SOKENDAI, Mitaka, Tokyo 181-8588}

\author{Dieu D. Nguyen}
\affil{National Astronomical Observatory of Japan, National Institutes 
of Natural Sciences (NINS), 2-21-1 Osawa, Mitaka, Tokyo 181-8588, Japan}

\author{Keiichi Wada}
\affil{Kagoshima University, Graduate School of Science and Engineering, 
Kagoshima 890-0065, Japan}
\affil{Ehime University, Research Center for Space and Cosmic Evolution, 
Matsuyama 790-8577, Japan}
\affil{Hokkaido University, Faculty of Science, Sapporo 060-0810, Japan}

\author{Yoshiaki Hagiwara}
\affil{Natural Science Laboratory, Toyo University, 5-28-20 Hakusan, 
Bunkyo-ku, Tokyo 112-8606, Japan}
\affil{National Astronomical Observatory of Japan, National Institutes 
of Natural Sciences (NINS), 2-21-1 Osawa, Mitaka, Tokyo 181-8588, Japan}

\author{Satoru Iguchi}
\affil{National Astronomical Observatory of Japan, National Institutes 
of Natural Sciences (NINS), 2-21-1 Osawa, Mitaka, Tokyo 181-8588, Japan}
\affil{Department of Astronomy, School of Science, The Graduate University 
for Advanced Studies, SOKENDAI, Mitaka, Tokyo 181-8588}

\author{Takuma Izumi}
\affil{National Astronomical Observatory of Japan, National Institutes 
of Natural Sciences (NINS), 2-21-1 Osawa, Mitaka, Tokyo 181-8588, Japan}
\affil{Department of Astronomy, School of Science, The Graduate University 
for Advanced Studies, SOKENDAI, Mitaka, Tokyo 181-8588}

\author{Nozomu Kawakatu}
\affil{Faculty of Natural Sciences, National Institute of Technology, 
Kure College, 2-2-11 Agaminami, Kure, Hiroshima 737-8506, Japan}

\author{Kouichiro Nakanishi}
\affil{National Astronomical Observatory of Japan, National Institutes 
of Natural Sciences (NINS), 2-21-1 Osawa, Mitaka, Tokyo 181-8588, Japan}
\affil{Department of Astronomy, School of Science, The Graduate University 
for Advanced Studies, SOKENDAI, Mitaka, Tokyo 181-8588}

\author{Kyoko Onishi}
\affil{Department of Space, Earth and Environment, Chalmers University 
of Technology, Onsala Space Observatory, SE-439 92 Onsala, Sweden}

\begin{abstract}
We present ALMA $\sim$0$\farcs$02-resolution observations of the nucleus 
of the nearby ($\sim$14 Mpc) type-2 AGN NGC 1068 at HCN/HCO$^{+}$/HNC J=3--2 
lines, as well as at their $^{13}$C isotopologue and vibrationally excited 
lines, to scrutinize the morphological/dynamical/chemical/physical 
properties of dense molecular gas in the putative dusty molecular torus 
around a mass-accreting supermassive black hole. 
We confirm almost east-west-oriented dense molecular gas emission both 
morphologically and dynamically, which we regard as coming from the torus.
Bright emission is compact ($\lesssim$3 pc), and low-surface-brightness 
emission extends out to 5--7 pc.
These dense molecular gas properties are not symmetric between the 
eastern and western torus.
The HCN J=3--2 emission is stronger than the HCO$^{+}$ J=3--2 emission 
within the $\sim$7 pc torus region, with an estimated dense molecular 
mass of (0.4--1.0)$\times$10$^{6}$M$_{\odot}$.
We interpret that HCN abundance is enhanced in the torus.
We detect signatures of outflowing dense molecular gas 
and a vibrationally excited HCN J=3--2 line.
Finally, we find that in the innermost ($\lesssim$1 pc) part of the 
torus, the dense molecular line rotation velocity, 
relative to the systemic velocity, is the opposite of that in the
outer ($\gtrsim$2 pc) part, in both the eastern and western torus.
We prefer a scenario of counter-rotating dense molecular gas with 
innermost almost-Keplerian-rotation and outer slowly rotating 
(far below Keplerian) components. 
Our high-spatial-resolution dense molecular line data reveal that 
torus properties of NGC 1068 are much more complicated than the 
simple axi-symmetrically rotating torus picture in the classical 
AGN unification paradigm.
\end{abstract}

\section{Introduction} 

Active galactic nuclei (AGNs) are galaxy nuclei that shine very 
brightly from compact regions.
The bright compact emission is believed to come from an actively 
mass-accreting supermassive black hole (SMBH) with mass of 
$>$10$^{6}$M$_{\odot}$. 
Optical spectroscopy of AGNs has found that some display broad 
($\gtrsim$2000 km s$^{-1}$)
emission lines (classified as type-1), while others show only 
narrow ($\lesssim$1000 km s$^{-1}$) emission lines (type-2).
It is now widely accepted that the presence of type-1 and -2 AGNs 
can naturally be explained by the presence of $\lesssim$a few 10 
parsec (pc)-scale toroidally distributed dust and gas, the so-called dusty 
molecular torus, which surrounds a UV-optical-continuum-emitting 
accretion disk around an SMBH and the sub-pc-scale broad-line-emitting 
regions photo-ionized by the continuum emission \citep{ant93,hon19}.
This AGN unification paradigm was originally proposed from 
observations of the nearby type-2 AGN, NGC 1068 
($z$ $\sim$ 0.0038, distance $\sim$ 14 Mpc, 1 arcsec is $\sim$70 pc) 
\citep{ant85}. 
Observationally scrutinizing the properties of the putative 
dusty molecular torus in NGC 1068 is an important first step toward 
understanding what the torus is and what its role is in AGNs. 
However, the apparent size of the putative $\lesssim$a few 10 pc-scale 
torus is small 
(e.g., 14 pc is 0$\farcs$2 at the distance of NGC 1068), 
so high-spatial-resolution observations are crucial to spatially resolving 
it clearly.
 
The use of ALMA has started to bring about significant advances in our understanding of 
the compact dusty molecular torus in NGC 1068. This is because 
high-spatial-resolution and high sensitivity data now can routinely be obtained 
at the (sub)millimeter wavelength, where rotational (J) transition lines of 
many abundant molecules are present.
\citet{gar14} conducted $\sim$0$\farcs$3--0$\farcs$5-resolution (sub)millimeter 
CO J=3--2/J=6--5 line and continuum observations of NGC 1068.
Molecular line emission at the putative torus position overlapped 
brighter, surrounding 
spatially extended (50--150 pc scale) emission in the host galaxy, 
but these authors constrained torus properties by modeling 
continuum flux at multiple frequencies and CO data 
(e.g., torus mass is M$_{\rm torus}$ $\sim$ a few $\times$ 10$^{5}$M$_{\odot}$ 
and size is r$_{\rm torus}$ $\sim$ 20 pc).
This torus mass is much smaller than the SMBH mass of NGC 1068, with 
M$_{\rm SMBH}$ $\sim$ 1 $\times$ 10$^{7}$M$_{\odot}$, which was estimated from 
$\sim$22 GHz ($\sim$1.4 cm) H$_{2}$O maser emission dynamics at 
$\lesssim$1 pc, constrained with Very Long Baseline Interferometry (VLBI)
high-spatial-resolution ($<$0$\farcs$01) observations \citep{gre96,hur02,lod03}.
\citet{ima16a} presented $\sim$0$\farcs$1--0$\farcs$2-resolution 
HCN/HCO$^{+}$ J=3--2 line observations (i.e., dense molecular gas tracers), 
isolated compact dense molecular emission at the location of the putative 
torus from spatially extended (50--150 pc scale) brighter molecular emission, 
and estimated torus mass and size to be M$_{\rm torus}$ $\sim$ 
several $\times$ 10$^{5}$M$_{\odot}$ and r$_{\rm torus}$ $\sim$ 10 pc, respectively.
The torus dynamics was still not clearly constrained owing to the lack 
of spatial resolution.
\citet{gar16} and \citet{gal16} conducted 
$\sim$0$\farcs$05--0$\farcs$08-resolution 
CO J=6--5 line observations of the NGC 1068 nucleus and 
revealed that molecular emission is morphologically elongated roughly 
along the east-west direction, but {\it observed} dynamical properties are 
dominated by almost north-south-oriented blueshifted and redshifted emission.

According to the classical torus picture, a torus is thought to rotate  
largely affected by the gravity of the central mass-dominating SMBH, 
because molecular viscosity is insufficient to remove angular 
momentum \citep[e.g.,][]{wad09,wad16}. 
In the case of NGC 1068, radio jet emission and optical [OIII] line emission 
photo-ionized by the AGN radiation extend to the north-south direction 
in the vicinity of the mass-accreting SMBH \citep{eva91,das06,gal04}.
Since such emission can easily escape along a direction almost 
perpendicular to the putative dusty molecular torus, the rotating torus 
in NGC 1068 is presumed to be located roughly along the east-west 
direction of the SMBH.
In this case, both the molecular emission morphology and dynamics 
are expected to be aligned to the almost east-west direction.   
The observed north-south oriented velocity structure \citep{gar16,gal16} 
is not easily reconciled with the classical torus picture. 
One possible scenario is that molecular outflow along the direction almost 
perpendicular to the torus is contributing greatly to the 
observed CO J=6--5 dynamics \citep{gal16}.
It is also possible that the highly turbulent torus can create the observed 
north-south oriented velocity structure \citep{gar16}. 
The classical torus picture was not unambiguously demonstrated from 
observations.

\citet{ima18a} conducted ALMA $\sim$0$\farcs$04--0$\farcs$07 
(2.8 pc $\times$ 4.9 pc)-resolution observations of the NGC 1068 torus 
using dense molecular tracers, HCN/HCO$^{+}$ J=3--2 lines, and revealed 
almost east-west-oriented emission both morphologically and dynamically.
This observational result conformed to what people had expected for the 
torus in NGC 1068.
However, it was also found that 
(1) the velocity difference between the blueshifted and redshifted 
components at the probed physical scale ($\sim$3 pc) is much smaller 
than that expected from Keplerian rotation dominated by the central 
SMBH (see also \citet{gar16}), 
and (2) the dense molecular emission is {\it blueshifted 
(redshifted)} at the {\it western (eastern)} part of the torus, while both 
(a) innermost ($\lesssim$1 pc) $\sim$22 GHz H$_{2}$O maser emission and 
(b) dense molecular emission in the host galaxy (50--150 pc scale) 
outside the torus, along the torus direction, 
showed {\it redshifted (blueshifted)} components at the 
{\it western (eastern)} part \citep{gre96,gal04,ima18a}.
Dense molecular gas in the putative torus looked to be counter-rotating 
with respect to the innermost ($\lesssim$1 pc) H$_{2}$O maser emission 
and outer (50--150 pc) host galaxy dense molecular gas.
Figure 1 illustrates the observed complex dynamics of dense molecular gas 
at the NGC 1068 nucleus.

\citet{imp19} obtained ALMA $\sim$0$\farcs$02 ($\sim$1.4 pc)-resolution 
HCN J=3--2 line data of the NGC 1068 torus and unveiled the presence 
of inner ($\lesssim$2 pc) and outer ($\gtrsim$2 pc) 
dynamically decoupled dense molecular gas components. 
The inner dense molecular gas displayed {\it redshifted (bluehifted)} 
motion at the {\it western (eastern)} part in a similar way to the 
innermost ($\lesssim$1 pc) H$_{2}$O maser emission \citep{gre96,gal04}, 
while the outer 
one showed {\it blueshifted (redshifted)} motion at the 
{\it western (eastern)} side, as previously observed in 
$\sim$0$\farcs$04--0$\farcs$07-resolution HCN/HCO$^{+}$ J=3--2 line 
data \citep{ima18a}.
\citet{imp19} argued that two, almost-Keplerian, counter-rotating 
dense molecular gas components can explain these observational results.
If this is the case, we are witnessing a very transient phase of the 
NGC 1068 torus, because such a counter-rotating molecular gas can quickly 
change to a co-rotating one.  
\citet{gar19} obtained $\sim$0$\farcs$03--0$\farcs$09-resolution multiple 
molecular line data of the NGC 1068 torus, and preferred an outflow scenario 
rather than the counter-rotating molecular gas model, based mainly on 
detailed dynamical modeling and the expected very short time scale of the 
counter-rotating torus.

The observed properties of the NGC 1068 torus have been found to be far more 
complex and interesting than the classical torus picture of simple 
rotation governed by the central SMBH's gravity.
Further high-spatial-resolution molecular line data obtained with ALMA will 
help improve and refine our understanding of the torus in NGC 1068. 
In particular, one intriguing question is ``Are the two 
dynamically decoupled torus molecular gas components found in HCN J=3--2 
\citep{imp19} also seen in other dense gas tracers?''. 
In this manuscript, we present the results of our ALMA
$\sim$0$\farcs$02-resolution observations of the NGC 1068 nucleus 
using dense gas tracers, HCN/HCO$^{+}$/HNC J=3--2 lines, 
as well as their $^{13}$C isotopologue and vibrationally excited lines.

Throughout this manuscript, ``$^{12}$C'' is simply described as ``C'', 
optical LSR velocity is adopted, and images of our newly taken ALMA 
data are displayed in ICRS coordinates, with north up and 
east to the left.

\section{Observations and Data Analyses} 

Our ALMA band 6 (211--275 GHz) observations of the NGC 1068 
nucleus were conducted through the Cycle 6 program 2018.1.00037.S 
(PI = M. Imanishi).
The observation details are summarized in Table 1.
Since our main interest is dense molecular gas properties of the 
compact torus, only long baselines were used.

We conducted two types of observations, one covering HCN/HCO$^{+}$ J=3--2 lines
as well as their vibrationally excited (HCN-VIB/HCO$^{+}$-VIB) J=3--2 lines 
(denoted as ``data-a'' in Table 1), and the other 
covering HNC, $^{13}$C isotopologue (H$^{13}$CN/H$^{13}$CO$^{+}$/HN$^{13}$C) 
and HNC-VIB J=3--2 lines (denoted as ``data-b''). 
In both observations, the phase center was the nucleus at 
(02$^{\rm h}$ 42$^{\rm m}$ 40.71$^{\rm s}$, $-$00$^{\circ}$ 00$'$ 47.94$''$)ICRS
and we employed the widest 1.875 GHz-width mode in each spectral window. 
For data-a, we took spectra at 
263.6--269.1 GHz, with three spectral windows to cover 
HCN J=3--2 ($\nu_{\rm rest}$ = 265.886 GHz) and 
HCO$^{+}$ J=3--2 ($\nu_{\rm rest}$ = 267.558 GHz) lines, as well as 
vibrationally excited HCN v$_{2}$=1, l=1f (HCN-VIB) J=3--2 
($\nu_{\rm rest}$ = 267.199 GHz) and HCO$^{+}$ v$_{2}$=1, l=1f (HCO$^{+}$-VIB) 
J=3--2 ($\nu_{\rm rest}$ = 268.689 GHz) lines.
For data-b, we took spectra at 256.7--260.5 GHz, with two spectral 
windows in LSB to cover H$^{13}$CN J=3--2 ($\nu_{\rm rest}$ = 259.012 GHz), 
H$^{13}$CO$^{+}$ J=3--2 ($\nu_{\rm rest}$ = 260.255 GHz),  
and HN$^{13}$C J=3--2 ($\nu_{\rm rest}$ = 261.263 GHz) lines.
In USB, we covered 270.0--273.7 GHz with two spectral windows to 
include HNC J=3--2 ($\nu_{\rm rest}$ = 271.981 GHz) and 
HNC v$_{2}$=1, l=1f (HNC-VIB) J=3--2 ($\nu_{\rm rest}$ = 273.870 GHz) lines.
Since HNC and $^{13}$C isotopologue J=3--2 lines are fainter than 
HCN/HCO$^{+}$ J=3--2 lines, we set the beam size requested in our 
program slightly larger for data-b ($\sim$0$\farcs$03) than that for 
data-a ($\sim$0$\farcs$02). 
 
For our data reduction, we used CASA (https://casa.nrao.edu) 
starting from calibrated data provided by ALMA.
We first determined a constant continuum flux level by using data 
that were not affected by emission and absorption lines, and used 
the CASA task ``uvcontsub'' to subtract the derived continuum emission 
from the data.
We then applied the task ``clean'' 
(Briggs-weighting; robust=0.5, gain $=$ 0.1) 
to the continuum-subtracted molecular line data as well as the 
extracted continuum data. 
The adopted velocity resolution was 10--20 km s$^{-1}$. 
The pixel scale was 0$\farcs$003 pixel$^{-1}$ and 
0$\farcs$005 pixel$^{-1}$ for data-a and -b, respectively.
The maximum recoverable scale (MRS) is 0$\farcs$24--0$\farcs$30 
and 0$\farcs$56--0$\farcs$62 for data-a and -b, respectively.
Emission with a spatial extent larger than these MRSs 
may be resolved out and may not be recovered in our ALMA data.
The absolute flux calibration uncertainty is expected to be $<$10\% 
at the observed frequencies in our ALMA Cycle 6 data.

\section{Results} 

Continuum emission at $\sim$260 GHz ($\sim$1.2 mm) at the NGC 1068 
nucleus taken in data-a is displayed as contours in Figure 2.
Table 2 summarizes the continuum emission properties at the 
NGC 1068 nucleus.
Although the continuum emission looks spatially compact, its 
flux is slightly ($\sim$26\%) larger in 
$\sim$0$\farcs$03-resolution data-b (8.2 mJy beam$^{-1}$) 
than in $\sim$0$\farcs$02-resolution data-a (6.5 mJy beam$^{-1}$), 
suggesting that continuum emission is spatially 
extended ($\gtrsim$0$\farcs$02 or $\gtrsim$1.4 pc). 
Spatially extended, radio-jet-related synchrotron emission \citep{gar19}, 
free-free emission \citep{imp19}, 
and/or dust emission in the torus at $\gtrsim$1.4 pc may contribute to the 
observed continuum flux of $\sim$0$\farcs$03-resolution data-b. 
We modified the beam size of data-a to match 
that of data-b, using the CASA task ``imsmooth'', and 
confirmed that the continuum peak flux of data-a became 
8.1 (mJy beam$^{-1}$) (54$\sigma$), which agrees well with that of 
data-b (8.2 mJy beam$^{-1}$). 

The integrated intensity (moment 0) maps of the HCN J=3--2 and
HCO$^{+}$ J=3--2 lines are shown in Figure 2 (top panels).
We adopt the $\sim$260 GHz continuum peak position as the location 
of the mass-accreting SMBH, because this position spatially 
agrees with continuum peaks derived in other high-spatial-resolution 
($<$0$\farcs$1) (sub)millimeter--centimeter observations 
at 5--700 GHz \citep{gal04,gar19}. 
Both the HCN J=3--2 and HCO$^{+}$ J=3--2 line emission are clearly 
detected along the almost east-west direction, relative to the active 
SMBH position, with the western side being significantly brighter than 
the eastern side, 
as previously seen in $\sim$0$\farcs$04 $\times$ 0$\farcs$07-resolution 
ALMA data \citep{ima18a}. 
We interpret this to mean that the emission originates in the putative torus 
in NGC 1068.
Table 3 summarizes more detailed properties of dense molecular line 
emission in the torus region.

The dense molecular line emission is particularly bright in a 
relatively compact area within $\sim$3 pc in both the west and east 
(Figure 2a,b).
In addition to this compact component, low-surface-brightness 
emission extends out to 5--7 pc, particularly in HCN J=3--2. 
Significantly detected emission of HCN J=3--2 is spatially more 
extended than that of HCO$^{+}$ J=3--2, despite larger noise 
in the former (Table 3, column 4).
The size of $\lesssim$7 pc is much smaller than the size of the 
CO J=3--2 and J=2--1 emitting region ($\sim$25--30 pc), which probes 
less dense and cooler molecular gas \citep{gar19}.
Dense and warm molecular gas that can excite HCN and HCO$^{+}$ to 
the J=3 level is mostly concentrated in $\lesssim$7 pc area 
in the NGC 1068 torus.

The intensity-weighted mean velocity (moment 1) maps of the HCN J=3--2 
and HCO$^{+}$ J=3--2 lines are also displayed in Figure 2 (bottom panels).
The velocity display range is set as the same as that adopted 
in the $\gtrsim$0$\farcs$04-resolution 
moment 1 maps of the same molecular lines by \citet{ima18a}, 
in order to compare velocity patterns between our new 
$\sim$0$\farcs$02-resolution and old $\gtrsim$0$\farcs$04-resolution 
data.
To discuss the dynamics of dense molecular emission in the torus, 
we adopt the systemic velocity of NGC 1068 to be 
V$_{\rm sys}$ = 1130 km s$^{-1}$ in the optical LSR velocity 
($z \sim$ 0.0038), because nuclear HCN J=3--2 and HCO$^{+}$ J=3--2 
emission lines showed velocity peaks around this value 
\citep{ima16a,ima18a,imp19}.
\citet{ima18a} found that both the HCN J=3--2 and HCO$^{+}$ J=3--2 
line emission are blueshifted in the western torus at $\lesssim$5 pc and 
redshifted in the eastern torus at $\lesssim$5 pc.
The implied rotation velocity was substantially slower than that 
expected for Keplerian rotation ($\pm$120 km s$^{-1}$ at $\sim$3 pc) 
for the central mass-dominating compact SMBH with mass of 
M$_{\rm SMBH}$ $\sim$ 1 $\times$ 10$^{7}$M$_{\odot}$ \citep{gre96,hur02,lod03}, 
unless we view the torus dense molecular gas rotation from a nearly face-on 
direction. This would be very unlikely for the type-2 AGN NGC 1068, 
because it is expected to be obscured by the 
fairly edge-on, east-west-oriented torus.
In our new $\sim$0$\farcs$02-resolution moment 1 map of the
HCN J=3--2 line (Figures 2c),  
western blueshifted and eastern redshifted motions are clearly 
confirmed at $\gtrsim$2 pc from the SMBH along the position angle 
PA $\sim$ 105--110$^{\circ}$ east of north \citep{ima18a}.
The velocity difference with respect to the systemic is only 
$\lesssim$50 km s$^{-1}$, which is still much smaller than that expected 
from Keplerian rotation.

It is also found that {\it innermost} ($\lesssim$1 pc) dense 
molecular emission is redshifted on the western side and blueshifted 
on the eastern side, the opposite of what is found in the outer 
torus ($\gtrsim$2 pc) emission. 
This pattern is clearly seen, particularly in the HCN J=3--2 line data 
(Figure 2c), and was previously detected by \citet{imp19} 
in the same line data.
Our new data reveal that western redshifted and eastern blueshifted 
{\it innermost} components are also seen in the HCO$^{+}$ J=3--2 line 
(Figure 2d), suggesting that this innermost torus 
dynamics is characteristic of dense molecular gas in NGC 1068.

Figure 3 (top panels) shows the intensity-weighted mean velocity 
(moment 1) maps of HCN J=3--2 and HCO$^{+}$ J=3—2, with a wider 
velocity range.
We see redshifted high-velocity components at the innermost western 
torus, in both the HCN J=3--2 and HCO$^{+}$ J=3--2 emission lines.  
Intensity-weighted velocity dispersion (moment 2) maps are 
displayed in Figure 3 (bottom panels). 
High velocity dispersion regions are seen in the western torus 
in both the lines, as previously found in $\gtrsim$0$\farcs$04-resolution 
maps \citep{ima18a}.
In the new $\sim$0$\farcs$02-resolution data, it becomes clear 
for the first time that the highest-velocity dispersion spot in the 
moment 2 map is shifted toward the outer northwestern side of the innermost 
redshifted high-velocity component in the moment 1 map.

Figure 4 (top panels) shows spectra, within the beam size, at the 
position of the continuum peak (hereafter, ``C-peak''; 
[02$^{\rm h}$ 42$^{\rm m}$ 40.709$^{\rm s}$, $-$00$^{\circ}$ 00$'$ 47.946$''$]), 
the HCN J=3--2 emission peak in the eastern torus (hereafter, ``E-peak'';
[02$^{\rm h}$ 42$^{\rm m}$ 40.710$^{\rm s}$, $-$00$^{\circ}$ 00$'$ 47.952$''$]), 
and 
the HCN J=3--2 emission peak in the western torus (hereafter, ``W-peak''; 
[02$^{\rm h}$ 42$^{\rm m}$ 40.708$^{\rm s}$, $-$00$^{\circ}$ 00$'$ 47.940$''$]), 
derived from Figure 2a.
Both HCN J=3--2 and HCO$^{+}$ J=3--2 lines are clearly detected 
in emission at the E- and W-peaks, but in absorption at the C-peak.
Gaussian fits to several clearly detected emission lines are shown 
in Figure 5, 
and estimated molecular line fluxes are tabulated in Table 3.
The molecular line widths are larger in the W-peak than in the 
E-peak, suggesting larger turbulence at the W-peak.
Figure 6 displays the velocity profiles of the HCN J=3--2 and 
HCO$^{+}$ J=3--2 lines at the C-peak.

To investigate overall molecular emission line properties in 
a larger volume of the torus, we also extract area-integrated spectra 
within 
(a) a 0$\farcs$2 (east-west) $\times$ 0$\farcs$1 (north-south) 
(14 pc $\times$ 7 pc) rectangular region (i.e., covering both 
the eastern and western torus),  
(b) a 0$\farcs$1 $\times$ 0$\farcs$1 (7 pc $\times$ 7 pc) 
square region of the eastern torus (hereafter, ``E-torus''), and 
(c) the same square region of the western torus (hereafter, ``W-torus'').
Figure 7 displays these area-integrated spectra.
Gaussian fits are applied to estimate the fluxes of these emission 
lines (Figure 8 and Table 4). 
The molecular line widths are larger in the W-torus than in the 
E-torus (Table 4), confirming that higher turbulence happens in the W-torus, 
as suggested from the higher velocity dispersion values there in moment 2 
maps (Figure 3c,d).

For the HCN J=3--2 line, \citet{imp19} obtained ALMA Cycle 5 data 
with spatial resolution ($\sim$0$\farcs$02) comparable to ours.
We combined our Cycle 6 and their Cycle 5 data and show several 
selected results in Appendix A. 
Main features detected in our Cycle 6 data are reproduced in the 
combined data as well.
To compare the observed properties of HCN J=3--2 and other lines 
in a uniform manner, we use the results obtained with our Cycle 6 
data only, because we want to avoid possible systematic uncertainties 
resulting from combining data taken with largely different observing 
parameters.

Beam-sized spectra of data-b ($^{13}$C isotopologue and 
HNC J=3--2 observations) at the same three 
positions (C-, E-, and W-peaks) are displayed in Figure 4d--4i.
Figure 9 displays the moment 0 maps of the H$^{13}$CN J=3--2, HNC J=3--2, 
and H$^{13}$CO$^{+}$ J=3--2 lines.
A $>$3$\sigma$ signal is detected either in emission or 
absorption in the torus region for H$^{13}$CN J=3--2 and HNC J=3--2 
(Table 3), while no significant signal is recognizable 
for H$^{13}$CO$^{+}$ J=3--2 ($<$3$\sigma$).
The H$^{13}$CN J=3--2 line velocity profiles at the C- and W-peaks are 
shown in Figure 10.

For the HCN J=3--2 and HCO$^{+}$ J=3--2 emission lines from 
the torus, we compare their fluxes in our Cycle 6 data and previously 
taken Cycles 4 and 2 data that include shorter baselines \citep{ima18a}. 
We found that their fluxes roughly agree (Appendix B), supporting our view 
that we recover virtually all dense molecular line emission flux from 
the torus in our Cycle 6 data. 
However, for spatially extended emission components in the host galaxy, 
the recovered fluxes of HCN J=3--2 and HCO$^{+}$ J=3--2 lines 
in the Cycle 6 data were found to be substantially smaller than 
those in the Cycles 4 and 2 data \citep{ima18a}.
We will thus discuss only compact torus emission.

\section{Discussion} 

\subsection{HCN-abundance-enhanced dense molecular torus}

It was found from lower-spatial-resolution ($\gtrsim$0$\farcs$04) 
molecular line observations that HCN emission is stronger than 
the HCO$^{+}$ emission in the NGC 1068 torus region within $<$10 pc 
from the central AGN 
\citep{gar14,vit14,ima16a,ima18a}.
Thanks to our high-spatial-resolution ($\sim$0$\farcs$02 or 
$\sim$1.4 pc) data, we can now investigate dense molecular emission line 
properties in detail in the eastern and western torus separately.
We find that the emission line flux of HCN J=3--2 is a factor of 
$\sim$1.5 higher than that of HCO$^{+}$ J=3--2 in both the W- and 
E-torus (Table 4).

Elevated HCN emission, relative to HCO$^{+}$ emission, is often found 
in luminous AGNs, compared with starburst-dominated regions
\citep[e.g.,][]{koh05,ima06b,ima07b,kri08,ima09,pri15,izu16,ima16c,ima18b,ima19}.
An enhanced HCN abundance by 
AGN radiation \citep[e.g.,][]{koh05,ima07b,yam07,izu16} 
and/or mechanical \citep{izu13,mar15} effects 
is proposed to be the physical origin.
We thus investigate an HCN-to-HCO$^{+}$ abundance ratio in the NGC 1068 torus 
based on our ALMA Cycle 6 data.

Since the emission lines of the abundant HCN and HCO$^{+}$ can be 
optically thick in the nuclear regions of active galaxies 
\citep[e.g.,][]{jia11,jim17},
it is not straightforward to convert from the observed HCN-to-HCO$^{+}$ 
J=3--2 flux ratio to an HCN-to-HCO$^{+}$ abundance ratio. 
We thus use the isotopologue H$^{13}$CN J=3--2 and H$^{13}$CO$^{+}$ J=3--2 
line data to constrain the HCN-to-HCO$^{+}$ abundance ratio, 
because the $^{13}$C isotopologue molecular abundance is usually $\gtrsim$10 
times smaller \citep{ala15} and so the isotopologue lines are more likely 
to be optically thin.
Unfortunately, these isotopologue H$^{13}$CN and H$^{13}$CO$^{+}$ 
emission lines are much fainter than the HCN and HCO$^{+}$ lines. 
We detect no clear isotopologue emission line in the W- and E-peaks 
($<$3$\sigma$), but identify a significant ($\sim$4.5$\sigma$)
negative absorption signal with $-$0.090 (Jy beam$^{-1}$ km s$^{-1}$) 
at the C-peak in the H$^{13}$CN J=3--2 moment 0 map (Figure 9a).
No significant absorption is recognizable at the C-peak in the 
H$^{13}$CO$^{+}$ J=3--2 moment 0 map 
($<$2$\sigma$ or $>$ $-$0.049 [Jy beam$^{-1}$ km s$^{-1}$] in Figure 9c), 
suggesting that the H$^{13}$CN column density is higher than H$^{13}$CO$^{+}$ 
in the molecular gas in front of the $\sim$260-GHz-continuum-emitting 
region.
Assuming that this absorbing molecular gas largely comes from the 
nearly edge-on dusty molecular torus at the foreground side of 
the central mass-accreting SMBH, the observed 
H$^{13}$CN-to-H$^{13}$CO$^{+}$ J=3--2 
absorption signal ratio with $>$1.8 suggests an enhanced HCN 
abundance, relative to HCO$^{+}$, in the NGC 1068 torus, 
where we assume that the H$^{13}$CN-to-H$^{13}$CO$^{+}$ column density 
ratio (= abundance ratio) is comparable to an HCN-to-HCO$^{+}$ 
abundance ratio. 

More precisely, an Einstein B-coefficient and excitation effect need 
to be taken into account to discuss a column density ratio 
(= an abundance ratio) based on an absorption strength. 
H$^{13}$CN J=3--2 and H$^{13}$CO$^{+}$ J=3--2 absorption features 
can be detected when millimeter $\sim$259 GHz and $\sim$260 GHz 
photons are absorbed by H$^{13}$CN and H$^{13}$CO$^{+}$ at the J=2 level, 
respectively, and are used for excitation to the J=3 level.
The B-coefficient from J=2 to J=3 (B$_{\rm 23}$) of H$^{13}$CO$^{+}$ 
is a factor of $\sim$1.7 higher than that of H$^{13}$CN, based on data 
from the Cologne Database of Molecular Spectroscopy (CDMS) \citep{mul05} 
via Splatalogue (http://www.splatalogue.net). 
Thus, for the same column density, an absorption strength can be 
a factor of $\sim$1.7 stronger for H$^{13}$CO$^{+}$ J=3--2 than 
for H$^{13}$CN J=3--2, as long as these lines are optically thin.
Additionally, the critical density of H$^{13}$CO$^{+}$ is a factor of 
$\sim$5 smaller than that of H$^{13}$CN for the same J-transition 
under the same temperature and line opacity \citep{shi15}.
We ran RADEX calculations \citep{van07} and found that for a volume number 
density of 10$^{4-6}$ (cm$^{-3}$), column density of 10$^{14-15}$ (cm$^{-2}$)
\footnote{
We assume the hydrogen column density in the NGC 1068 torus to be 
N$_{\rm H}$ $\sim$ 10$^{24-25}$ cm$^{-2}$, because the estimated high value 
of N$_{\rm H}$ $\sim$ 10$^{25}$ cm$^{-2}$ \citep{mat97,bau15,mari16} is toward a 
compact X-ray emitting source, and N$_{\rm H}$ in other directions 
can be smaller in the case of a clumpy molecular gas structure 
(see $\S$4.5). 
Adopting a factor of 4--5 enhanced HCN-to-H$_{2}$ abundance ratio 
of 10$^{-8}$, compared to the canonical ratio of a few $\times$ 10$^{-9}$ 
observed in active galaxies \citep[e.g.,][]{mar06,sai18}
and a $^{12}$C-to-$^{13}$C abundance ratio of $\sim$30--100 
\citep[e.g.,][]{hen93a,hen93b,mar10,hen14}, 
we obtain the H$^{13}$CN and H$^{13}$CO$^{+}$ column density in the NGC 1068 
torus to be $\sim$10$^{14-15}$ (cm$^{-2}$)}, 
kinetic temperature of 30--100 (K), and line width of 200--300 (km s$^{-1}$), 
H$^{13}$CO$^{+}$ is excited to J=2 more than H$^{13}$CN is.
Considering these two factors (i.e., B-coefficient and excitation), 
we derive a column density of H$^{13}$CN that is a factor of $>$3 
($>$1.8 $\times$ 1.7) higher than that of H$^{13}$CO$^{+}$, or 
an HCN-to-HCO$^{+}$ abundance ratio of $>$3, in the NGC 1068 torus.
Theoretically, an HCN abundance enhancement around a luminous 
AGN is predicted in some, but not all, parameter ranges, and 
so is not uncontroversially predicted \citep{mei05,mei07,har13,vit17}.
However, observations of molecular gas in the vicinity of luminous AGNs 
suggest an enhanced HCN abundance in almost all cases 
\citep[e.g.,][]{ste94,ala15,nak18,ima18b,sai18,kam20,tak20}. 
We argue that this is also the case for the NGC 1068 torus.

In the widely accepted molecular gas model in galaxies \citep{sol87}, 
a molecular cloud consists of compact dense clumps with a small 
volume-filling factor.
Each clump can be optically thick for HCN J=3--2 and HCO$^{+}$ J=3--2 
lines and is randomly moving.
Molecular emission lines from clumps at the far side of the molecular 
cloud can escape toward Earth if their velocity is different from the 
foreground clumps inside the cloud.
If an abundance is higher for HCN than HCO$^{+}$, flux attenuation 
by line opacity of foreground clumps is likely to be higher for 
HCN J=3--2 than HCO$^{+}$ J=3--2.
In a more highly turbulent region, the HCN J=3--2 flux can increase 
more than the HCO$^{+}$ J=3--2 flux, because of larger reduction of 
line opacity by foreground clumps in the former.
In fact, HCN-to-HCO$^{+}$ J=3--2 flux ratio at the highly turbulent 
W-peak ($\sim$1.5; Table 3) is higher than that ($\sim$1) at the 
less turbulent E-peak.
Figure 11 displays an HCN-to-HCO$^{+}$ J=3--2 flux ratio in the overall 
bright compact ($\lesssim$3 pc) torus region. 
The flux ratio tends to be systematically higher in the more turbulent 
W-torus than the E-torus, as expected from the enhanced HCN abundance 
scenario.

The observed wider spatial extent of HCN J=3--2 than HCO$^{+}$ 
J=3--2 in the NGC 1068 torus region (Figures 2a,b) can also be a 
natural consequence of the enhanced HCN abundance. 
Because the critical density of HCO$^{+}$ J=3--2 is a factor of $\sim$5 
smaller than that of HCN J=3--2 \citep{shi15}, under the same abundance, 
the HCO$^{+}$ J=3--2 line emission shows higher flux than 
HCN J=3--2 everywhere, and the emission size of HCO$^{+}$ is usually 
larger than HCN at the same J-transition line 
\citep{ima07b,sai15,sai18,ima19}.
Emission from a highly abundant molecule (i.e., HCN) is still 
detectable from less dense and cooler molecular gas outside the 
compact ($\lesssim$3 pc) main torus of NGC 1068 \citep{gar19}.

Summarizing, various observational results described in this section 
consistently suggest an enhanced HCN abundance, relative to HCO$^{+}$, 
in the NGC 1068 torus. 

\subsection{Estimated torus dense molecular mass}

We can estimate HCN J=3--2 and HCO$^{+}$ J=3--2 emission line luminosity 
from their fluxes within the 
14 pc $\times$ 7 pc torus region (Table 4), using the following 
equations \citep{sol05},
\begin{eqnarray}
\left(\frac{L_{line}}{\rm L_{\odot}}\right) = 1.04 \times 10^{-3} \left(\frac{\nu_{rest}}{\rm GHz}\right) (1+z)^{-1} \left(\frac{D_{L}}{\rm Mpc}\right)^2 \left(\frac{S \Delta V}{\rm Jy\ km\ s^{-1}}\right) 
\end{eqnarray}
and
\begin{eqnarray}
\left(\frac{L'_{line}}{\rm K\ km\ s^{-1}\ pc^{2}}\right) = 3.25 \times 10^{7} \left(\frac{\nu_{rest}}{\rm GHz}\right)^{-2} (1+z)^{-1} \left(\frac{D_{L}}{\rm Mpc}\right)^2 \left(\frac{S \Delta V}{\rm Jy\ km\ s^{-1}}\right), 
\end{eqnarray}
where S$\Delta$V is the Gaussian-fit, velocity-integrated line flux 
and D$_{L}$ is the luminosity distance.
Table 5 summarizes the derived luminosity.
Assuming that HCN and HCO$^{+}$ emission in the dense and warm NGC 1068 
torus is almost thermalized at up to J=3 and 
optically thick at J=3--2 and lower J-transitions, then the J=1--0 
luminosity within the same area 
(in K km s$^{-1}$ pc$^{2}$) is expected to be comparable to the J=3--2 
luminosity; namely, HCN J=1--0 and HCO$^{+}$ J=1--0 emission line 
luminosity is 
L$'$$_{\rm HCN\ J=1-0}$ $\sim$ 3.1 $\times$ 10$^{5}$ (K km s$^{-1}$ pc$^{2}$) 
and 
L$'$$_{\rm HCO^+\ J=1-0}$ $\sim$ 2.0 $\times$ 10$^{5}$ (K km s$^{-1}$ pc$^{2}$), 
respectively.
Using the conversion from (optically thick) HCN J=1--0 luminosity 
to dense molecular hydrogen (H$_{2}$) mass, 
M$_{\rm dense-H_2}$(HCN) = 6--12 $\times$ HCN J=1--0 luminosity 
(M$_{\odot}$ [K km s$^{-1}$ pc$^{2}$]$^{-1}$) \citep{gao04,kri08,ler17}, 
we obtain HCN-derived dense H$_{2}$ mass within the 
14 pc $\times$ 7 pc torus region to be 
M$_{\rm dense-H_2}$(HCN) = 2--4 $\times$ 10$^{6}$M$_{\odot}$.
The dense H$_{2}$ mass can also be derived from HCO$^{+}$ data 
to be M$_{\rm dense-H_2}$(HCO$^{+}$) = 0.4--1.0 $\times$ 10$^{6}$M$_{\odot}$, 
where the relation of 
M$_{\rm dense-H_2}$(HCO$^{+}$) = 2--5 $\times$ HCO$^{+}$ J=1--0 
luminosity (M$_{\odot}$ [K km s$^{-1}$ pc$^{2}$]$^{-1}$) in an 
optically thick regime \citep{ler17} is adopted.  
The derived masses are also summarized in Table 5.
These H$_{2}$ masses include the compact ($\lesssim$3 pc) main torus 
component with bright, high-surface-brightness dense molecular line 
emission (Figure 2a,b) and a spatially extended (3--7 pc) 
low-surface-brightness emission component just outside the main torus.

The dense H$_{2}$ mass derived from HCN is a factor of $\sim$4--5 
higher than the one derived from HCO$^{+}$, where a standard HCN-to-H$_{2}$ 
and HCO$^{+}$-to-H$_{2}$ abundance ratio observed in galaxies is implicitly 
assumed. 
The HCN J=3--2 line flux is also a factor of $\sim$1.5 higher than 
the HCO$^{+}$ J=3--2 line flux in all the area-integrated torus spectra 
in Figure 8 (Table 4).
To explain an HCN-to-HCO$^{+}$ J-transition line flux ratio significantly 
higher than unity in the vicinity of a luminous AGN, an 
HCN-to-HCO$^{+}$ {\it abundance} ratio much higher than its {\it flux} ratio 
is required theoretically, because rotational (J) excitation 
is generally more difficult in HCN, due to higher critical density, 
than in HCO$^{+}$ \citep{yam07,izu16}. 
The fact that (1) the HCN-derived 
dense H$_{2}$ mass is a factor of $\sim$4--5 higher than 
the HCO$^{+}$-derived mass and 
(2) the HCN J=3--2 emission line flux 
is a factor of $\sim$1.5 higher than that of HCO$^{+}$ J=3--2 can naturally 
be explained by the enhanced HCN abundance scenario 
\footnote{
If a molecular cloud consists of randomly moving, compact dense 
clumps with a small volume-filling factor \citep{sol87}, 
increased abundance of a certain molecule will result in the 
increase of its emission line flux from a molecular cloud, 
even if individual clumps are optically thick.
This is because the $\tau$=1 sphere in each clump moves outward and 
an area-filling factor of line-emitting regions in the molecular cloud 
will increase \citep{ima07b}.
Practically, there can be a wide spread in a clump's line opacity. 
Clumps which are initially optically thin for a certain molecular 
line under non-enhanced abundance can radiate stronger molecular 
line emission when the abundance of that molecule is enhanced.
It is very likely that HCN and HCO$^{+}$ emission line flux 
increases with increasing abundance in actual molecular clouds 
in galaxies.} 
in the NGC 1068 torus ($\S$4.1).
By considering the suggested HCN overabundance and the resulting 
possible overestimatation of HCN-derived H$_{2}$ mass,  
we adopt the HCO$^{+}$-derived torus dense H$_{2}$ mass of 
M$_{\rm dense-H_2}$(torus) $\sim$ 0.4--1.0 $\times$ 10$^{6}$M$_{\odot}$ 
for our subsequent discussion.  

In this HCO$^{+}$-based torus dense molecular mass estimate, it is 
assumed that HCO$^{+}$ is excited by H$_{2}$ molecules in a wide volume 
number density range \citep{ler17}, 
while it is often assumed that HCO$^{+}$ is primarily excited by 
H$_{\rm 2}$ molecules whose volume number density is as high as  
the critical density.
If we simply regard that the HCO$^{+}$ J=3--2 emission preferentially probes 
dense molecular gas with $\sim$10$^{6}$ cm$^{-3}$, close to the 
HCO$^{+}$ J=3--2 critical density \citep{shi15}, the emitting size of 
the 0.4--1.0 $\times$ 10$^{6}$M$_{\odot}$ gas is estimated be $\sim$3--6 pc, 
where a spherical distribution with a dense molecular clump volume-filling 
factor of $\sim$0.01--0.1 \citep{wad07,wad12} is assumed 
(see also $\S$4.5).
This estimated size is roughly comparable to the observed size of the 
high-surface-brightness HCO$^{+}$ J=3--2 emission in the main torus 
component (Figure 2b).
This independent, simple comparison also suggests that the estimated 
torus dense molecular mass is reasonable. 
Our adopted torus mass of 
M$_{\rm dense-H_2}$(torus) $\sim$ 0.4--1.0 $\times$ 10$^{6}$M$_{\odot}$ 
also roughly agrees with the recent updated estimate by \citet{gar19} 
based on high-spatial-resolution (0$\farcs$03--0$\farcs$09) ALMA 
data ($\sim$0.3 $\times$ 10$^{6}$M$_{\odot}$), 
and is significantly smaller than the central SMBH mass with 
M$_{\rm SMBH}$ $\sim$ 1 $\times$ 10$^{7}$M$_{\odot}$ \citep{gre96,hur02,lod03}.

\subsection{Redshifted dense molecular emission from the innermost 
northwestern torus} 

The moment 1 maps of HCN J=3--2 and HCO$^{+}$ J=3--2 in Figure 3a,b 
show a redshifted, high-velocity emission component 
at the innermost ($\lesssim$1 pc) part of the northwestern torus, with this 
signature being much clearer for HCN J=3--2 than HCO$^{+}$ J=3--2.
Redshifted $\sim$22 GHz ($\sim$1.4 cm) H$_{2}$O maser emission was also 
detected with VLBI very-high-spatial-resolution observations, 
at 0.6--1.1 pc on the northwestern side of the mass-accreting SMBH 
\citep{gre96,gre97}.
It is interesting to compare our ALMA-detected, redshifted dense 
molecular line emission with the VLBI-detected H$_{2}$O maser emission. 

The $\sim$22 GHz ($\sim$1.4 cm) H$_{2}$O maser emission shows a declining 
rotation curve, indicative of an almost edge-on rotating disk, and 
is fitted with a slightly sub-Keplerian motion of V $\propto$ r$^{-0.31}$, 
with the estimated central SMBH mass of M$_{\rm SMBH}$ $\sim$ 
1 $\times$ 10$^{7}$M$_{\odot}$ \citep{gre96,gre97}, where V is velocity and 
r is the distance from the central mass-dominating SMBH.
Figure 12 displays velocity properties of the HCN J=3--2 and 
HCO$^{+}$ J=3--2 emission lines along the northwest direction from 
the SMBH, overplotted with 
(1) Keplerian rotation (V $\propto$ r$^{-0.5}$) dominated by a point 
source with mass of 1 $\times$ 10$^{7}$M$_{\odot}$, and 
(2) the slightly sub-Keplerian motion (V $\propto$ r$^{-0.31}$) 
derived from the H$_{2}$O maser emission \citep{gre97}. 
Our ALMA data demonstrate that HCN J=3--2 and HCO$^{+}$ J=3--2 emission 
line dynamics roughly agrees with the VLBI-detected H$_{2}$O maser dynamics 
at $<$0.7 pc.
It is thus reasonable to interpret that H$_{2}$O, HCN J=3--2, and 
HCO$^{+}$ J=3--2 emission lines probe the same spatial and dynamical 
gas components at the innermost northwestern torus.
  
The corresponding blueshifted emission components at the innermost 
southeastern torus, expected from (sub-)Keplerian rotation, are much 
weaker both for HCN J=3--2 and HCO$^{+}$ J=3--2 lines (Figure 3a,b).
This is also the case for H$_{2}$O maser emission \citep{gre96,gre97}.
In the spectra of the torus (14 pc $\times$ 7 pc) and western torus 
(7 pc $\times$ 7 pc) in Figure 8, an emission excess (compared with 
the single Gaussian fit) is seen at the redshifted part, at 
V $\sim$ 1300--1500 km s$^{-1}$ (V$_{\rm sys}$ $+$ [170--370] km s$^{-1}$) 
in both HCN J=3--2 and HCO$^{+}$ J=3--2.
This is most likely originating in the dynamically decoupled, 
redshifted high velocity dense molecular line emission from the 
innermost northwestern torus (Figure 3a,b).
A similarly strong excess emission at V $\sim$ 760--960 km s$^{-1}$ 
(V$_{\rm sys}$ $-$ [170--370] km s$^{-1}$) is not clearly seen in the 
spectra at the eastern and whole torus (Figure 8).
This could be understood if the amount of blueshifted HCN, HCO$^{+}$, and 
H$_{2}$O molecules at the innermost southeastern torus is smaller than 
that of redshifted molecules at the innermost northwestern torus, 
because it seems very unlikely that all of these molecular emission  
lines at 22--260 GHz (1.2--14 mm) are strongly flux attenuated by dust 
extinction, free-free absorption, and/or line opacity ($\S$4.1) 
only at the innermost southeastern side.

Figure 13 shows position velocity diagrams of HCN J=3--2 and 
HCO$^{+}$ J=3--2 lines along the position angle of 
PA = 115$^{\circ}$ east of north.
This position angle is in between the 
innermost ($\lesssim$0.7 pc) dense molecular line emission with 
PA $\sim$ 135$^{\circ}$ and outer torus (2--5 pc) emission with 
PA = 105--110$^{\circ}$ \citep{ima18a}.
Redshifted HCN J=3--2 emission with V $\sim$ 1300--1500 km s$^{-1}$ 
(V$_{\rm sys}$ $+$ [170--370] km s$^{-1}$) is clearly detected at the 
innermost northwestern torus (right side of the SMBH in Figure 13a), 
while the corresponding blueshifted HCN J=3--2 emission with 
V $\sim$ 760--960 km s$^{-1}$ (V$_{\rm sys}$ $-$ [170--370] km s$^{-1}$) 
at the innermost southeastern torus is much less clear.
This reinforces the above scenario that dense molecular gas distribution 
at the innermost torus, just outside the mass-accreting 
SMBH, is asymmetric in such a way that there is much more gas at the 
northwestern side than at the southeastern side.

Summarizing, we suggest that the spatial distribution of 
H$_{2}$O, HCN, and HCO$^{+}$ at the innermost ($\lesssim$1 pc) torus 
is asymmetric in that they are found in larger amounts on the 
northwestern side. 
The dynamics of these northwestern molecular emission follows almost 
Keplerian rotation at $\lesssim$0.7 pc.

\subsection{Dynamical properties of the torus}

While the redshifted emission component at the northwestern 
torus roughly follows an almost-Keplerian rotation at the 
innermost part at $\lesssim$0.7 pc, it starts to deviate from that, 
rotating far below the Keplerian at $\gtrsim$0.7 pc (Figure 12).
It was found that the outer part of the northwestern torus 
at 2--5 pc showed blueshifted dense molecular emission, 
contrary to the innermost redshifted emission 
\citep{ima18a,imp19}. 
We perform a detailed dynamical investigation of the dense molecular 
emission in the W- and E-torus separately, because the emission properties 
are highly asymmetric.

Figure 14 plots the peak velocity and velocity width of 
HCN J=3--2 and HCO$^{+}$ J=3--2 emission, based on the Gaussian fit 
of beam-sized spectra extracted at various distances from the central 
SMBH along PA = 115$^{\circ}$.
At 2--6 pc, the dense molecular emission is blueshifted and redshifted, 
with respect to the systemic, in the northwestern and southeastern 
torus, respectively (Figure 14a).
These velocity patterns are the opposite of those at the innermost 
($\lesssim$0.7 pc) torus, both in the west and east, supporting the 
presence of counter-rotating dense molecular gas \citep{imp19}. 

The suggested counter-rotating dense molecular gas can also naturally 
explain the positional shift of the highest velocity dispersion 
spot in the northwestern torus in the moment 2 map, relative to the 
innermost redshifted high-velocity (V $\gtrsim$ 1300 km s$^{-1}$ 
or V$_{\rm sys}$ $+$ [$\gtrsim$170] km s$^{-1}$)
line-emitting region; the former is located just outside the latter 
(Figure 3). 
This is also confirmed in Figures 14a and 14b.
In a counter-rotating molecular gas, the highest velocity dispersion 
value can be observed at the border between the innermost redshifted 
and outer blueshifted components, owing to beam dilution and/or 
increased turbulence created through instability.

The implied rotation velocity of outer ($\gtrsim$2 pc) 
torus dense molecular gas looks much slower than Keplerian motion 
governed by the central SMBH mass (Figure 14a), as seen in previously 
taken ALMA data \citep{gar16,ima18a}.
The intrinsic rotation velocity can be larger than the observed 
one by a factor of 1/sin({\it i}), where {\it i} is an 
inclination angle ({\it i} = 90$^{\circ}$ for an edge-on view). 
As described in $\S$3, since NGC 1068 is a type-2 AGN where 
the central mass-accreting SMBH is believed to be obscured by 
a dusty molecular torus, it is very likely that the torus 
inclination is {\it i} $\sim$ 90$^{\circ}$ (edge-on), rather than 
{\it i} $\sim$ 0$^{\circ}$ (face-on).
In fact, recent high-spatial-resolution ($<$0$\farcs$1) observations 
in the infrared--(sub)millimeter wavelength region have consistently 
supported a nearly edge-on ({\it i} $>$ 60$^{\circ}$) view of the 
NGC 1068 torus \citep{gar19,gra20,lop20}.
\citet{gar16} estimated that the torus inclination is 
{\it i} $\sim$ 90$^{\circ}$ at the innermost part at $\lesssim$1 pc,  
but {\it i} = 34--66$^{\circ}$ at the outer part at $\gtrsim$2--4 pc.
Even adopting the small value of {\it i} = 34--66$^{\circ}$, 
the intrinsic rotation velocity can be higher than the observed one 
only by a factor of 1/sin({\it i}) = 1.1--1.8 and is still significantly 
slower than the Keplerian rotation.

Summarizing, based on our ALMA data, we prefer a scenario in which 
an outer ($\gtrsim$2 pc) dense molecular gas is rotating slowly 
(much slower than Keplerian) and counter-rotating with respect 
to the innermost dense gas component.
This slowly rotating outer ($\gtrsim$2 pc) torus scenario is different 
from the argument of almost Keplerian rotation there by \citet{imp19}.

\subsection{Vibrationally excited dense molecular emission lines}

A luminous AGN can emit a strong mid-infrared (3--30 $\mu$m) continuum 
due to AGN-heated hot ($>$100 K) dust.
HCN molecules can be vibrationally excited to the v$_{2}$=1 level 
($>$1000 K) by absorbing mid-infrared $\sim$14 $\mu$m photons 
\citep{aal95,ran11}.
In fact, an HCN v$_{2}$=1, l=1f (HCN-VIB) J=3--2 emission line 
was detected in several luminous buried AGNs, surrounded by 
a large column density of dense molecular gas
\citep{sak10,aal15a,aal15b,cos15,mart16,ima16b}.
NGC 1068 contains a luminous AGN 
(L$_{\rm AGN}$ $>$ 4 $\times$ 10$^{44}$ erg s$^{-1}$) \citep{boc00,gra20}, 
surrounded by Compton-thick (N$_{\rm H}$ $>$10$^{24}$ cm$^{-2}$) obscuring 
material \citep{koy89,uen94,mat97,bau15,mari16,zai20}.
Hence, it is quite possible that the HCN-VIB J=3--2 emission line 
is strongly radiated in the NGC 1068 torus where dense molecular 
gas is illuminated by the AGN-origin mid-infrared $\sim$14 $\mu$m photons.
However, this HCN-VIB J=3--2 emission line has not been clearly detected 
in NGC 1068 at the systemic velocity (V$_{\rm sys}$ = 1130 km s$^{-1}$) 
in previously taken ALMA spectra \citep{ima16a,ima18a}.

In Figure 6b, an absorption feature is seen at 
V $\sim$ 1400--1700 km s$^{-1}$ 
(V$_{\rm sys}$ $+$ [270--570] km s$^{-1}$) 
in the HCO$^{+}$ J=3--2 line spectrum.
If this originates in HCO$^{+}$ J=3--2 absorption, the redshifted 
$\gtrsim$270 km s$^{-1}$ velocity seems 
too large for infalling gas, unless it is located very close to the 
central SMBH and is strongly affected by its gravitational force. 
In such close proximity to a luminous AGN, HCN abundance is expected 
to be higher than that of HCO$^{+}$, owing to AGN radiation and/or 
mechanical effects ($\S$4.1),  
but we see no clear absorption signature at V $\sim$ 1400--1700 km s$^{-1}$ 
(V$_{\rm sys}$ $+$ [270--570] km s$^{-1}$) 
for plausibly more abundant HCN (Figure 6a). 
It seems unlikely that infalling HCO$^{+}$ gas is the origin 
of the detected V $\sim$ 1400--1700 km s$^{-1}$ absorption feature. 
The frequency of this absorption feature coincides with the HCN-VIB J=3--2 
line at V$_{\rm sys}$ = 1130 km s$^{-1}$ (Figure 4a).
We pursue this possibility. 
In fact, an HCN-VIB absorption feature at J=4--3 has recently been detected 
with ALMA in a nearby bright obscured AGN, NGC 1052 \citep{kam20}.
Applying a Gaussian fit to this absorption feature 
(velocity integrated flux $=$ $-$130$\pm$32 mJy beam$^{-1}$ km s$^{-1}$) 
and adopting the continuum flux level of 6.5 (mJy beam$^{-1}$) 
in data-a (Table 2), we obtain the absorption equivalent width 
to be EW $\sim$ 2 $\times$ 10$^{6}$ (cm s$^{-1}$).
We can convert this EW to the absorbing column density using the formula   
\begin{eqnarray}
EW = \frac{\lambda^{3} A_{ul} g_{u}}{8 \pi g_{l}} \times N 
\end{eqnarray}
\citep{ryb79,gon14}, where 
EW is in units of (cm s$^{-1}$); 
$\lambda$ is wavelength (in cm); 
A$_{ul}$ is the Einstein A coefficient for spontaneous emission from the
upper (u) level to the lower (l) level in units of (s$^{-1}$); 
g$_{u}$ and g$_{l}$ are the statistical weights (= 2J $+$ 1) at the upper 
and lower levels, respectively; and N is the column density in the lower
energy level, in units of (cm$^{-2}$).
The A$_{ul}$ value for HCN-VIB J=3--2 is 7.3 $\times$ 10$^{-4}$ (s$^{-1}$)  
at Splatalogue \citep{mul05}. 
We obtain the column density of vibrationally excited HCN at J=2 to be 
N$_{\rm HCN-VIB\ J=2}$ $\sim$ 3.5 $\times$ 10$^{13}$ cm$^{-2}$.
Even if we adopt an enhanced HCN-to-H$_{2}$ abundance ratio of 
$\sim$10$^{-8}$ ($\S$4.1),
we obtain the column density of H$_{2}$ co-existing with 
vibrationally excited HCN to be 
as high as N$_{\rm H2}$ $\sim$ a few $\times$ 10$^{21}$ cm$^{-2}$.
This is a lower limit because vibrationally excited HCN can be populated 
at other J-transitions.  
If the central luminous AGN is surrounded by this column density of 
mid-infrared-pumped molecular gas in all torus directions, we may expect 
to detect the HCN-VIB line in {\it emission}.
We thus search for this emission signature in our sensitive ALMA 
spectrum.

In the W-peak spectrum in Figure 4c, at the lower frequency side of 
the HCO$^{+}$ J=3--2 emission line, we see a distinct emission line 
at $\nu_{\rm obs}$ $\sim$ 266.0 GHz (denoted as ``HCN-VIB J=3--2 (red)''), 
which corresponds to the HCN-VIB J=3--2 line at V $\sim$ 1340 km s$^{-1}$
(V$_{\rm sys}$ $+$ 210 km s$^{-1}$).
A Gaussian fit to this line is shown in Figure 5e. 
The estimated peak velocity and velocity-integrated emission line 
flux are 1339$\pm$7 (km s$^{-1}$) and 0.041$\pm$0.009 (Jy km s$^{-1}$), 
respectively (Table 3). 
Using Equations (1) and (2), 
the HCN-VIB J=3--2 emission line luminosity is estimated to be 
L$_{\rm HCN-VIB\ J=3-2}$ $\sim$ 2.2 (L$_{\odot}$) or 
L$'$$_{\rm HCN-VIB\ J=3-2}$ $\sim$ 3.6 $\times$10$^{3}$ (K km s$^{-1}$ pc$^{2}$).
Since the HCN-VIB J=3--2 emission line is detected clearly only in the 
beam-sized W-peak spectrum, we use the luminosity as that from the NGC 1068 
torus.
The HCN-VIB J=3--2 emission line luminosity is about two orders of 
magnitude smaller than that of the HCN J=3--2 emission line from the 
whole torus region ($\S$4.2).
This $\sim$1\% HCN-VIB-to-HCN luminosity ratio is much smaller than 
several HCN-VIB-line-detected 
buried AGNs (and such candidates) in nearby luminous infrared galaxies, 
where the HCN-VIB-to-HCN J=3--2 luminosity ratios are $>$5\% 
\citep{sak10,aal15a,aal15b,cos15,mart16,ima16b}.

In NGC 1068, since no directly transmitted X-ray emission is seen 
even at $\sim$60--100 keV \citep{mat97,bau15,mari16}, the absorbing column 
density toward the X-ray emitting region around the central luminous 
AGN is likely to be as high as 
N$_{\rm H}$ $\gtrsim$10$^{25}$ cm$^{-2}$ (i.e., heavily Compton thick).
If the central luminous AGN is surrounded by material with this 
high column density in virtually all directions, HCN-VIB J=3--2 emission 
line luminosity can be high, relative to the HCN J=3--2 emission line 
luminosity, through the greenhouse effect \citep{gon19}.
It may be that the volume- (and area-) filling factor of dense absorbing 
gas clumps in the NGC 1068 torus is substantially smaller than the 
HCN-VIB-line-detected infrared luminous galaxies, which are mostly 
major mergers.
The physical scale of the mid-infrared hot dust-emitting region around 
a luminous AGN (larger than the innermost torus dust sublimation radius; 
$\gtrsim$0.1 pc for L$_{\rm AGN}$ $>$ 4 $\times$ 10$^{44}$ [erg s$^{-1}$] 
\citep{bar87}) is much larger than that of the X-ray emitting region 
($\lesssim$10 Schwarzschild radius or $\lesssim$10$^{-5}$ pc for 
the 1 $\times$ 10$^{7}$M$_{\odot}$ SMBH).
Thus, it is possible that the X-ray emission is completely blocked 
by foreground dense clumps, while a large fraction of the mid-infrared 
emission can escape without being absorbed by the surrounding 
obscuring material.
This could explain the observed low HCN-VIB-to-HCN J=3--2 emission line 
luminosity ratio in the NGC 1068 torus. 
In fact, the mid-infrared 3--13 $\mu$m spectra toward the NGC 1068 
nucleus, excluding the bright ring-shaped ($\sim$10--15$''$ radius) 
starburst ring emission, show modestly strong 3.4 $\mu$m carbonaceous 
dust and 9.7 $\mu$m silicate dust absorption features 
\citep{roc91,ima97,rhe06,mas06,geb09}. However, their optical depths are 
much smaller than several buried AGNs, whose mid-infrared continua are 
thought to be almost fully blocked by the foreground obscuring dust 
\citep[e.g.,][]{imd00,ima06a,lev07,spo07,ima07a,ima08,ima10}, 
suggesting a large fraction of leaked mid-infrared continuum emission 
through the NGC 1068 torus.

It is reasonable that the HCN-VIB J=3--2 emission is detected as 
the redshifted (V $\gtrsim$ 1300 km s$^{-1}$ or V$_{\rm sys}$ $+$ 
[$\gtrsim$170] km s$^{-1}$) component in the beam-sized 
W-peak spectrum (Figure 4c) for two reasons. 
First, the amount of HCN is larger at the redshifted innermost northwestern 
torus than it is at the blueshifted innermost southeastern torus ($\S$4.3). 
Secondly, there should be plenty of mid-infrared 
$\sim$14 $\mu$m photons that were emitted from AGN-heated hot 
dust grains. 
Both of these conditions can naturally produce strong HCN-VIB J=3--2 
emission from the redshifted HCN gas at the innermost northwestern 
torus, which is covered in the beam-sized W-peak spectrum, but not in the 
beam-sized E-peak spectrum.
We create a moment 0 map of HCN-VIB J=3--2 line, by integrating signals 
of redshifted components (V = 1170--1430 km s$^{-1}$ 
or V$_{\rm sys}$ $+$ [40--300] km s$^{-1}$), which 
is shown in Figure 15.
The HCN-VIB J=3--2 emission line is detected with $\sim$4.8$\sigma$ 
at the northwestern side of the C-peak (close to the W-peak), supporting 
the scenario in which the HCN-VIB J=3--2 emission comes from the redshifted 
dense molecular gas at the innermost northwestern torus.

No clear absorption feature is seen at the expected frequency 
of HCO$^{+}$-VIB J=3--2 line for V$_{\rm sys}$ = 1130 km s$^{-1}$ 
in the spectrum toward the C-peak (Figure 4a).  
In the W-peak spectrum (Figure 4c), some emission line signature 
is seen at the frequency of the HCO$^{+}$-VIB J=3--2 line with 
V $\sim$ 1340 km s$^{-1}$ (V$_{\rm sys}$ $+$ 210 km s$^{-1}$; 
denoted as ``HCO$^{+}$-VIB J=3--2 (red)'').
If this line identification is correct, this may be the first detection of 
an HCO$^{+}$-VIB J=3--2 line in {\it emission} in an external galaxy. 
Only HCO$^{+}$-VIB line detection in {\it absorption} has been 
reported at J=4--3 in an obscured AGN \citep{kam20}.
HCO$^{+}$ can be vibrationally excited to the v$_{2}$=1 level by 
absorbing mid-infrared $\sim$12 $\mu$m photons. 
Since there is plenty of AGN-heated hot dust at the innermost 
northwestern torus, strong HCO$^{+}$-VIB J=3--2 emission may be produced. 
However, our Gaussian fit of this possible emission line provides 
detection significance of $<$3$\sigma$. 
Additionally, the observed peak frequency of this possible emission 
line corresponds also to HOC$^{+}$ J=3--2 ($\nu_{\rm rest}$ = 268.451 GHz) 
at V$_{\rm sys}$ $\sim$ 1130 km s$^{-1}$ (Figure 4c).
This HOC$^{+}$ J=3--2 emission line can be strong in molecular gas 
illuminated by strong X-ray radiation, such as an AGN torus  
\citep{use04,qiu20}, and was actually detected in a few luminous 
obscured AGNs and such candidates \citep{aal15a,aal15b,ima16b,ala18}. 
The HCO$^{+}$-VIB J=3--2 emission line is still elusive, even in 
the nearby luminous AGN, NGC 1068. 
The detection of HCN-VIB line and no clear detection of HCO$^{+}$ 
line are consistent with the enhanced HCN abundance scenario in the torus 
(discussed in $\S$4.1).

HNC can also be vibrationally excited to the v$_{2}$=1 level by absorbing 
mid-infrared $\sim$21.5 $\mu$m photons, but we see neither a clear emission 
nor an absorption signature of the HNC-VIB J=3--2 line in torus spectra 
(Figure 4g,h,i). 
The HNC J=3--2 emission from the torus is much weaker than 
HCN J=3--2 and HCO$^{+}$ J=3--2 emission (Figure 4), suggesting a low 
HNC abundance there.
The temperature of molecular gas and dust in the torus in the 
vicinity of a luminous AGN is higher than that in the 
general inter-stellar medium in the host galaxy. 
Since an HNC abundance is known to decrease in Galactic high-temperature 
molecular clouds around active regions \citep{sch92,hir98,gra14}, 
this may be the case also for the NGC 1068 torus.
Under low HNC abundance, the HNC-VIB J=3--2 emission line can be weak.

Summarizing, HCN-VIB J=3--2 line signatures are seen in absorption 
in the beam-sized spectrum toward the C-peak and in emission 
originating from the redshifted molecular gas at the innermost 
northwestern torus (detected in the beam-sized W-peak spectrum).
The observed small HCN-VIB-to-HCN J=3--2 emission line flux ratio 
suggests that the NGC 1068 torus is highly clumpy, and a large fraction 
of mid-infrared $\sim$14 $\mu$m photons emitted from AGN-heated hot 
dust grains escape without being absorbed by HCN molecules in the torus.
Absorption and emission signatures of the HCO$^{+}$-VIB J=3--2 
and HNC-VIB J=3--2 lines are much less clear.

\subsection{Molecular emission in the eastern torus}

Although velocity-integrated emission line fluxes of both HCN J=3--2 
and HCO$^{+}$ J=3--2 are significantly higher at the W-peak 
than at the E-peak in moment 0 maps (Figure 2a,b) and beam-sized spectra 
(Table 3, column 9), those within the 7 pc $\times $7 pc 
(0$\farcs$1 $\times$ 0$\farcs$1) region are higher in the E-torus 
than in the W-torus (Table 4, column 6), suggesting that 
a larger amount of spatially extended 
low-surface-brightness molecular line emission outside the compact 
($\lesssim$3 pc) main torus is present in the E-torus. 
In fact, in the moment 0 and 1 maps of HCN J=3--2, we recognize 
spatially extended, low-surface-brightness emission outside 
the main E-torus as far east as R.A. = 02$^{\rm h}$ 42$^{\rm m}$ 40.714$^{\rm s}$ 
(Figure 2a,c).
\citet{ima16a,ima18a} also marginally detected 
signatures of bridging HCN J=3--2 emission between the E-torus and 
massive molecular clouds in the host galaxy, at the eastern side of 
the E-torus.
\citet{mul09} also argued the presence of molecular gas falling 
toward the E-torus from the eastern side, based on near-infrared 
2.2 $\mu$m observations of very hot ($>$1000 K) molecular gas 
(but see \citet{bar14,may17,gar19}). 

Molecular gas in the host galaxy at $\sim$50--150 pc 
on the eastern side of the torus is located on a far side \citep{gar19} and 
shows blueshifted (V $\sim$ 1000--1100 km s$^{-1}$ 
or V$_{\rm sys}$ $-$ [30--130] km s$^{-1}$) emission
\citep{sch00,kri11,gar14,gar16,ima16a,ima18a,gar19}. 
The velocity of the dense molecular line emission just on the eastern side 
of the E-torus (V $\sim$ 1100--1130 km s$^{-1}$ 
or V$_{\rm sys}$ $-$ [0--30] km s$^{-1}$
at $\sim$5 pc from the SMBH; 
R.A. = 02$^{\rm h}$ 42$^{\rm m}$ 40.714$^{\rm s}$ shown as a black upward 
arrow in Figure 2c) is between that of the eastern host galaxy 
(V $\sim$ 1000--1100 km s$^{-1}$ 
or V$_{\rm sys}$ $-$ [30--130] km s$^{-1}$) and that of the 
redshifted, outer portion of the compact main E-torus 
(V $\sim$ 1140--1150 km s$^{-1}$ or V$_{\rm sys}$ $+$ [10--20] km s$^{-1}$ 
at 2--3 pc on the eastern side of the SMBH; 
R.A. = 02$^{\rm h}$ 42$^{\rm m}$ 40.712$^{\rm s}$ indicated with the 
leftmost red downward arrow in Figure 2c).  
Figure 16 illustrates the above geometry and dynamics.
Velocity at the southeastern torus along PA = 115$^{\circ}$ 
also displays a sudden change at $\sim$7 pc from inner redshifted motion  
to outer blueshifted motion (Figure 14). 
This bridging emission at 5--7 pc east of the SMBH could be 
related to molecular gas being transported from the eastern host galaxy 
(far side) to the E-torus (front side) and may contribute to 
an increase in the dense molecular line flux in the 7 pc $\times $7 pc 
area-integrated E-torus spectrum.

\subsection{Outflowing dense molecular gas}

We now investigate the signatures of molecular outflow from 
the NGC 1068 nucleus, whose detection was reported 
in several papers \citep{gar14,gal16,imp19,gar19}. 
Searches for massive molecular outflows have usually targeted 
the presence of a very broad emission line wing component, in addition to 
a main molecular emission component determined by galaxy 
dynamics \citep[e.g.,][]{mai12,cic14,gar15,ima17,per18,lut20}.
In the molecular line spectra from the torus (Figure 8), 
besides the redshifted emission excess most likely originating from 
the redshifted high velocity component at the innermost northwestern 
torus ($\S$4.3), a broad emission component is not clearly 
detected.

We instead investigate the beam-sized spectra at the C-peak, 
which show absorption features.
The velocity profile of the HCN J=3--2 line in Figure 6a shows 
a broad absorption tail at the blueshifted 
side (V = 700--1100 km s$^{-1}$ or V$_{\rm sys}$ $-$ [30--430] km s$^{-1}$). 
This can be explained by outflow activity \citep{imp19}.
A similar blueshifted absorption tail is much weaker in the 
HCO$^{+}$ J=3--2 line spectrum in Figure 6b.
Since the beam-size ($\sim$0$\farcs$02 $\times$ 0$\farcs$02) 
is not sufficiently smaller than the separation between the central 
SMBH position and the innermost dense molecular emission from the 
E- and W-torus, the beam-sized spectra at the C-peak (Figure 6) can 
be contaminated by emission from the innermost torus.
We attempt to correct for this contamination by averaging the spectra 
at the 0$\farcs$02 northeast 
and southwest sides of the SMBH (i.e., perpendicular to the 
northwest- and southeast-oriented innermost torus emission) 
and subtracting the average from the observed spectrum at the C-peak, 
following \citet{imp19}.
The corrected HCN J=3--2 and HCO$^{+}$ J=3--2 line spectra 
at the C-peak are displayed in Figure 6 (red dotted line).
For HCN J=3--2, the peak absorption depth in the corrected spectrum 
is slightly smaller than that estimated by \citet{imp19}, but this 
difference can be caused by different beam patterns resulting in
different correction degrees of the contamination. 
What is more important is that the blueshifted absorption tail of 
HCN J=3--2 becomes clearer, and that of HCO$^{+}$ J=3--2 is also 
seen, supporting the presence of dense molecular outflow coming 
toward us from the $\sim$260-GHz-continuum-emission-peak position 
around the mass-accreting SMBH.  
We interpret that this outflow is still confined to the compact 
nuclear region of NGC 1068, because an enhanced HCN abundance 
affected by a luminous AGN ($\S$4.1) can naturally explain 
an outflow feature in HCN J=3--2 that is stronger than in HCO$^{+}$ J=3--2.
No signature of an outflow-origin blueshifted absorption tail is 
seen in HNC J=3--2 (Figure 4g), as expected from low HNC abundance 
in molecular gas in the vicinity of the AGN ($\S$4.5). 
 
We apply a Gaussian fit to the broad blueshifted HCN J=3--2 
absorption wing in the raw beam-sized spectrum at the C-peak 
in Figure 6a, and obtain the central velocity of 
991$\pm$27 (km s$^{-1}$), FWHM = 295$\pm$45 (km s$^{-1}$), 
peak absorption dip of $-$0.51$\pm$0.08 (mJy beam$^{-1}$), and 
velocity integrated flux of $-$162$\pm$35 (mJy beam$^{-1}$ km s$^{-1}$). 
Adopting the continuum flux of 6.5 (mJy beam$^{-1}$) (Table 2), 
we obtain an absorption dip of $\sim$8\% ($\tau$ $\sim$ 0.08; i.e., 
optically thin) and absorption equivalent width of 
EW$_{\rm HCN32\ outflow}$ = 2.5$\pm$0.5 $\times$ 10$^{6}$ (cm s$^{-1}$). 
Using Equation (3), we estimate the HCN J=2 column density of the 
outflow to be 3.7$\pm$0.8 $\times$ 10$^{13}$ (cm$^{-2}$), where 
A$_{\rm ul}$ = 8.4 $\times$ 10$^{-4}$ (s$^{-1}$) is adopted for HCN J=3--2 
\citep{mul05}. 
If these outflowing HCN molecules spatially distribute in the beam size 
(0$\farcs$019 $\times$ 0$\farcs$017 or 1.3 pc $\times$ 1.2 pc; Table 2), 
the derived column density corresponds to (44$\pm$10) $\times$ 10$^{49}$ 
HCN molecules, or even larger if we take into account HCN at other 
J-transitions.
Assuming a high HCN-to-H$_{2}$ abundance ratio of 10$^{-8}$ ($\S$4.1) because 
the outflow is supposed to be confined to the nuclear region, we obtain the 
outflowing dense H$_{2}$ mass to be M$_{\rm outflow}$ $\gtrsim$ 75M$_{\odot}$. 
This mass corresponds to outflow within the 1.3 pc $\times$ 1.2 pc 
area in front of the $\sim$260-GHz-continuum-emitting region. 
If the outflowing gas is clumpy and the area-filling factor of clumps, 
which show the absorption line, is 50\% (10\%), each clump should show an 
absorption dip of $\sim$16\% ($\sim$80\%) or 
$\tau$ $\sim$ 0.17 ($\tau$ $\sim$ 1.6), respectively, where continuum flux 
attenuation by dust extinction is neglected at $\sim$260 GHz ($\sim$1.2 mm). 
For a very small area-filling factor ($<<$100\%), the clump will be 
in an optically thick non-linear regime for the absorption line strength, 
so the outflow mass will increase.
Furthermore, if the outflow is spatially more extended in the transverse 
direction than the 1.3 pc $\times$ 1.2 pc area, the actual outflow mass 
will be even larger.
Adopting the outflow velocity of V$_{\rm outflow}$ $\sim$ 140 (km s$^{-1}$)
(V$_{\rm sys}$ $-$ 991 km s$^{-1}$), 
the outflow energy is estimated to be E$_{\rm outflow}$ $\gtrsim$ 1.4 $\times$ 
10$^{42}$ (J).

Such an energetic outflow can disturb the dynamics of the innermost torus 
molecular gas and could reproduce a dynamically decoupled innermost 
($\lesssim$0.7 pc) emission component from the outer ($\gtrsim$2 pc) 
torus emission, as observed in Figure 3a,b. 
In fact, \citet{gar19} prefer an outflow scenario, rather than the 
counter-rotating molecular gas scenario, to explain their obtained 
$\gtrsim$0$\farcs$03-resolution molecular gas dynamics of 
the NGC 1068 torus.
However, the dynamical similarity between the innermost HCN J=3--2 
and almost-Keplerian rotating H$_{2}$O maser emission ($\S$4.3) 
leads us to prefer the counter-rotating molecular gas scenario 
as the origin of the redshifted dense molecular emission from the 
innermost northwestern torus, because there is no strong reason 
why the dynamical properties of the H$_{2}$O maser rotating 
disk and outflowing molecular gas agree.  

\subsection{Schematic picture of the NGC 1068 torus}

After considering all observational results, we provide a 
schematic picture of dense molecular gas in the NGC 1068 torus 
in Figure 17.
There are two components of dense molecular gas that are counter-rotating
(i.e., an innermost [$\lesssim$0.7 pc] component characterized by 
almost-Keplerian rotation and an outer [$\gtrsim$2 pc] component 
characterized by slow rotation [much slower than Keplerian rotation]).
Outflowing dense molecular gas from the central mass-accreting SMBH 
in our direction and redshifted high velocity (V = 1300--1500 km s$^{-1}$ 
or V$_{\rm sys}$ $+$ [170--370] km s$^{-1}$)
HCN-VIB J=3--2, HCN J=3--2, and HCO$^{+}$ J=3--2 emission originating 
in the innermost northwestern torus are present.

Counter-rotating dense molecular gas in the compact torus can be 
quickly relaxed dynamically \citep{gar19}.
To produce such a transient, short-time-scale counter-rotation 
within the $\lesssim$7 pc torus scale, some external force seems 
required.
We propose a simple scenario that a
compact massive gas clump fell into the pre-existing torus 
and collided with the western torus from the far side.
Such an infalling gas clump may originate in minor galaxy merger 
\citep{fur16,tan17} or streaming motion \citep{mul09}.
The western part of the pre-existing torus had been redshifted 
before the collision, as seen in the innermost ($\lesssim$0.7 pc) 
H$_{2}$O maser dynamics.
The collision changed the dynamics of the bulk of the western 
torus to the blueshifted motion. 
The observed rotation, much slower than the Keplerian at $\gtrsim$2 pc, 
could be explained in this transient phase.
Dense molecular gas at the innermost ($\lesssim$0.7 pc) northwestern torus 
could remain redshifted if it is gravitationally affected more strongly 
by the central SMBH mass than by the outer collision-induced 
turbulence.
Such a colliding clump could make significant turbulence in the 
W-torus and explain the fact that the observed dense molecular emission 
line widths are larger
in the W-torus than in the E-torus (Figure 8 and Table 4).
This may also be related to the asymmetric dense molecular gas 
distribution at the innermost torus, where there is a larger amount of 
gas on the W-torus side ($\S$4.3).

In order for this scenario to work quantitatively, the angular 
momentum of the colliding compact massive gas clump must be 
as large as that of the pre-existing torus.
Since the pre-existing torus mass can be as high as 
$\gtrsim$a few $\times$ 10$^{5}$M$_{\odot}$ ($\S$4.2), 
the clump should be as massive as $\gtrsim$a few $\times$ 10$^{5}$M$_{\odot}$, 
in the case that the infalling velocity is comparable to the 
pre-collisional torus rotation velocity at the impact point.
In our Galaxy and nearby galaxies, the bulk of giant molecular 
clouds (GMCs) are as massive as $\gtrsim$a few $\times$ 10$^{5}$M$_{\odot}$ 
\citep{sol87,fuk10}. 
Recent ALMA high-spatial-resolution observations have revealed the 
presence of a very compact, massive GMC 
($>$5 $\times$ 10$^{6}$M$_{\odot}$ within $<$25 pc) \citep{joh15}.
Assuming the same gas volume number density, a GMC with 
$\gtrsim$a few $\times$ 10$^{5}$M$_{\odot}$ can be as small as $\lesssim$8 pc. 
If we adopt the relation between SMBH mass and bulge velocity 
dispersion established in the local universe \citep[e.g.,][]{fer00,kor13}, 
the SMBH mass of $\sim$1 $\times$ 10$^{7}$M$_{\odot}$ in NGC 1068 corresponds 
to the bulge velocity dispersion of $\sim$100 km s$^{-1}$. 
On the other hand, the Keplerian velocity at $\sim$5 pc for 
M$_{\rm SMBH}$ $\sim$ 1 $\times$ 10$^{7}$M$_{\odot}$ is $\sim$90 km s$^{-1}$.
Thus, in terms of angular momentum, it is possible that the collision 
of a compact, highly dense, massive GMC with mass of 
$\gtrsim$a few $\times$ 10$^{5}$M$_{\odot}$ and infalling velocity of 
$\sim$100 km s$^{-1}$ changed the rotation direction of the 
pre-existing, almost-Keplerian rotating torus at a $\gtrsim$2-pc 
scale on the western side.

In a standard torus model, without invoking an external turbulence, 
velocity dispersion of only $<$20 km s$^{-1}$ can be produced at 
$\lesssim$7 pc scale \citep{kaw20}.
In the HCN J=3--2 and HCO$^{+}$ J=3--2 moment 2 maps (Figure 3c,d), 
the highest velocity dispersion value in the W-torus is 
$\gtrsim$100 km s$^{-1}$. 
Although the border of the counter-rotating molecular gas could 
create an apparently large velocity dispersion due to beam dilution 
($\S$4.4), the fact that the observed HCN-to-HCO$^{+}$ J=3--2 flux ratio 
is higher at the W-peak than at the E-peak (Table 3, Figure 11) 
strongly suggests that the intrinsic turbulence is also higher 
at the W-peak ($\S$4.1).
It is possible that the GMC collision contributes to the increased 
turbulence there. 
The HCN J=3--2 and HCO$^{+}$ J=3--2 emission line flux in the beam-sized 
spectrum at the highly turbulent W-peak is 0.45$\pm$0.04 and 
0.30$\pm$0.04 (Jy km s$^{-1}$), respectively (Table 3, column 9). 
Using the same conversion as applied in $\S$4.2, we obtain dense 
molecular mass close to the highly turbulent W-peak to be 
$\sim$10$^{5}$M$_{\odot}$, where the factor of 4--5 HCN overabundance is 
taken into account.
If a GMC with mass of $\gtrsim$a few $\times$ 10$^{5}$M$_{\odot}$ infalls 
to the W-torus with $\sim$100 km s$^{-1}$ velocity, the observed 
turbulence close to the W-peak could be explained energetically.

Once the collision-induced strong turbulence with $\sigma$ 
$\gtrsim$40 km s$^{-1}$ happens in the W-torus at $\sim$2-pc scale 
(Figures 3c,d, 14b), the turbulence can propagate to the E-torus 
with a timescale of $\sim$0.1 Myr 
(= $\pi$ $\times$ 2 [pc] / $\gtrsim$40 [km s$^{-1}$]), 
possibly explaining the observed counter-rotation in the E-torus as well.
Forthcoming papers will present detailed numerical simulation results 
of the collision scenario (Wada et al. 2020, in preparation) as well as 
more detailed dynamical properties of dense molecular gas in the 
torus region (Nguyen et al. 2020, in preparation).

Our very-high-spatial-resolution ($\sim$0$\farcs$02) ALMA data 
have revealed very interesting and complicated dense molecular 
gas properties in the torus of the nearby well-studied 
type-2 AGN, NGC 1068.
In a classical torus model, a simple, almost Keplerian 
rotation is assumed unless some turbulence is added, 
because molecular viscosity is too small to significantly 
extract the angular momentum of the torus in a short timescale. 
The torus is a structure where a large amount of dense 
molecular gas distributes in a compact area, outside an 
accretion disk around a central SMBH.
Angular momentum removal of molecular gas in the torus 
can realize mass transport into the accretion disk and 
ignite AGN activity.
The counter-rotating molecular torus is one way 
to achieve this \citep{imp19} and may be the origin 
of the observed luminous AGN activity in NGC 1068 
(L$_{\rm AGN}$ $>$ 4 $\times$ 10$^{44}$ erg s$^{-1}$) 
\citep{boc00,gra20}. 
Future ALMA very-high-spatial-resolution molecular line 
observations of NGC 1068 and other nearby AGNs will further 
improve our understanding of the torus and its relation to 
luminous AGN activity in our universe.

\section{Summary} 

We have conducted ALMA very-high-spatial-resolution ($\sim$0$\farcs$02 
or $\sim$1.4 pc) observations of the nucleus of the nearby ($\sim$14 Mpc) 
well-studied type-2 AGN, NGC 1068, at HCN, HCO$^{+}$, and HNC J=3--2 lines, 
as well as at their $^{13}$C isotopologue and vibrationally excited lines.
These lines are suitable to observationally constrain the 
morphological/dynamical/chemical/physical properties of the dense 
molecular gas which is believed to be abundant in the putative 
torus around a mass-accreting SMBH.
We found the following main results. 

\begin{enumerate}

\item We confirmed almost east-west-oriented HCN J=3--2 and HCO$^{+}$ 
J=3--2 emission both morphologically and dynamically, as expected 
for the torus in NGC 1068.  
The primary bright emission comes from a $\lesssim$3 pc region, 
with the western side being significantly brighter than the eastern 
side.
Low-surface-brightness dense molecular line emission is also 
seen outside the compact main torus component, out to $\sim$7 pc, 
roughly along the east-west direction.
The total dense molecular mass within a 14 pc (east-west) $\times$ 
7 pc (north-south) torus region is estimated to be 
M$_{\rm torus}$ = 0.4--1.0 $\times$ 10$^{6}$M$_{\odot}$, which is 
substantially smaller than the central SMBH mass with 
M$_{\rm SMBH}$ $\sim$ 1 $\times$ 10$^{7}$M$_{\odot}$.

\item An outer torus at $\gtrsim$2 pc is blueshifted in the western part 
and redshifted in the eastern part, as previously seen \citep{ima18a}.
However, the innermost torus at $\lesssim$1 pc is redshifted in the 
western part and blueshifted in the eastern part, the 
opposite of the outer torus, in both the HCN J=3--2 and 
HCO$^{+}$ J=3--2 lines.
The dense molecular gas in the torus appears counter-rotating between 
the innermost and outer parts.

\item A redshifted high-velocity (V $\sim$ 1300--1500 km s$^{-1}$ 
or V$_{\rm sys}$ $+$ [170--370] km s$^{-1}$)  
emission component is seen at the innermost northwestern torus, 
both in the HCN J=3--2 and HCO$^{+}$ J=3--2 lines.
The velocity of this component at $\lesssim$0.7 pc is comparable to 
that of a VLBI-detected, almost-Keplerian rotating $\sim$22 GHz H$_{2}$O 
maser emission disk at the northwestern side of the central mass-accreting 
SMBH, suggesting 
that we are probing the same spatial and dynamical component there.
The redshifted HCN J=3--2 and HCO$^{+}$ J=3--2 emission components are 
much brighter than the corresponding blueshifted components at the 
southeastern side, as seen in the H$_{2}$O maser emission.
It is suggested that dense molecular gas spatial distribution at 
the innermost torus is not symmetric in that the amount of gas is larger 
on the northwestern side than the southeastern side.

\item The western compact ($\lesssim$3 pc) main torus displays a larger 
velocity dispersion than the eastern main torus.
The maximum velocity dispersion spot in the western torus is 
just outside the innermost redshifted high velocity component.
This positional shift can also be explained by counter-rotating dense 
molecular gas because the velocity dispersion can be very large at the border 
of counter-rotating components owing to beam dilution and/or actual high 
turbulence caused by instability. 

\item The HCN J=3--2 line in the spectrum toward the $\sim$260 GHz continuum 
emission peak displays a broad absorption wing at the blueshifted side of 
the systemic velocity (V$_{\rm sys}$ = 1130 km s$^{-1}$), which was 
interpreted to be resulting from dense molecular outflow. 
This blueshifted broad absorption signature is stronger in HCN J=3--2 
than HCO$^{+}$ J=3--2.

\item The vibrationally excited (v$_{2}$=1, l=1f) HCN J=3--2 line (HCN-VIB) 
is detected in absorption in the spectrum toward the continuum peak 
and in redshifted (V $\sim$ 1340 km s$^{-1}$ 
or V$_{\rm sys}$ $+$ 210 km s$^{-1}$)
emission at the innermost northwestern torus.
The HCN-VIB-to-HCN J=3--2 emission line flux ratio is significantly smaller 
than those of other buried AGNs with detected HCN-VIB J=3--2 emission lines, 
possibly 
because of small volume- and area-filling factors of dense molecular clumps 
in the NGC 1068 torus.

\item The observed flux of HCN J=3--2 emission is systematically 
higher than that of HCO$^{+}$ J=3--2 emission in the torus 
within a $\sim$7-pc scale from the central SMBH.
We attribute this to higher abundance of HCN than HCO$^{+}$ 
in the torus region around a luminous AGN. 
Based on the isotopologue H$^{13}$CN J=3--2 and H$^{13}$CO$^{+}$ J=3--2 
absorption strengths in the spectrum toward the continuum peak, 
we derive a factor of $>$3 higher abundance of HCN than HCO$^{+}$.
This enhanced-HCN-abundance scenario is supported by various 
other observational results, including 
(1) higher HCN-to-HCO$^{+}$ J=3--2 emission line flux ratio in the 
more turbulent western compact main torus,
(2) wider spatial distribution of significantly detected 
line emission in the torus region for HCN J=3--2 than for HCO$^{+}$ J=3--2, 
despite a factor of $\sim$5 larger critical density for the former, 
(3) larger torus mass estimated from HCN than HCO$^{+}$, 
(4) stronger emission from the innermost northwestern 
torus and absorption toward the continuum peak for HCN-VIB J=3--2 
than for HCO$^{+}$-VIB J=3--2.
Radiative and/or mechanical effects of a luminous AGN may enhance 
HCN abundance more than HCO$^{+}$ in the torus molecular gas.

\item An area-integrated ($\sim$7 pc) spectra demonstrated that molecular 
emission in the western torus shows a wider line profile. 
Together with the suggested counter-rotation of dense molecular gas 
in the torus, we conjecture that some external turbulence happened 
in the western torus, which may be related to the observed high AGN 
luminosity in NGC 1068, because angular momentum in the torus can 
be removed efficiently and mass accretion onto the 
central SMBH could be enhanced.

\end{enumerate}

Our high-spatial-resolution ALMA data revealed that dense molecular 
gas properties in the torus in the nearby well-studied AGN, NGC 1068, 
are much more complex than the simple axi-symmetric, 
almost-Keplerian rotating torus picture.
Further ALMA very-high-spatial-resolution observations of NGC 1068 
and other nearby AGNs may help discover more interesting torus features 
and obtain a clearer view about the role of the dusty molecular torus in AGNs.

\acknowledgments

We thank Dr. Kazuya Saigo for his support to analyze ALMA data, 
and the anonymous referee for his/her valuable comment which 
helped significantly improve the clarify of this manuscript.
This paper makes use of the following ALMA data: 
ADS/JAO.ALMA\#2018.1.00037.S, \#2017.1.01666.S, \#2016.1.00052.S, and 
\#2013.0.00188.S.
ALMA is a partnership of ESO (representing its member states), NSF (USA) 
and NINS (Japan), together with NRC (Canada) and NSC and ASIAA (Taiwan), 
in cooperation with the Republic of Chile. 
The Joint ALMA Observatory is operated by ESO, AUI/NRAO and NAOJ.
Data analysis was in part carried out on the open use data analysis
computer system at the Astronomy Data Center, ADC, of the National
Astronomical Observatory of Japan. 
This research has made use of NASA's Astrophysics Data System and the
NASA/IPAC Extragalactic Database (NED) which is operated by the Jet
Propulsion Laboratory, California Institute of Technology, under
contract with the National Aeronautics and Space Administration. 

%

\vspace{5mm}
\facilities{ALMA}






\appendix

\section{ALMA Cycles 6 and 5 data combination for HCN J=3--2 line}

\citet{imp19} presented the results of ALMA Cycle 5 
$\sim$0$\farcs$02-resolution HCN J=3--2 line observations of NGC 1068 
(2017.1.01666.S).
We combined our Cycle 6 and their Cycle 5 data, after excluding their 
short baseline (44--83 m) data, in order not to degrade the image size 
of the final data product. 
Figure 18 displays moment 0, 1, and 2 maps and a position velocity diagram 
of the HCN J=3--2 line for the combined data. 
Beam-sized spectra at the E-, W-, and C-peaks are shown in Figure 19.
The main features detected in our Cycle 6 data in Figures 2, 3, 5, 
6, and 13 are well reproduced in the combined data.

\section{Comparison with previously taken data for the torus emission}

HCN J=3--2 and HCO$^{+}$ J=3--2 emission lines were significantly 
detected in previously taken (Cycles 4 and 2) lower-spatial-resolution 
($\gtrsim$0$\farcs$04) data at the compact ($<$10 pc) torus 
\citep{ima16a,ima18a}.
Figure 20 compares HCN J=3--2 and HCO$^{+}$ J=3--2 emission line 
profiles from the torus measured in a 0$\farcs$15- or $\sim$10-pc radius 
circular aperture, taken in Cycle 6 data, Cycle 4 data with long baseline 
only, and Cycles 4 and 2 combined data including both long and short 
baselines \citep{ima18a}. 
Emission profiles of both lines are comparable in all data, within 
the possible absolute calibration uncertainty of individual ALMA 
observations.



\clearpage


\begin{deluxetable}{llccc|ccc}
\tabletypesize{\scriptsize}
\tablecaption{Log of ALMA Cycle 6 Observations \label{tbl-1}} 
\tablewidth{0pt}
\tablehead{
\colhead{Data} & \colhead{Date} & \colhead{Antenna} & 
\colhead{Baseline} & \colhead{Integration} & \multicolumn{3}{c}{Calibrator} \\ 
\colhead{} & \colhead{(UT)} & \colhead{Number} & \colhead{(m)} &
\colhead{(min)} & \colhead{Bandpass} & \colhead{Flux} & \colhead{Phase}  \\
\colhead{(1)} & \colhead{(2)} & \colhead{(3)} & \colhead{(4)} &
\colhead{(5)} & \colhead{(6)} & \colhead{(7)}  & \colhead{(8)} 
}
\startdata 
data-a & 2019 June 6 & 42 & 237--15238 & 47 & 
J0423$-$0120 & J0423$-$0120 & J0239$-$0234 \\  
(HCN/HCO$^{+}$ J=3--2) & 2019 June 6 & 42 & 237--15238 & 47 
& J0423$-$0120 & J0423$-$0120 & J0239$-$0234 \\  
 & 2019 June 6 & 41 & 237--15238 & 47 & J0423$-$0120 & J0423$-$0120 & 
J0239$-$0234 \\  
 & 2019 June 8 & 44 & 83--16196 & 47 & J0423$-$0120 & J0423$-$0120 & 
J0239$-$0234 \\  
 & 2019 June 8 & 44 & 83--16196 & 47 & J0423$-$0120 & J0423$-$0120 & 
J0239$-$0234\\  
 & 2019 June 8 & 44 & 83--16196 & 47 & J0423$-$0120 & J0423$-$0120 & 
J0239$-$0234 \\  
 & 2019 June 9 & 44 & 83--16196 & 47 & J0006$-$0623 & J0006$-$0623 
& J0239$-$0234 \\  
data-b & 2019 July 13 
& 46 & 111--12645 & 47 & J0423$-$0120 & J0423$-$0120 & J0239$-$0234  \\  
($^{13}$C isotopologue/HNC J=3--2)  & 2019 July 13 & 46 & 111--12645 & 47 
& J0423$-$0120 & J0423$-$0120 & J0239$-$0234  \\  
 & 2019 July 14 & 43 & 111--11436 & 47 & J0006$-$0623 & J0006$-$0623 
& J0239$-$0234  \\  
\enddata

\tablecomments{Col.(1): Data. Data-a include 
HCN/HCO$^{+}$ J=3--2 and their vibrationally excited 
(HCN-VIB/HCO$^{+}$-VIB) J=3--2 lines.
Data-b include 
HNC/H$^{13}$CN/H$^{13}$CO$^{+}$/HN$^{13}$C/HNC-VIB J=3--2 lines.
Col.(2): Observing date in UT. 
Col.(3): Number of antennas used for observations. 
Col.(4): Baseline length in meters. Minimum and maximum baseline lengths are 
shown.  
Col.(5): Net on source integration time in minutes.
Cols.(6), (7), and (8): Bandpass, flux, and phase calibrator for the 
target source, respectively.}

\end{deluxetable}

\begin{deluxetable}{lccccc}
\tabletypesize{\scriptsize}
\tablecaption{Continuum Emission \label{tbl-3}}
\tablewidth{0pt}
\tablehead{
\colhead{Data} & \colhead{Frequency} & \colhead{Peak flux} & 
\colhead{Peak coordinate} & \colhead{rms} & \colhead{Synthesized beam} \\
\colhead{} & \colhead{(GHz)} & \colhead{(mJy/beam)} & 
\colhead{(RA,DEC)ICRS} & \colhead{(mJy/beam)} & 
\colhead{($''$ $\times$ $''$) ($^{\circ}$)} \\  
\colhead{(1)} & \colhead{(2)} & \colhead{(3)}  & \colhead{(4)}  &
\colhead{(5)} & \colhead{(6)} 
}
\startdata 
data-a & 263.6--269.1 & 6.5 (87$\sigma$) & 
(02$^{\rm h}$ 42$^{\rm m}$ 40.709$^{\rm s}$, $-$00$^{\circ}$ 00$'$ 47.946$''$) & 0.075 
& 0.019$\times$0.017 (63$^{\circ}$) \\
data-b & 256.7--260.5, 270.0--273.7 & 8.2 (162$\sigma$) & 
(02$^{\rm h}$ 42$^{\rm m}$ 40.709$^{\rm s}$, $-$00$^{\circ}$ 00$'$ 47.945$''$) & 0.050 &  
0.034$\times$0.026 (53$^{\circ}$) \\
\enddata

\tablecomments{Col.(1): Data.
Col.(2): Frequency range (in GHz) used to extract continuum emission.
Frequencies of obvious emission and absorption lines are excluded.
Col.(3): Flux (in mJy beam$^{-1}$) at the emission peak.
Value at the highest flux pixel (0$\farcs$003 pixel$^{-1}$ or 
0$\farcs$005 pixel$^{-1}$) is adopted. 
The detection significance relative to the root mean square (rms) noise 
is shown in parentheses. 
Possible systematic uncertainties coming from absolute flux
calibration ambiguity in individual ALMA observations and 
choice of frequency range to determine the continuum level 
are not taken into account. 
Col.(4): Coordinate of the continuum emission peak in ICRS.
Col.(5): The rms noise level (1$\sigma$) (in mJy beam$^{-1}$), derived
from the standard deviation of sky signals.
Col.(6): Synthesized beam (in arcsec $\times$ arcsec) and position angle
(in degrees). 
The position angle is 0$^{\circ}$ along the north--south direction
and increases with the counterclockwise direction.}

\end{deluxetable}

\begin{deluxetable}{ll|ccc|cccc}
\rotate
\tabletypesize{\scriptsize}
\tablecaption{Properties of detected molecular lines in the torus region
\label{tbl-3}}  
\tablewidth{0pt}
\tablehead{
\colhead{Position} & \colhead{Line} & 
\multicolumn{3}{c}{Integrated intensity (moment 0) map} &  
\multicolumn{4}{c}{Gaussian line fit} \\  
\colhead{} & \colhead{} &\colhead{Peak} &
\colhead{rms} & \colhead{Beam} & \colhead{Velocity} 
& \colhead{Peak} & \colhead{FWHM} & \colhead{Flux} \\ 
\colhead{} & \colhead{} & 
\multicolumn{2}{c}{(Jy beam$^{-1}$ km s$^{-1}$)} & 
\colhead{($''$ $\times$ $''$) ($^{\circ}$)} &
\colhead{(km s$^{-1}$)} & \colhead{(mJy)} & \colhead{(km s$^{-1}$)} &
\colhead{(Jy km s$^{-1}$)} \\  
\colhead{(1)} & \colhead{(2)} & \colhead{(3)} & \colhead{(4)} & 
\colhead{(5)} & \colhead{(6)} & \colhead{(7)} & \colhead{(8)} &
\colhead{(9)} 
}
\startdata 
E-peak & HCN J=3--2 & 0.26 (7.3$\sigma$) & 0.036 & 0.021 $\times$ 0.018 (59) 
& 1099$\pm$7 & 1.1$\pm$0.1 & 193$\pm$18 & 0.23$\pm$0.03 \\ 
 & HCO$^{+}$ J=3--2 & 0.29 (10$\sigma$) & 0.028 
& 0.021 $\times$ 0.018 (60) 
& 1094$\pm$9 & 0.90$\pm$0.08 & 225$\pm$34 & 0.22$\pm$0.04 \\
W-peak & HCN J=3--2 & 0.45 (13$\sigma$) & 0.036 & 0.021 $\times$ 0.018 (59) 
& 1160$\pm$12 & 1.0$\pm$0.1 & 408$\pm$30 & 0.45$\pm$0.04 \\ 
 & & & & & 1088$\pm$10, 1192$\pm$19 & 0.53$\pm$0.12, 0.83$\pm$0.08 
& 104$\pm$33, 456$\pm$37 & 0.46$\pm$0.06 \\ 
 & HCO$^{+}$ J=3--2 & 0.33 (11$\sigma$) & 0.028 & 0.021 $\times$ 0.018 (60) 
& 1179$\pm$15 & 0.72$\pm$0.05 & 393$\pm$42 & 0.30$\pm$0.04 \\ 
 & & & & & 1109, 1223 (fix)$\tablenotemark{A}$ & 0.49$\pm$0.15, 0.52$\pm$0.07 
& 86, 498 (fix)$\tablenotemark{A}$ & 0.32$\pm$0.04 \\
 & HCN-VIB J=3--2 & 0.049(4.8$\sigma$)$\tablenotemark{B}$ & 
0.010 & 0.021 $\times$ 0.018 (60) & 
1339$\pm$7 & 0.35$\pm$0.05 & 109$\pm$20 & 0.041$\pm$0.009 \\ \hline
Close to C-peak$\tablenotemark{C}$ & H$^{13}$CN J=3--2 
& $-$0.090 (4.5$\sigma$)$\tablenotemark{C}$ & 0.020 
& 0.035 $\times$ 0.029 (60) & & & &  \\ 
Close to W-peak$\tablenotemark{D}$ & HNC J=3--2 
& 0.091 (3.6$\sigma$)$\tablenotemark{D}$ & 0.025 & 
0.036 $\times$ 0.026 (54) 
& & & & \\
\enddata

\tablenotetext{A}{Fixed to the best fit value.}

\tablenotetext{B}
{The highest positive value in the western torus (Figure 15) is found at 
(02$^{\rm h}$ 42$^{\rm m}$ 40.708$^{\rm s}$, $-$00$^{\circ}$ 00$'$ 47.937$''$).}

\tablenotetext{C}{The most negative signal close to the 
C-peak (Figure 9a) is found at 
(02$^{\rm h}$ 42$^{\rm m}$ 40.710$^{\rm s}$, $-$00$^{\circ}$ 00$'$ 47.945$''$).
This spatially agrees with the C-peak within the position  
determination accuracy of (beam-size)/(signal-to-noise-ratio) 
$\sim$ 0$\farcs$007.
}

\tablenotetext{D}{The highest positive value in the western 
torus (Figure 9b) is found at 
(02$^{\rm h}$ 42$^{\rm m}$ 40.708$^{\rm s}$, $-$00$^{\circ}$ 00$'$ 47.945$''$).}

\tablecomments{Col.(1): Position. 
C-peak (continuum emission peak): 
(02$^{\rm h}$ 42$^{\rm m}$ 40.709$^{\rm s}$, $-$00$^{\circ}$ 00$'$ 47.946$''$), 
E-peak (HCN J=3--2 emission peak in the eastern torus): 
(02$^{\rm h}$ 42$^{\rm m}$ 40.710$^{\rm s}$, $-$00$^{\circ}$ 00$'$ 47.952$''$).
W-peak (HCN J=3--2 emission peak in the western torus): 
(02$^{\rm h}$ 42$^{\rm m}$ 40.708$^{\rm s}$, $-$00$^{\circ}$ 00$'$ 47.940$''$). 
Col.(2): Molecular line.
Col.(3): Integrated intensity (in Jy beam$^{-1}$ km s$^{-1}$) at the 
emission peak. 
The detection significance relative to the rms noise (1$\sigma$) in the 
moment 0 map is shown in parentheses. 
Possible systematic uncertainty is not included. 
Col.(4): The rms noise (1$\sigma$) level in the moment 0 map 
(in Jy beam$^{-1}$ km s$^{-1}$), derived from signals in the annular region
with 2$\farcs$4--6$\farcs$0 in radius centered at 
(02$^{\rm h}$42$^{\rm m}$40.70$^{\rm s}$, $-$00$^{\circ}$00$'$48.5$''$) 
to avoid areas of significant molecular line emission.
Col.(5): Beam size (in arcsec $\times$ arcsec) and position angle 
(in degrees). Position angle is 0$^{\circ}$ along the north-south direction, 
and increases counterclockwise. 
Cols.(6)--(9): Gaussian fit of clearly detected emission lines in 
beam-sized spectra.
At the W-peak, two Gaussian fits are applied because a very broad
component is observed, in addition to a narrow component.  
Col.(6): Optical LSR velocity of emission peak (in km s$^{-1}$). 
Col.(7): Peak flux (in mJy). 
Col.(8): Observed FWHM (in km s$^{-1}$) in the spectra of Fig. 5.  
Col.(9): Gaussian-fit, velocity-integrated flux (in Jy km s$^{-1}$).
At the W-peak, the single and two Gaussian fits provide comparable 
flux values both for HCN J=3--2 and HCO$^{+}$ J=3--2 lines.
We adopt the single Gaussian fit results for our discussion.}

\end{deluxetable}

\begin{deluxetable}{ll|cccc}
\tabletypesize{\small}
\tablecaption{Properties of HCN J=3--2 and HCO$^{+}$ J=3--2 emission 
lines in area-integrated torus spectra 
\label{tbl-4}}  
\tablewidth{0pt}
\tablehead{
\colhead{Region} & \colhead{Line} & 
\multicolumn{4}{c}{Gaussian line fit} \\  \hline
\colhead{} & \colhead{} & \colhead{Velocity} 
& \colhead{Peak} & \colhead{FWHM} & \colhead{Flux} \\ 
\colhead{} & \colhead{} & 
\colhead{(km s$^{-1}$)} & \colhead{(mJy)} & \colhead{(km s$^{-1}$)} &
\colhead{(Jy km s$^{-1}$)} \\  
\colhead{(1)} & \colhead{(2)} & \colhead{(3)} & \colhead{(4)} & 
\colhead{(5)} & \colhead{(6)} 
}
\startdata 
Torus (0$\farcs$2 $\times$ 0$\farcs$1) & HCN J=3--2 & 1135$\pm$2 & 17$\pm$1 
& 190$\pm$7 & 3.4$\pm$0.1 \\
 & HCO$^{+}$ J=3--2 & 1129$\pm$4 & 11$\pm$1 & 189$\pm$9 & 2.2$\pm$0.1 \\
E-torus (0$\farcs$1 $\times$ 0$\farcs$1) & HCN J=3--2 & 1137$\pm$2 & 
10$\pm$1 & 171$\pm$6 & 1.9$\pm$0.1 \\
 & HCO$^{+}$ J=3--2 & 1133$\pm$4 & 6.8$\pm$0.3 & 170$\pm$10 & 1.2$\pm$0.1 \\
W-Torus (0$\farcs$1 $\times$ 0$\farcs$1) & HCN J=3--2 & 1137$\pm$5 & 
6.6$\pm$0.3 & 236$\pm$14 & 1.6$\pm$0.1 \\
 & HCO$^{+}$ J=3--2 & 1122$\pm$7 & 4.6$\pm$0.3 & 213$\pm$24 & 1.0$\pm$0.1 \\
\enddata

\tablecomments{Col.(1): Region. 
Col.(2): Line.
Cols.(3)--(6): Gaussian fit of emission lines in area-integrated torus 
spectra.
Col.(3): Optical LSR velocity of emission peak (in km s$^{-1}$). 
Col.(4): Peak flux (in mJy). 
Col.(5): Observed FWHM (in km s$^{-1}$).  
Col.(6): Gaussian-fit, velocity-integrated flux (in Jy km s$^{-1}$).}

\end{deluxetable}

\begin{deluxetable}{lccc}
\tabletypesize{\small}
\tablecaption{HCN J=3--2 and HCO$^{+}$ J=3--2 emission line luminosity and 
estimated dense molecular mass in the 14 pc $\times$ 7 pc torus region
\label{tbl-5}}  
\tablewidth{0pt}
\tablehead{
\colhead{Line} & \colhead{L} & \colhead{L$'$} & \colhead{M$_{\rm dense}$} \\ 
\colhead{} & \colhead{10$^{2}$ (L$_{\odot}$)} & 
\colhead{10$^{5}$ (K km s$^{-1}$ pc$^{2}$)} & \colhead{10$^{6}$ (M$_{\odot}$)} \\  
\colhead{(1)} & \colhead{(2)} & \colhead{(3)} & \colhead{(4)} 
}
\startdata 
HCN J=3--2 & 1.8 & 3.1 & 2--4 \\
HCO$^{+}$ J=3--2 & 1.2 & 2.0 & 0.4--1.0 \\
\enddata

\tablecomments{
Col.(1): Line.
Col.(2): Luminosity (in L$_{\odot}$).
Col.(3): Luminosity (in K km s$^{-1}$ pc$^{2}$). 
Col.(4): Estimated dense molecular mass (in M$_{\odot}$).
}

\end{deluxetable}

\clearpage

\begin{figure}
\begin{center}
\includegraphics[angle=0,scale=.35]{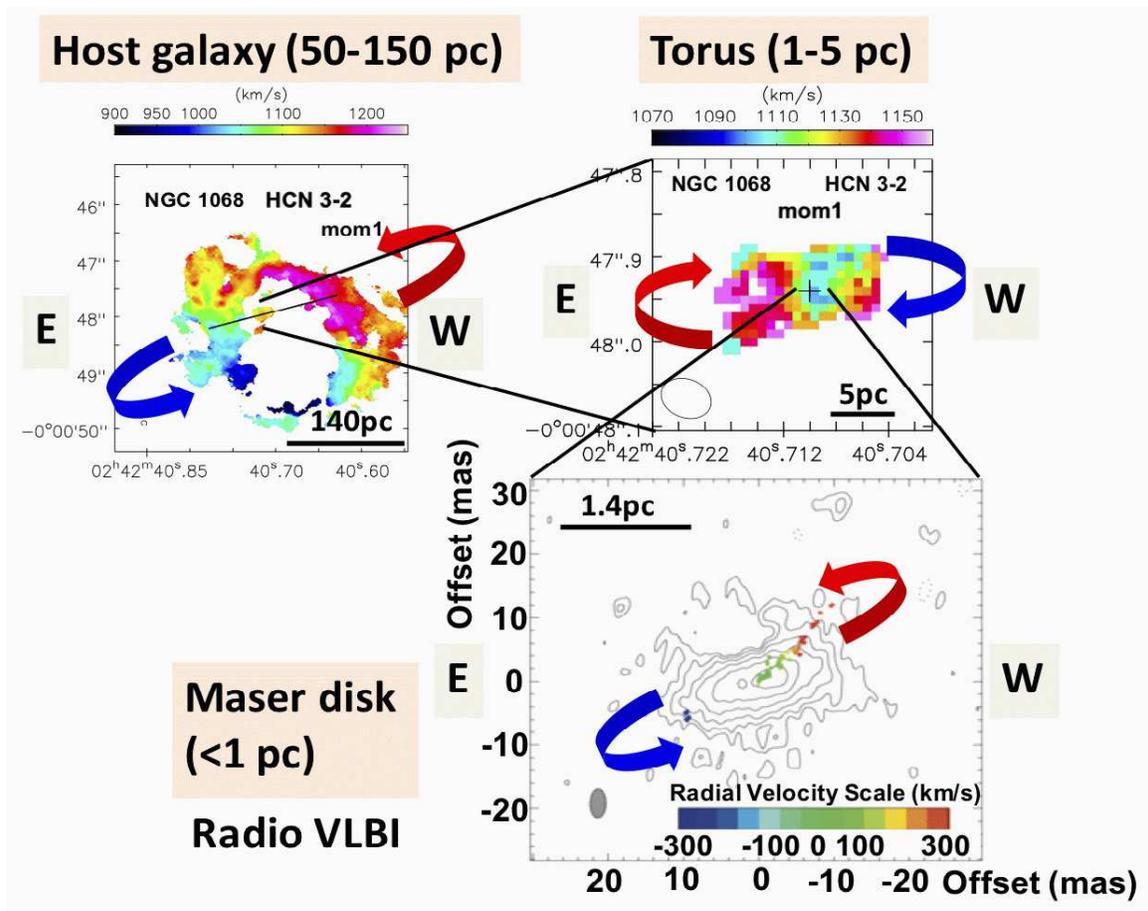} \\
\end{center}
\caption{
Observed dynamical properties of dense molecular and H$_{2}$O maser emission 
at the inner part of NGC 1068.
{\it (Top Left)}: Intensity-weighted rotation velocity (moment 1) map of 
HCN J=3--2 emission line in the host galaxy (50--150 pc) scale 
\citep{ima18a}. 
The thin solid line crossing the torus at the center indicates the 
torus axis (PA $\sim$ 105$^{\circ}$).
{\it (Top Right)}: Moment 1 map of HCN J=3--2 in the torus (1--5 pc) scale 
\citep{ima18a}.
{\it (Bottom)}: H$_{2}$O maser emission dynamics at the innermost 
($\lesssim$1 pc) part. Modified from \citet{gal04}. 
Velocity relative to their adopted systemic velocity is color coded.  
The coordinates are offset (in mas) from the VLBA 5 GHz continuum 
shown as contours.
North is up and east is to the left in all plots.
The length of the thick horizontal bar corresponds to 140 pc, 
5 pc, 1.4 pc in the top-left, top-right, bottom panel, respectively.
Thick curved blue and red arrows indicate blueshifted and redshifted 
motion relative to the systemic velocity of NGC 1068, respectively. 
The western side is {\it redshifted} for maser emission at $\lesssim$1 pc 
{\it (bottom)} and dense molecular emission in the host galaxy 
at 50--150 pc scale {\it (top left)}, but is {\it blueshifted} 
for the torus dense molecular emission at 1--5 pc scale {\it (top right)}.
}
\end{figure}

\begin{figure}
\begin{center}
\includegraphics[angle=0,scale=.3]{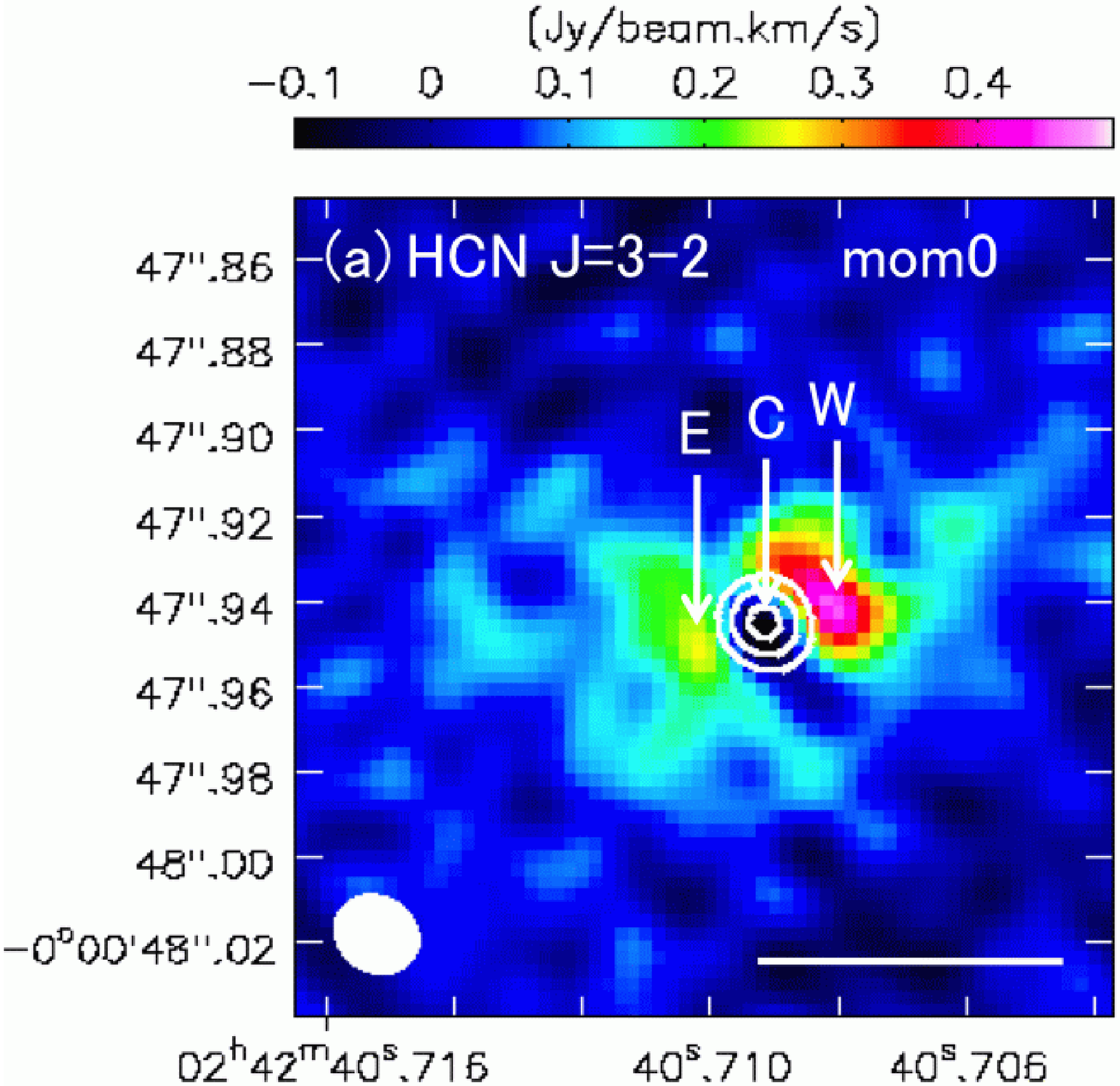} 
\includegraphics[angle=0,scale=.3]{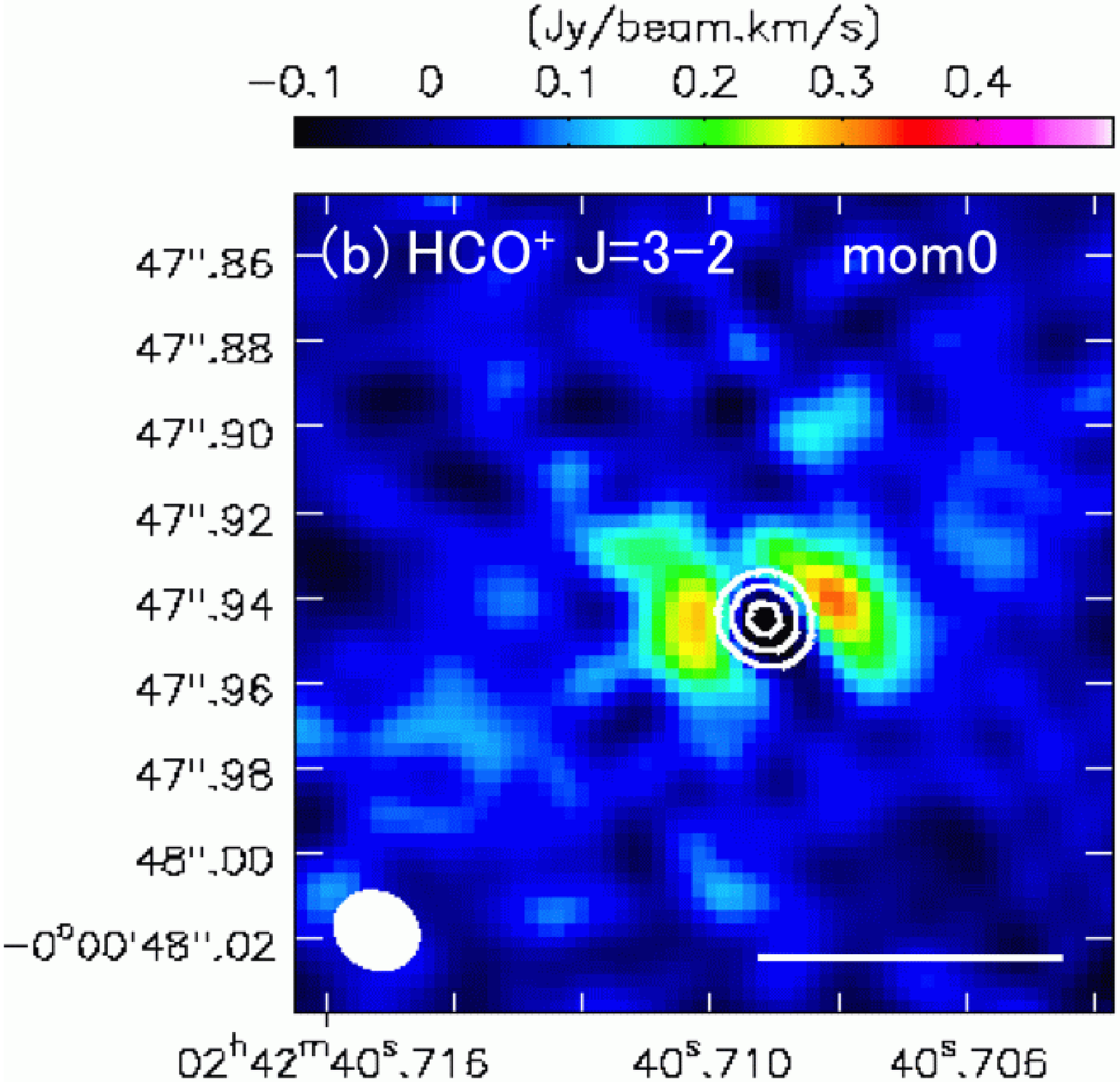} \\ 
\includegraphics[angle=0,scale=.3]{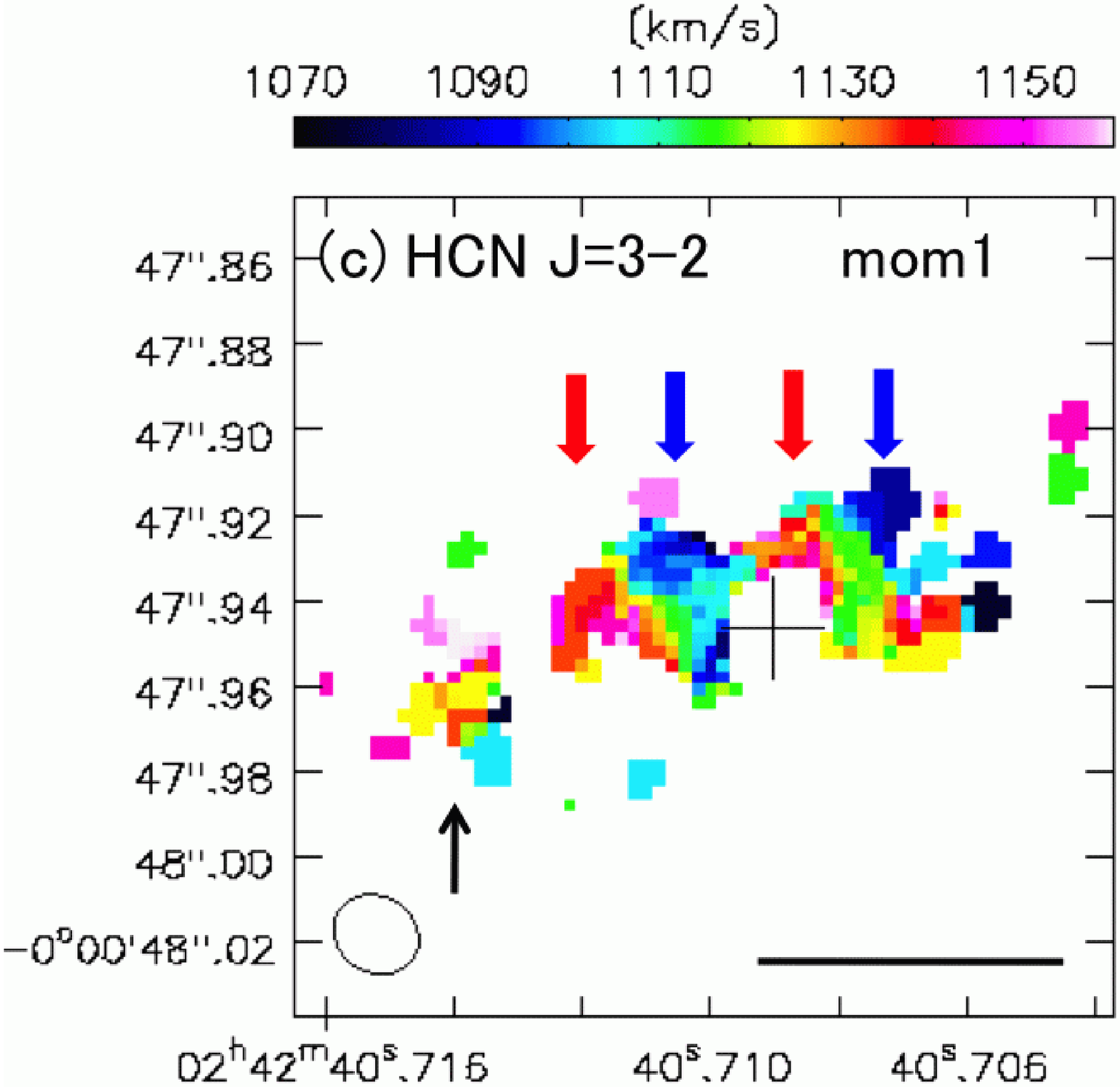} 
\includegraphics[angle=0,scale=.3]{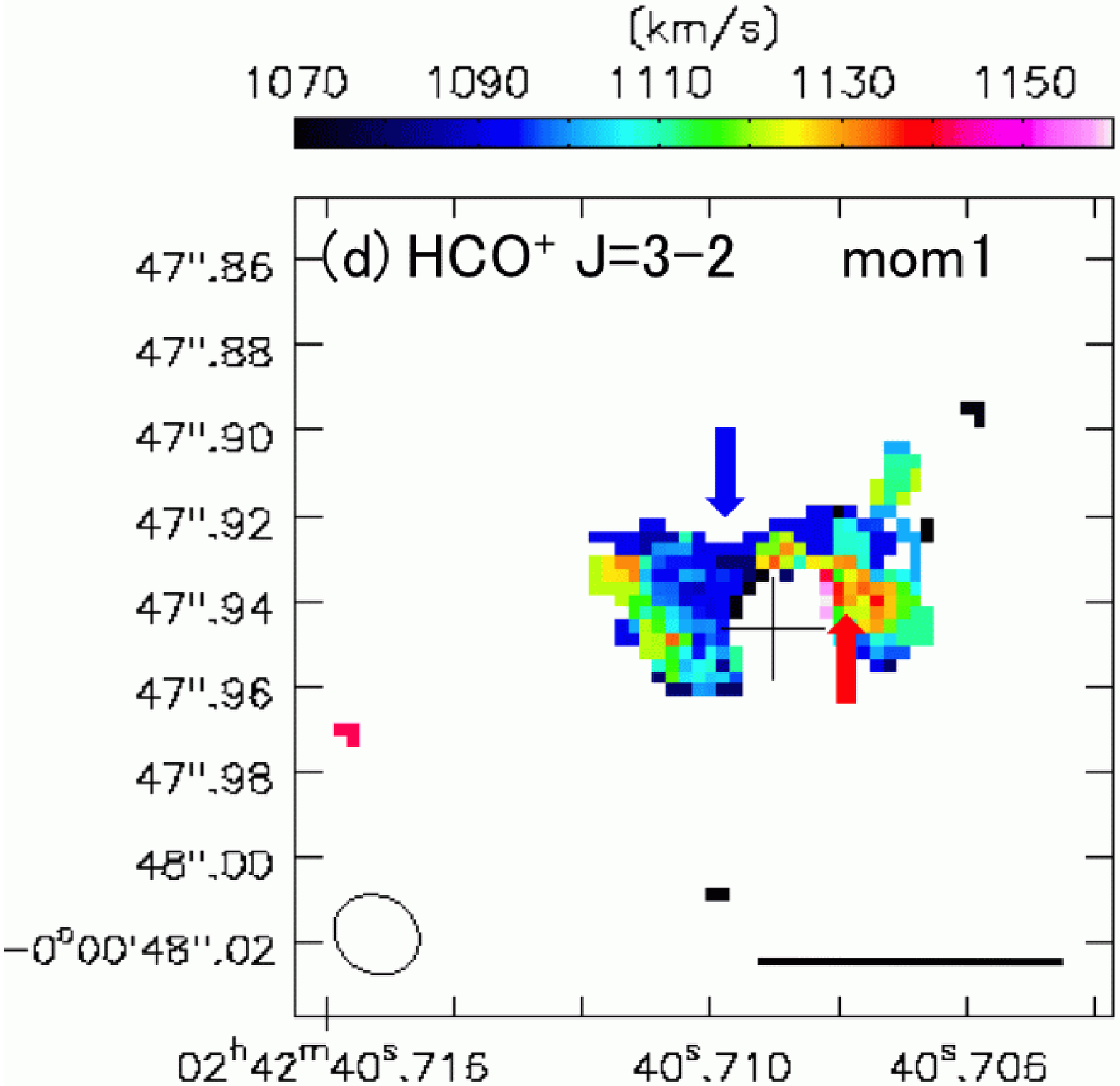} \\ 
\end{center}
\caption{
{\it (Top)}: Integrated intensity (moment 0) maps of {\it (a)} HCN J=3--2 
and {\it (b)} HCO$^{+}$ J=3--2 lines, overlaid on the contours 
of $\sim$260 GHz ($\sim$1.2 mm) continuum emission 
(40, 60, and 80$\sigma$; 1$\sigma$ = 0.075 mJy beam$^{-1}$).
The color scale is the same between these two lines 
($-$0.1 to $+$0.5 Jy beam$^{-1}$ km s$^{-1}$).  
In {\it (a)}, E-, C-, and W-peaks (Table 3) are indicated with thin white 
downward arrows, with the notes of ``E'', ``C'', and ``W'', respectively. 
The horizontal white bar at the lower right part of each figure 
indicates 5 pc at the distance of NGC 1068.     
Beam size is shown as a filled circle in the lower-left region.
{\it (Bottom)}: Intensity-weighted mean velocity (moment 1) maps of 
{\it (c)} HCN J=3--2 and {\it (d)} HCO$^{+}$ J=3--2 emission lines.  
The black cross denotes the location of the continuum emission 
peak (02$^{\rm h}$ 42$^{\rm m}$ 40.709$^{\rm s}$, $-$00$^{\circ}$ 00$'$ 47.946$''$), 
which is taken as the mass-accreting SMBH location. 
The velocity display range (optical LSR velocity V$_{\rm opt}$[LSR] = 
1070--1160 km s$^{-1}$ or from V$_{\rm sys}$ $-$ 60 
to V$_{\rm sys}$ $+$ 30 km s$^{-1}$, where V$_{\rm sys}$ is the systemic velocity 
of 1130 km s$^{-1}$) is set in the same way as adopted by \citet{ima18a}, 
to see blueshifted and redshifted velocity components significantly 
slower than the Keplerian rotation at 2--5 pc from the mass-dominating SMBH 
(M$_{\rm SMBH}$ $\sim$1 $\times$ 10$^{7}$M$_{\odot}$).
For the HCN J=3--2 emission, inner redshifted and outer blueshifted emission 
components in the western torus are marked with thick red and blue downward 
arrows, respectively.
Inner blueshifted and outer redshifted emission components in the eastern 
torus are also indicated as thick blue and red downward arrows, respectively.
Eastern emission at RA = 02$^{\rm h}$ 42$^{\rm m}$ 40.714$^{\rm s}$ 
(discussed in $\S$4.6) is indicated as a black upward arrow.
For the HCO$^{+}$ J=3--2 emission, inner redshifted and blueshifted emission 
in the western and eastern torus, respectively, are indicated as thick 
red upward and blue downward arrows.
The horizontal black bar at the lower right part of each figure 
indicates 5 pc at the distance of NGC 1068.     
Beam size is shown as an open circle in the lower-left region.
An appropriate cutoff ($\sim$4$\sigma$) is applied to moment 1 maps 
in {\it (c)} and {\it (d)} so that they are not dominated by noise. 
}
\end{figure}

\clearpage

\begin{figure}
\begin{center}
\includegraphics[angle=0,scale=.3]{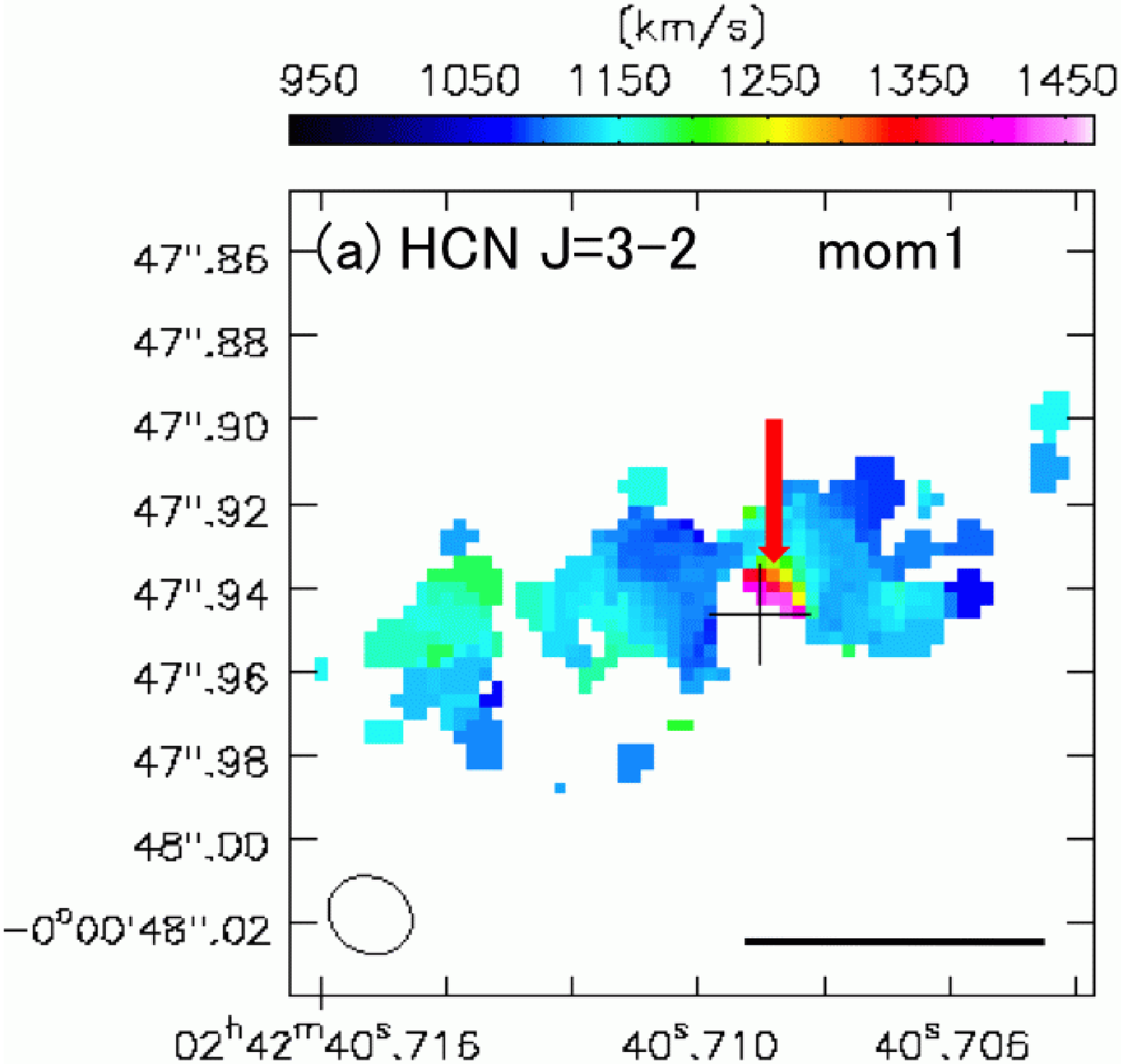} 
\includegraphics[angle=0,scale=.3]{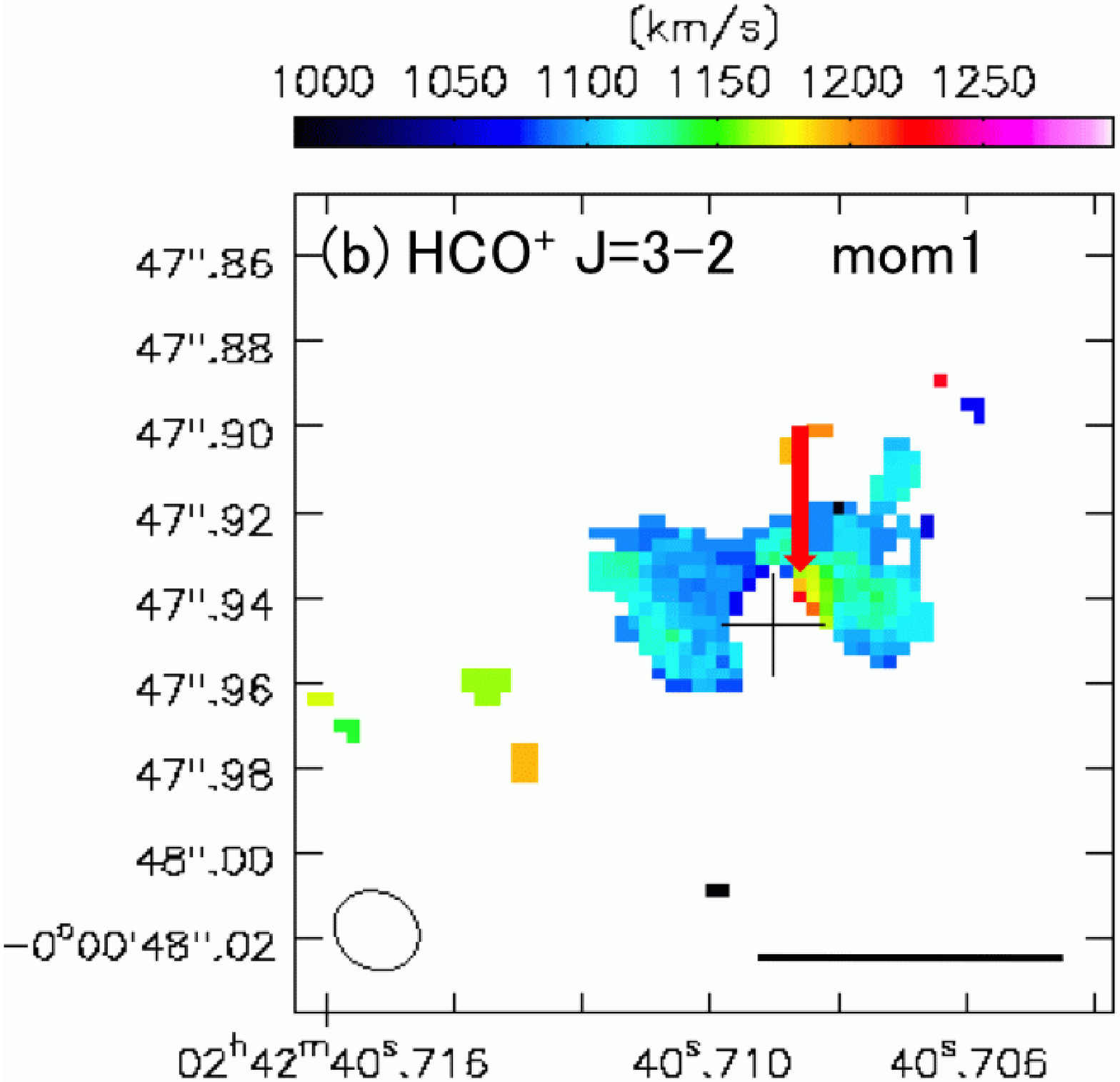} \\ 
\includegraphics[angle=0,scale=.3]{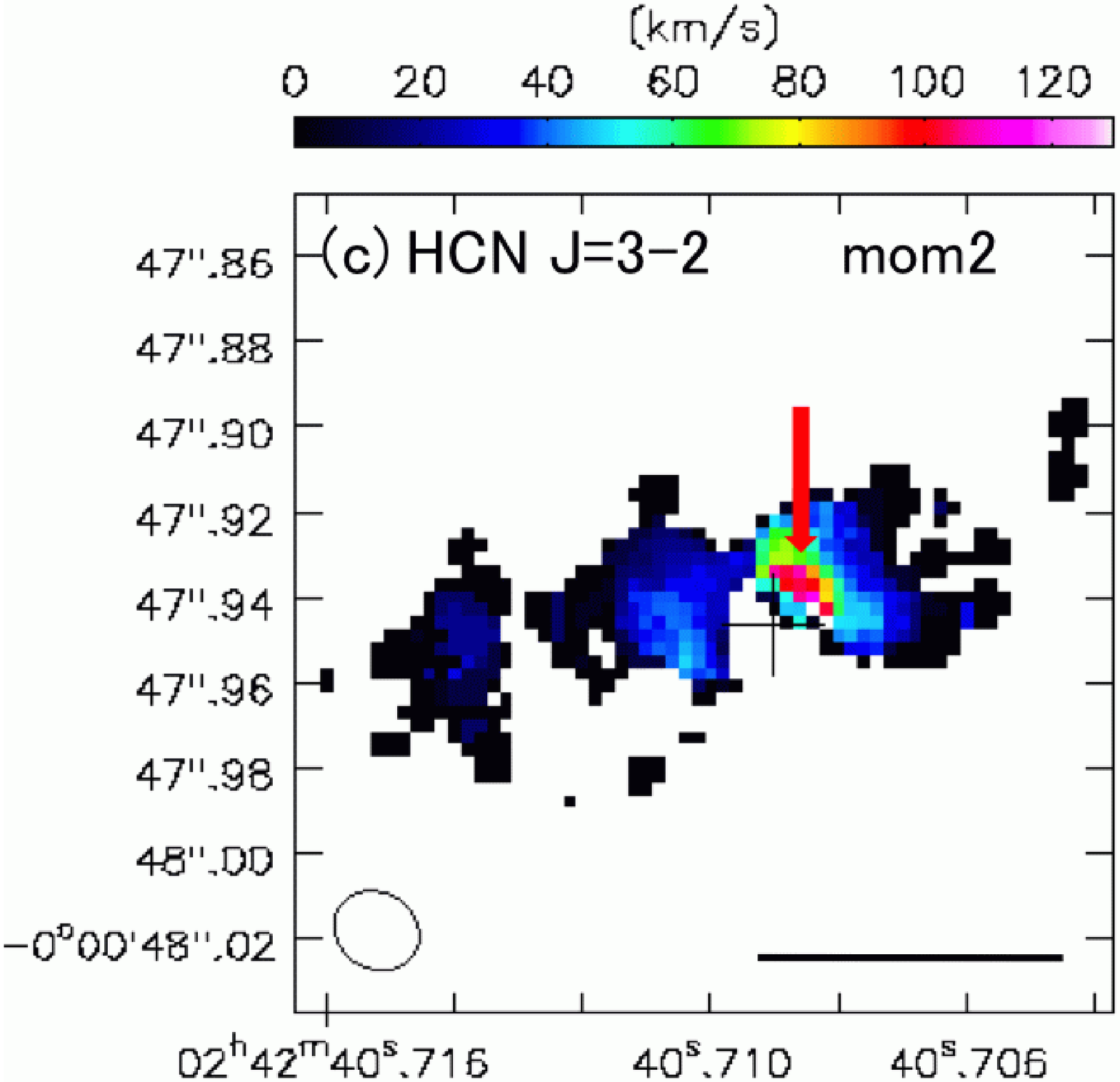} 
\includegraphics[angle=0,scale=.3]{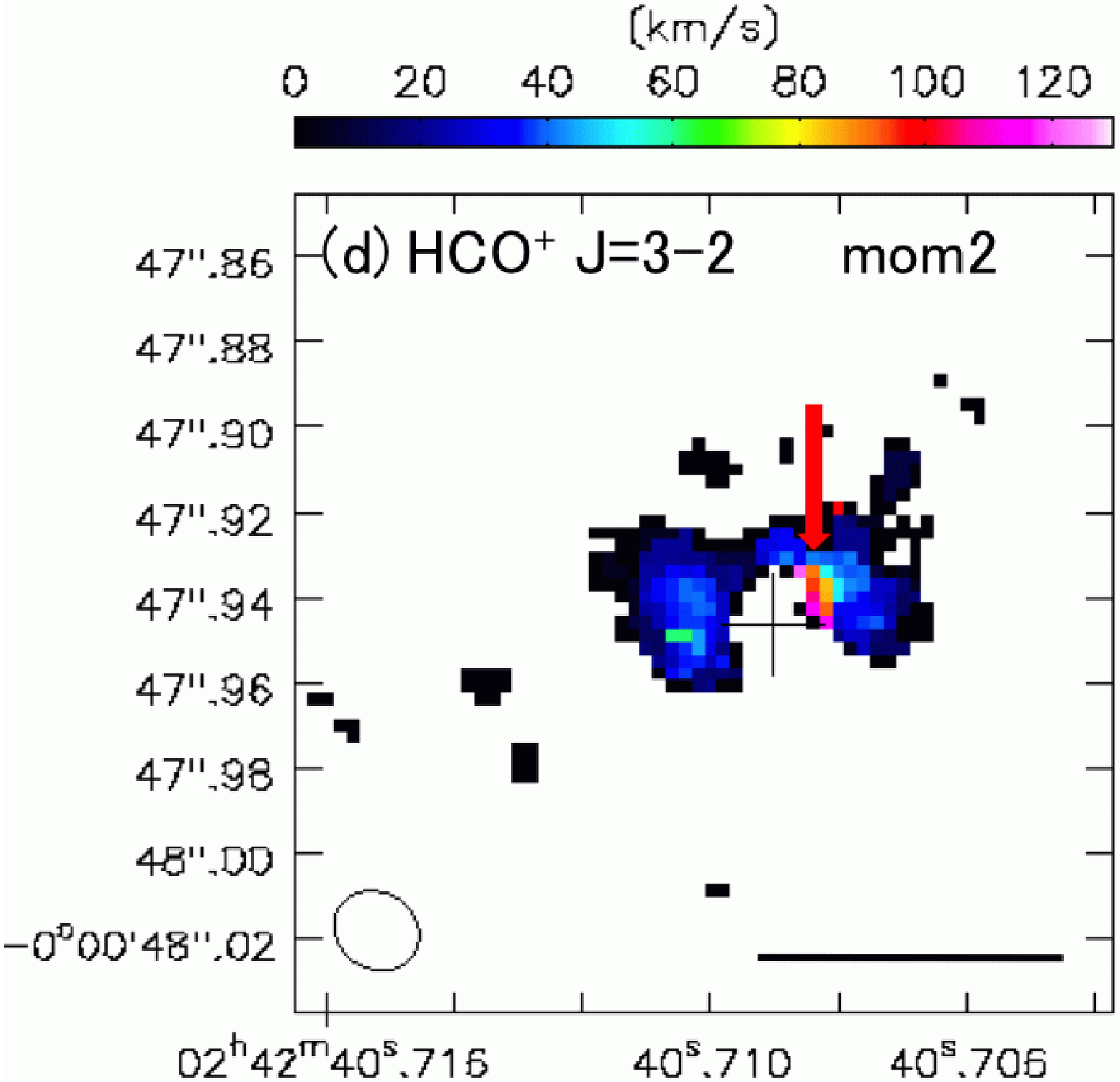} \\ 
\end{center}
\caption{
{\it (Top)}: Intensity-weighted mean velocity (moment 1) maps  
of {\it (a)} HCN J=3--2 and {\it (b)} HCO$^{+}$ J=3--2 emission lines.
The velocity display range is 930--1470 km s$^{-1}$ 
(from V$_{\rm sys}$ $-$ 200 to V$_{\rm sys}$ $+$ 340 km s$^{-1}$) 
for {\it (a)} HCN J=3--2,  
to display a redshifted high velocity ($\gtrsim$1300 km s$^{-1}$ 
or V$_{\rm sys}$ $+$ [$\gtrsim$170] km s$^{-1}$) 
emission component at the innermost western torus.
The range is 990--1300 km s$^{-1}$ (from V$_{\rm sys}$ $-$ 140 to 
V$_{\rm sys}$ $+$ 170 km s$^{-1}$)
for {\it (b)} HCO$^{+}$ J=3--2.
The black cross and horizontal black bar indicate the location of 
the continuum emission peak (= SMBH position) and 5 pc at the distance 
of NGC 1068, respectively.     
The innermost redshifted component in the western torus is 
indicated as a thick red downward arrow.
Beam size is shown as an open circle in the lower-left region.
{\it (Bottom)}: Intensity-weighted velocity dispersion (moment 2) 
maps of {\it (c)} HCN J=3--2 and {\it (d)} HCO$^{+}$ J=3--2 emission 
lines.  
The velocity display range is 0--130 km s$^{-1}$ for both lines.
The black cross, horizontal black bar, and open circle have the same 
meanings as in the top panels.
The high velocity dispersion region in the western torus is 
indicated as a thick red downward arrow.
For all maps in {\it (a--d)}, an appropriate cutoff ($\sim$4$\sigma$) 
is applied to prevent them from being overwhelmed by noise.
}
\end{figure}
 
\clearpage
 
\begin{figure}
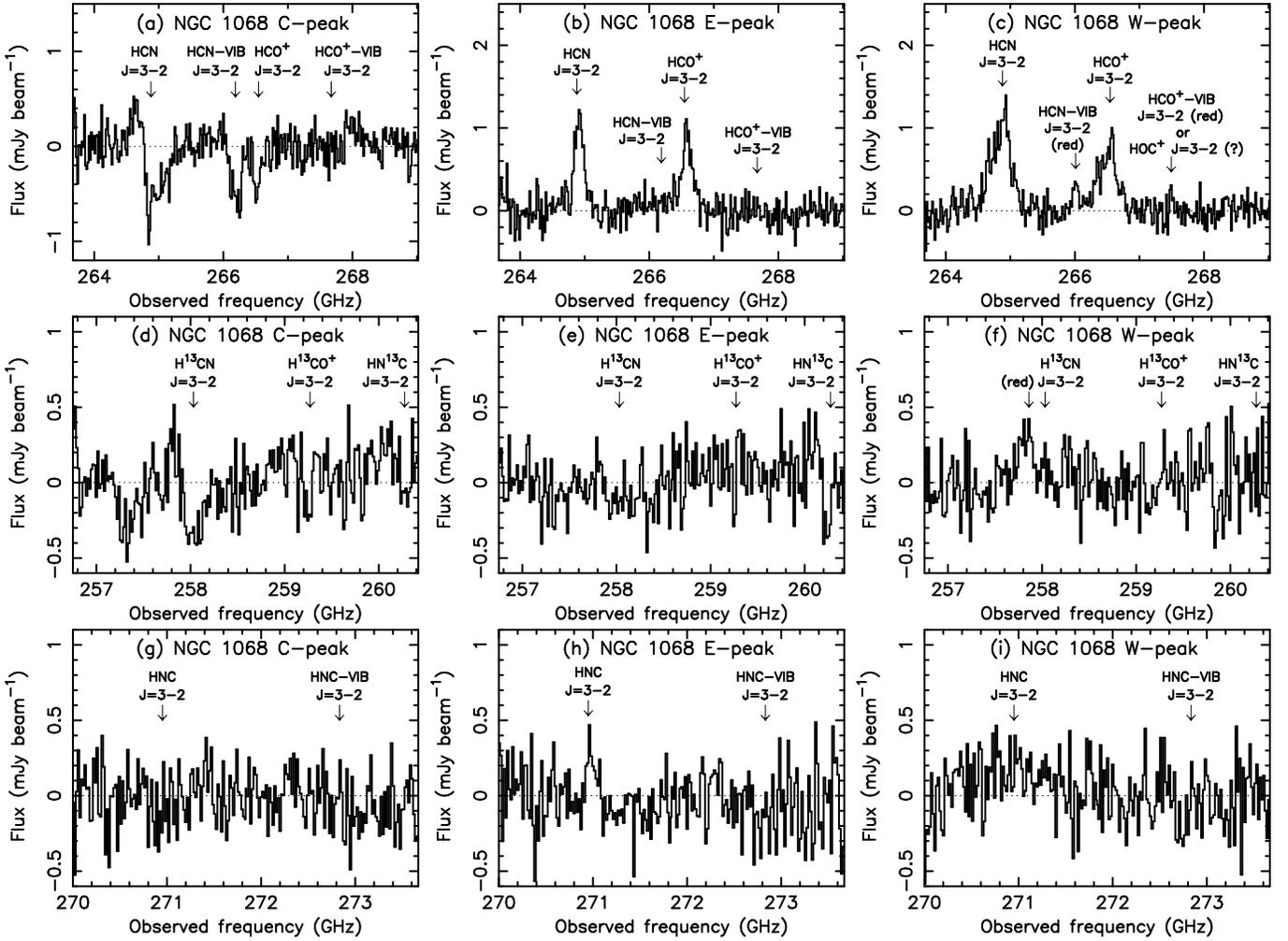

\begin{center}
\includegraphics[angle=-90,scale=.263]{f4a.eps} 
\includegraphics[angle=-90,scale=.263]{f4b.eps} 
\includegraphics[angle=-90,scale=.263]{f4c.eps} \\
\includegraphics[angle=-90,scale=.263]{f4d.eps} 
\includegraphics[angle=-90,scale=.263]{f4e.eps} 
\includegraphics[angle=-90,scale=.263]{f4f.eps} \\
\includegraphics[angle=-90,scale=.263]{f4g.eps} 
\includegraphics[angle=-90,scale=.263]{f4h.eps} 
\includegraphics[angle=-90,scale=.263]{f4i.eps} \\
\end{center}
\caption{
{\it (a), (b), (c)}: Spectra of data-a (HCN/HCO$^{+}$ J=3--2 observations), 
within the beam size ($\sim$0$\farcs$02), at the C-peak, E-peak, and 
W-peak, respectively. 
Some targeted lines are indicated with downward arrows at 
V$_{\rm sys}$ = 1130 km s$^{-1}$.
In the spectrum at {\it (c)} W-peak, downward arrows are added at 
V = 1340 km s$^{-1}$ (V$_{\rm sys}$ $+$ 210 km s$^{-1}$)
for HCN-VIB J=3--2 and 
HCO$^{+}$-VIB J=3--2 lines with the note of ``(red)'', to represent 
redshifted emission component at the innermost western torus 
(see $\S$4.3 and 4.5).
{\it (d), (e), (f)}: Lower frequency part of the spectrum of data-b 
($^{13}$C isotopologue/HNC J=3--2 observations), 
within the beam size ($\sim$0$\farcs$03), 
at the C-peak, E-peak, and W-peak, respectively. 
In {\it (f)}, a downward arrow is added at V = 1340 km s$^{-1}$ 
(V$_{\rm sys}$ $+$ 210 km s$^{-1}$)
for H$^{13}$CN J=3--2 with the note of ``(red)''.
{\it (g), (h), (i)}: Higher frequency part of the 
beam-sized ($\sim$0$\farcs$03) spectrum of data-b 
at the C-peak, E-peak, and W-peak, respectively. 
Note that the beam size is slightly larger for data-b (see Tables 2 and 3).
In all plots, the abscissa is observed frequency (in GHz) and 
the ordinate is flux density (in mJy beam$^{-1}$).
The thin dotted horizontal straight line indicates the zero flux level.
}
\end{figure}

\clearpage

\begin{figure}
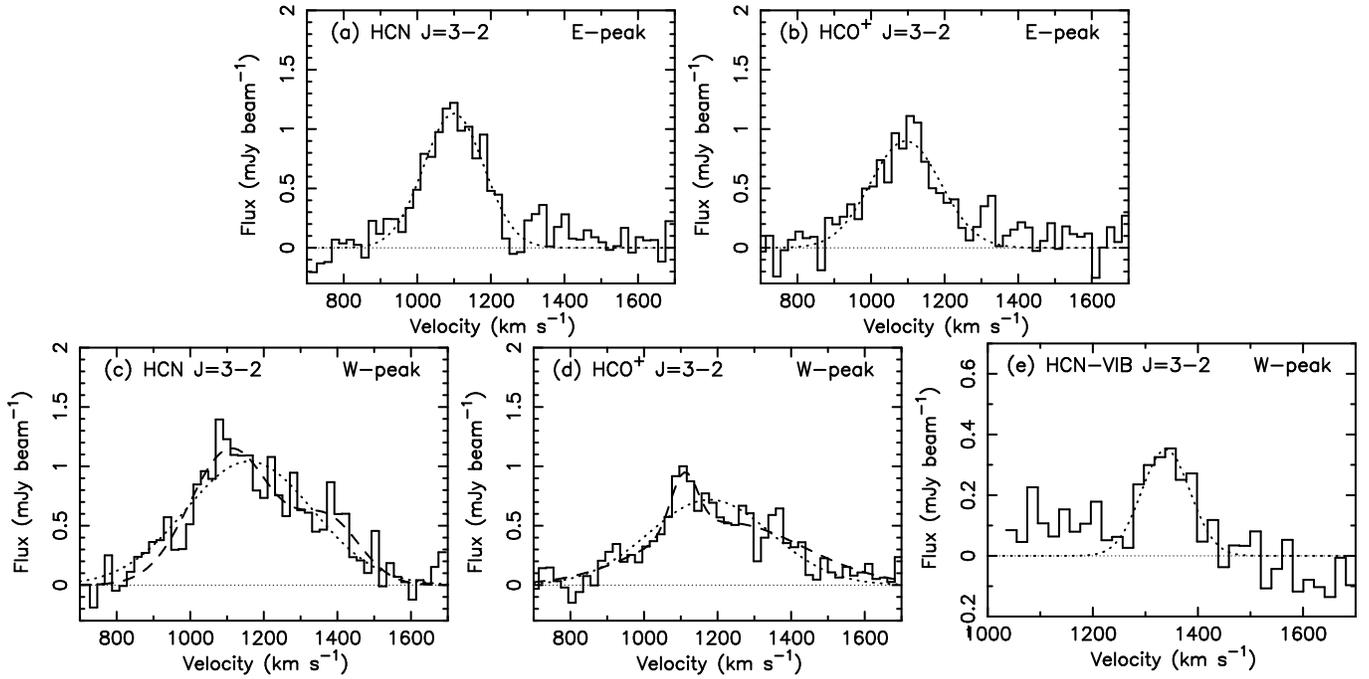

\begin{center}
\includegraphics[angle=-90,scale=.262]{f5a.eps} 
\includegraphics[angle=-90,scale=.262]{f5b.eps} \\
\includegraphics[angle=-90,scale=.262]{f5c.eps} 
\includegraphics[angle=-90,scale=.262]{f5d.eps} 
\includegraphics[angle=-90,scale=.262]{f5e.eps} 
\end{center}
\caption{
Gaussian fit of detected molecular emission line in the beam-sized 
spectrum. 
{\it (a)}: HCN J=3--2 at the eastern torus peak (E-peak). 
{\it (b)}: HCO$^{+}$ J=3--2 at the E-peak. 
{\it (c)}: HCN J=3--2 at the western torus peak (W-peak). 
{\it (d)}: HCO$^{+}$ J=3--2 at the W-peak. 
{\it (e)}: HCN-VIB J=3--2 at the W-peak. 
The abscissa is optical LSR velocity (in km s$^{-1}$) and the ordinate 
is flux density (in mJy beam$^{-1}$).
Single Gaussian fit is shown as curved dotted line.
For the HCN J=3--2 and HCO$^{+}$ J=3--2 emission lines at the W-peak, 
two Gaussian fit is also added with curved dashed line.
The thin dotted horizontal straight line indicates the zero flux level.
}
\end{figure}

\begin{figure}
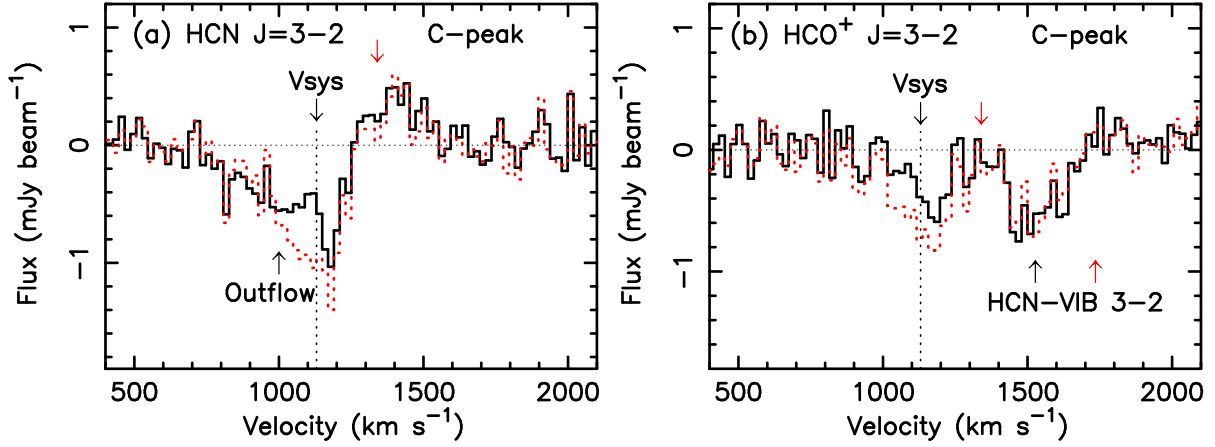

\begin{center}
\includegraphics[angle=-90,scale=.35]{f6a.eps} 
\includegraphics[angle=-90,scale=.35]{f6b.eps} 
\end{center}
\caption{
Velocity profile of {\it (a)} HCN J=3--2 and {\it (b)} HCO$^{+}$ J=3--2 
lines at the C-peak.
The black solid line indicates an observed spectrum within the beam size.
The red dotted line is a spectrum after an attempt to correct 
for the possible emission contamination from the innermost parts of the 
western and eastern torus (see $\S$4.7).
The abscissa is optical LSR velocity (in km s$^{-1}$) and the ordinate 
is flux density (in mJy beam$^{-1}$).
The thin dotted horizontal straight line indicates the zero flux level.
The systemic velocity (1130 km s$^{-1}$) is shown as 
a vertical dotted line with the note of ``V$_{\rm sys}$''. 
A red downward arrow is added at the velocity of V = 1340 km s$^{-1}$ 
(V$_{\rm sys}$ $+$ 210 km s$^{-1}$)
to indicate redshifted emission at the innermost western torus 
(see $\S$4.5).
In {\it (a)}, a blueshifted broad absorption feature (discussed in $\S$4.7) 
is indicated with a black upward arrow with the note of ``Outflow''.  
In {\it (b)}, a black and red upward arrow with the note of ``HCN-VIB 3--2'' 
indicates HCN-VIB J=3--2 line at V$_{\rm sys}$ = 1130 km s$^{-1}$ and 
V = 1340 km s$^{-1}$ (V$_{\rm sys}$ $+$ 210 km s$^{-1}$), respectively.
}
\end{figure}

\begin{figure}
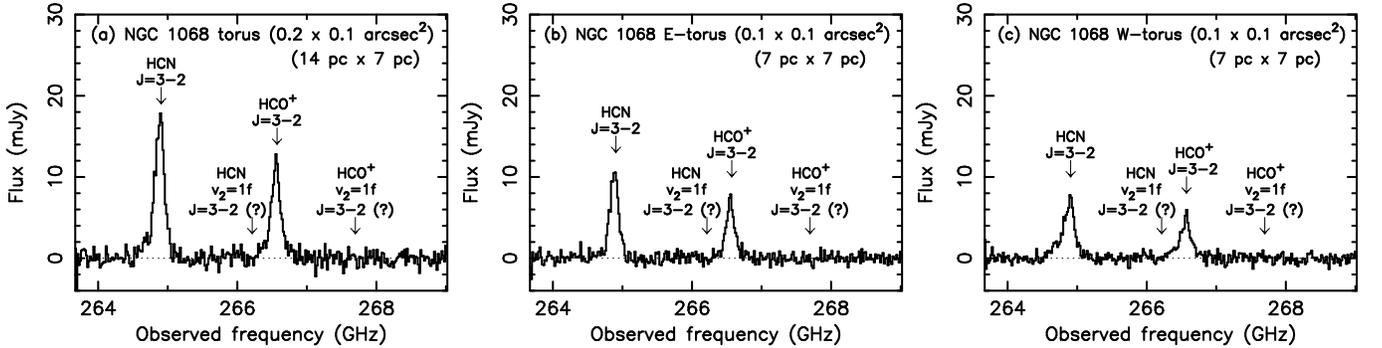

\begin{center}
\includegraphics[angle=-90,scale=.265]{f7a.eps} 
\includegraphics[angle=-90,scale=.265]{f7b.eps} 
\includegraphics[angle=-90,scale=.265]{f7c.eps}  
\end{center}
\caption{
{\it (a)}: An area-integrated spectrum of HCN J=3--2 and 
HCO$^{+}$ J=3--2 emission lines with 0$\farcs$2 east-west and 
0$\farcs$1 north-south (14 pc $\times$ 7 pc) rectangular 
region centered at the C-peak. 
{\it (b)}: An area-integrated spectrum of 0$\farcs$1 $\times$ 
0$\farcs$1 square region in the eastern torus 
(E-torus; eastern half of the rectangular region).
{\it (c)}: An area-integrated spectrum of 0$\farcs$1 $\times$ 
0$\farcs$1 square region in the western torus 
(W-torus; western half of the rectangular region).
The abscissa is observed frequency (in GHz) and the ordinate 
is flux density (in mJy).
The thin dotted horizontal straight line indicates the zero flux level.
}
\end{figure}

\begin{figure}
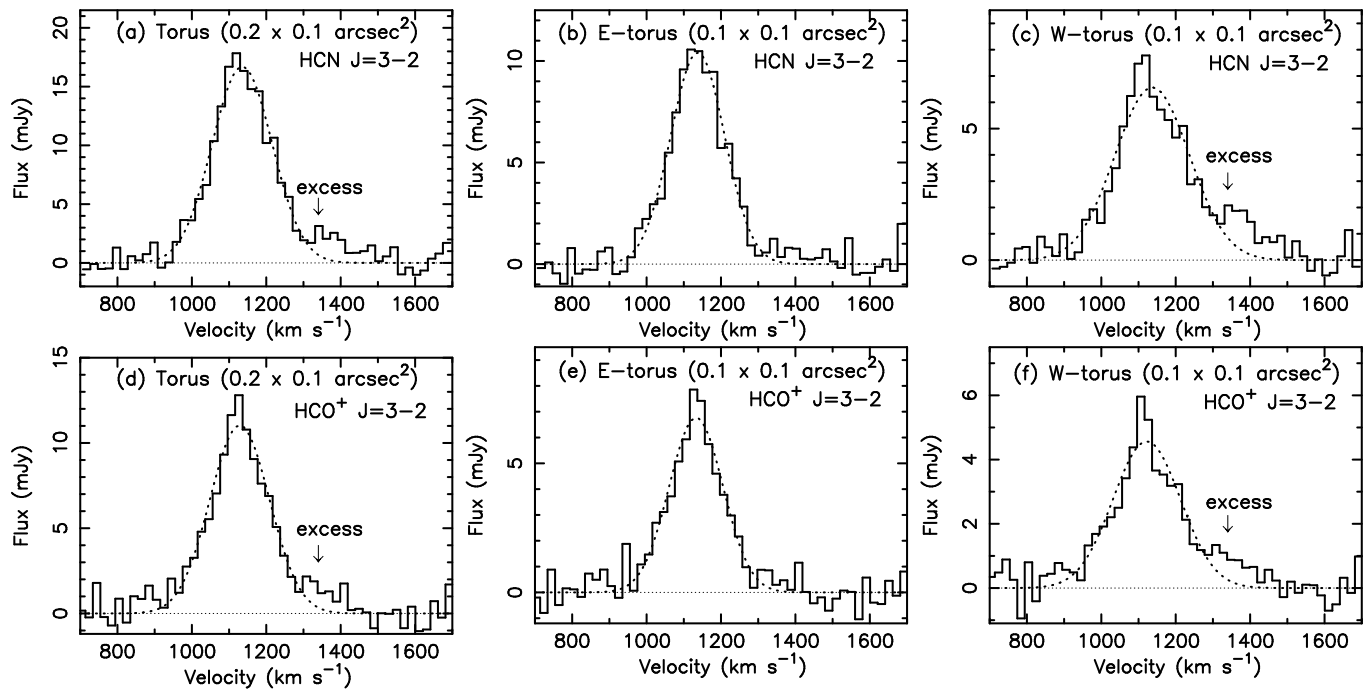

\begin{center}
\includegraphics[angle=-90,scale=.265]{f8a.eps} 
\includegraphics[angle=-90,scale=.265]{f8b.eps} 
\includegraphics[angle=-90,scale=.265]{f8c.eps} 
\includegraphics[angle=-90,scale=.265]{f8d.eps} 
\includegraphics[angle=-90,scale=.265]{f8e.eps} 
\includegraphics[angle=-90,scale=.265]{f8f.eps} 
\end{center}
\caption{
Gaussian fits of {\it (Top)} HCN J=3--2 and 
{\it (Bottom)} HCO$^{+}$ J=3--2 emission lines in area-integrated spectra 
shown in Figure 7.
{\it (a), (b), (c)}: HCN J=3--2 emission from the torus 
(0$\farcs$2 $\times$ 0$\farcs$1 [= 14 pc $\times$ 7 pc] rectangular region), 
E-torus (0$\farcs$1 $\times$ 0$\farcs$1 square region), and 
W-torus (0$\farcs$1 $\times$ 0$\farcs$1 square region), respectively. 
{\it (d), (e), (f)}: HCO$^{+}$ J=3--2 emission from the torus, E-torus,
and W-torus, respectively.
The abscissa is optical LSR velocity (in km s$^{-1}$) and the ordinate 
is flux density (in mJy).
The thin dotted horizontal straight line indicates the zero flux level.
Single Gaussian fit is shown with a curved dotted line.
Redshifted excess emission, relative to the Gaussian component 
(discussed in $\S$4.3), is indicated with 
a downward arrow with the note of ``excess''.
}
\end{figure}

\begin{figure}
\begin{center}
\includegraphics[angle=0,scale=.249]{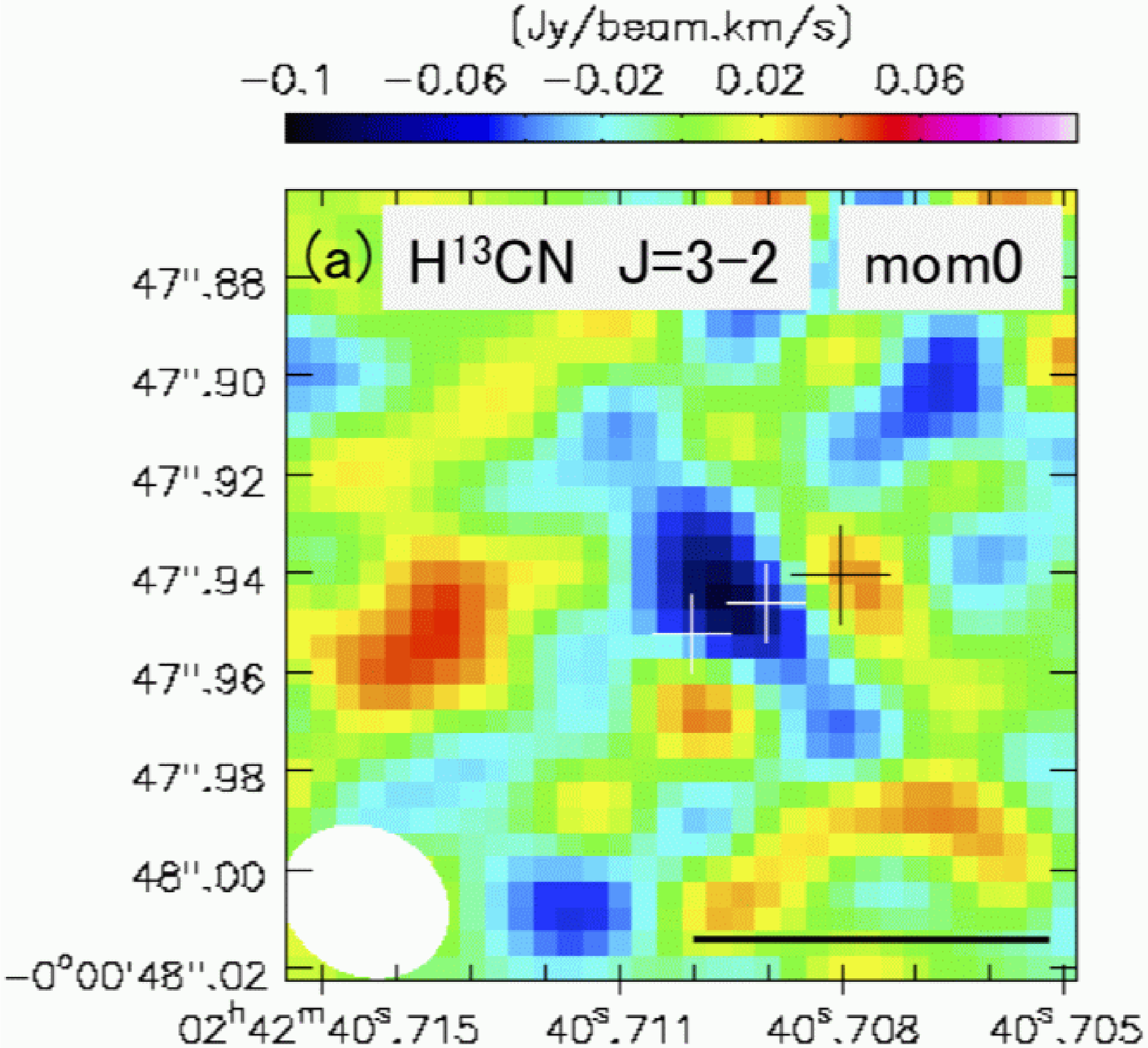} 
\includegraphics[angle=0,scale=.249]{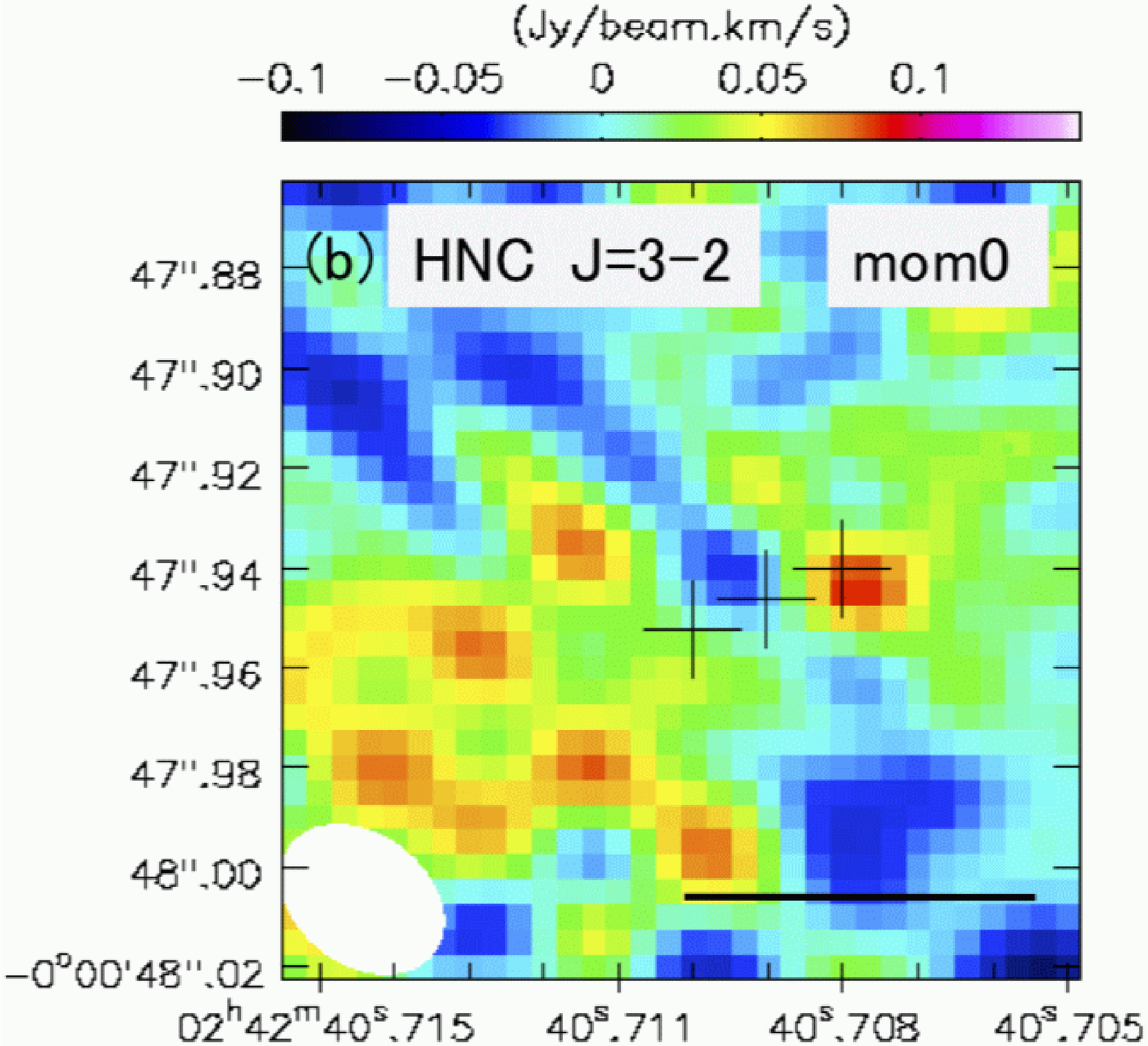} 
\includegraphics[angle=0,scale=.249]{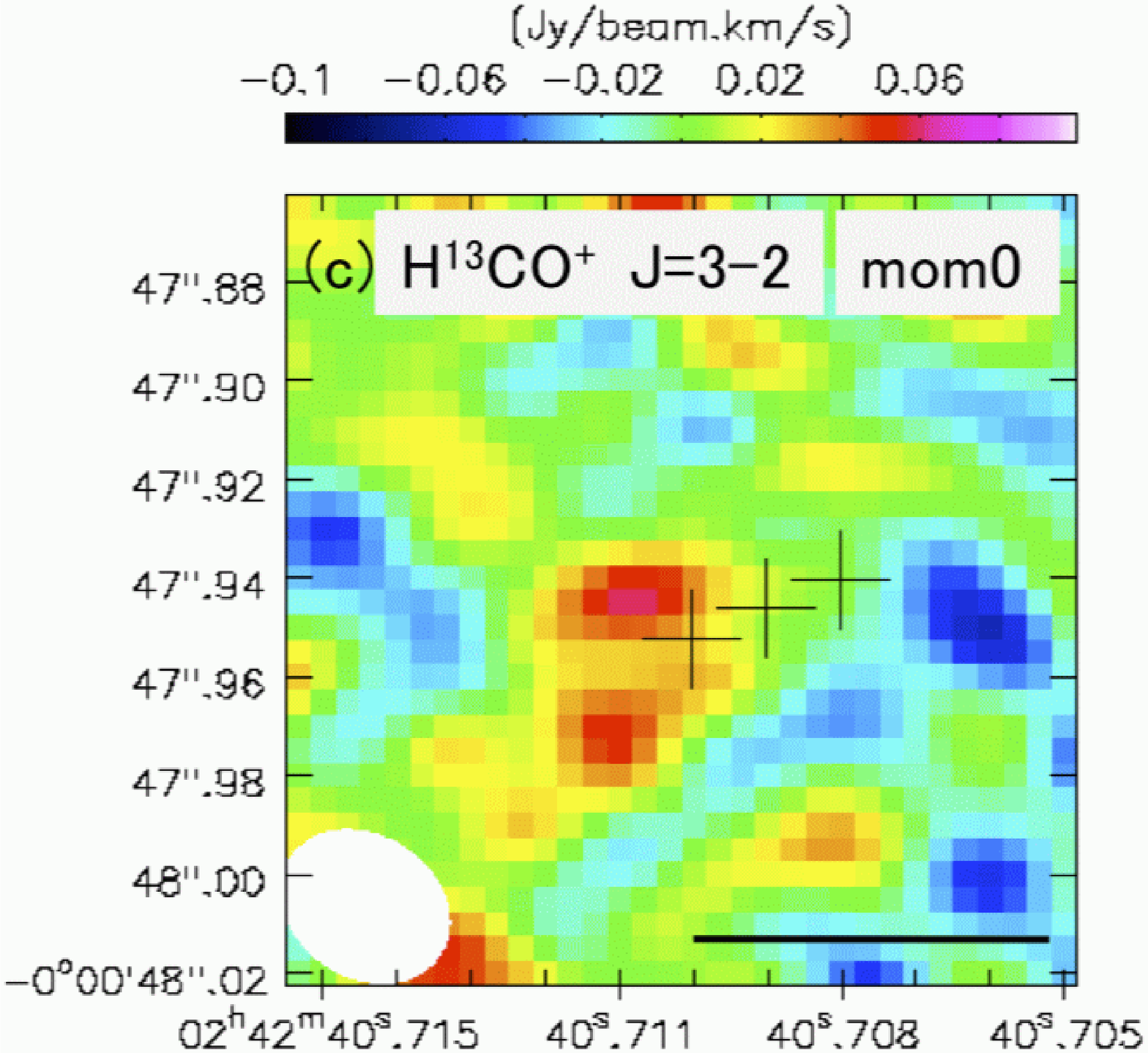} 
\end{center}
\caption{
Integrated intensity (moment 0) map of {\it (a)} H$^{13}$CN J=3--2,  
{\it (b)} HNC J=3--2, and {\it (c)} H$^{13}$CO$^{+}$ J=3--2. 
Three cross marks are added at the E-peak, C-peak, and W-peak (Table 3) 
from the left to the right.
In all maps, signals in the same velocity range (920--1320 km s$^{-1}$ 
or from V$_{\rm sys}$ $-$ 210 to V$_{\rm sys}$ $+$ 190 km s$^{-1}$) 
are integrated.
The rms noise level is 0.020, 0.025, and 0.024 (Jy beam$^{-1}$ km s$^{-1}$) 
for {\it (a)} H$^{13}$CN J=3--2, {\it (b)} HNC J=3--2, and 
{\it (c)} H$^{13}$CO$^{+}$ J=3--2, respectively.
For {\it (a)} H$^{13}$CN J=3--2, a clear negative (i.e., absorption) signal 
(4.5$\sigma$; blue colored) is recognizable close to the C-peak 
(middle cross). 
For {\it (b)} HNC J=3--2, an emission signal (red colored) is seen close to 
the W-peak (right cross) with 3.6$\sigma$.
For {\it (a)} and {\it (b)}, any other possible emission/absorption 
signatures are insignificant ($<$3$\sigma$).
For {\it (c)} H$^{13}$CO$^{+}$ J=3--2, neither emission nor absorption signal 
is significant in the torus region ($<$3$\sigma$).
The horizontal black bar at the lower-right side indicates 5 pc 
at the distance of NGC 1068.     
Beam size is shown as a filled circle in the lower-left region.
}
\end{figure}

\begin{figure}
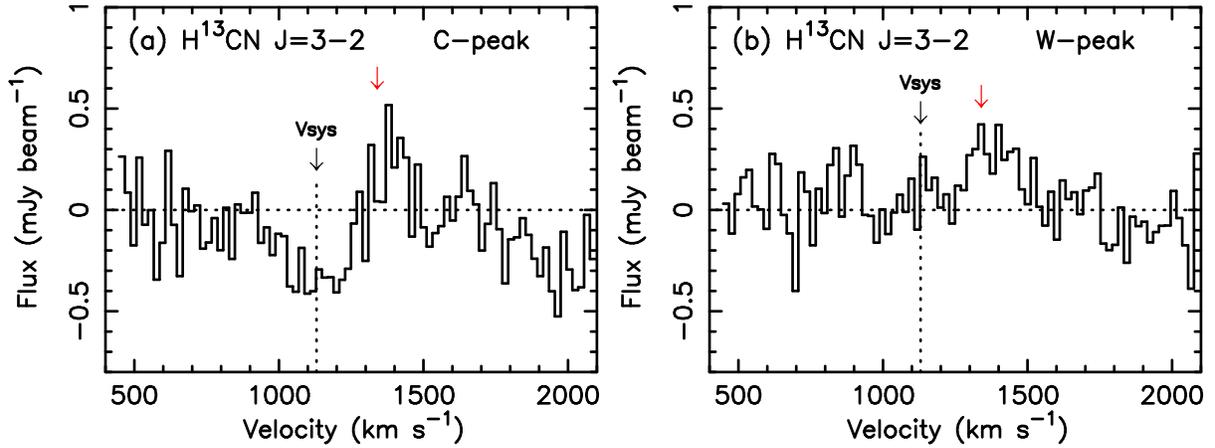

\begin{center}
\includegraphics[angle=-90,scale=.35]{f10a.eps} 
\includegraphics[angle=-90,scale=.35]{f10b.eps} 
\end{center}
\caption{
Velocity profile of H$^{13}$CN J=3--2 line at the {\it (a)} C-peak
and {\it (b)} W-peak. 
The abscissa is optical LSR velocity (in km s$^{-1}$) and the ordinate 
is flux density (in mJy beam$^{-1}$).
The thin dotted horizontal straight line indicates the zero flux level.
The systemic velocity (1130 km s$^{-1}$) is shown as a vertical 
dotted line with the note of ``V$_{\rm sys}$''. 
A red downward arrow is added at V = 1340 km s$^{-1}$ 
(V$_{\rm sys}$ $+$ 210 km s$^{-1}$), as in Figure 6.
}
\end{figure}

\begin{figure}
\begin{center}
\includegraphics[angle=0,scale=0.75]{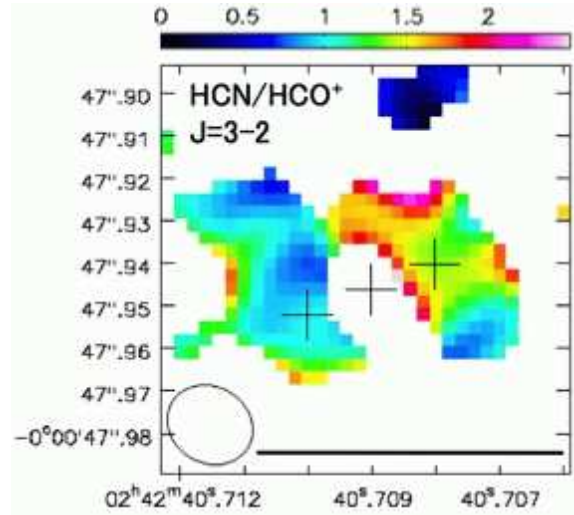} 
\end{center}
\caption{
The ratio of HCN J=3--2 flux to HCO$^{+}$ J=3--2 flux 
(in Jy beam$^{-1}$ km s$^{-1}$).
Only pixels with $>$3$\sigma$ HCO$^{+}$ J=3--2 detection (i.e., denominator) 
in the moment 0 map in Figure 2b are shown. 
Thus, this ratio map does not preferentially pick up elevated HCN emission 
regions only. 
Three cross marks are added at the E-peak, C-peak, and W-peak (Table 3) 
from the left to the right.
The horizontal black bar at the lower-right side indicates 5 pc 
at the distance of NGC 1068.     
Beam size is shown as an open circle in the lower-left region.
}
\end{figure}

\begin{figure}
\begin{center}
\includegraphics[angle=-90,scale=.35]{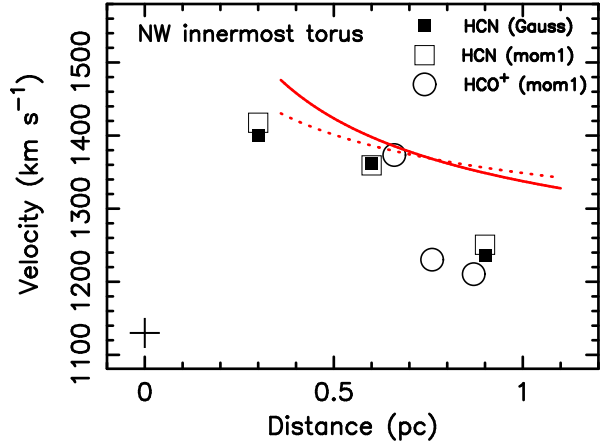} 
\end{center}
\caption{
Velocity of the redshifted HCN J=3--2 and HCO$^{+}$ J=3--2 
emission lines at the innermost northwestern torus.
The abscissa is projected distance from the central SMBH (in pc), and 
the ordinate is velocity (in km s$^{-1}$).  
The plus mark indicates the systemic velocity (V$_{\rm sys}$ = 1130 km s$^{-1}$) 
at the SMBH position.
Filled square: HCN J=3--2 velocity based on the Gaussian fit of 
beam-sized spectrum extracted at each distance from the central SMBH.
Open square: HCN J=3--2 intensity-weighted velocity derived from 
moment 1 map (Figure 3a).
Open circle: HCO$^{+}$ J=3--2 intensity-weighted velocity from moment 1 map 
(Figure 3b).
HCN J=3--2 line velocity is measured along PA = 135$^{\circ}$ east of north.
HCO$^{+}$ J=3--2 line velocity is measured along PA = 105--120$^{\circ}$, 
because strong HCO$^{+}$ J=3--2 emission is detected along this direction.
The Gaussian fit of the HCO$^{+}$ J=3--2 emission line is not shown, 
owing to large uncertainty caused by a strong absorption feature 
at 1400--1700 km s$^{-1}$ (or V$_{\rm sys}$ $+$ [270--570] km s$^{-1}$)
seen in Figure 6b.
The red solid and dotted lines mean, respectively, 
Keplerian motion (V $\propto$ r$^{-0.5}$) with the central SMBH mass 
of $\sim$ 1 $\times$ 10$^{7}$M$_{\odot}$ \citep{gre96} and 
inclination $\it i$ = 90$^{\circ}$ (edge-on), and sub-Keplerian motion with 
V $\propto$ r$^{-0.31}$ ($\it i$ = 90$^{\circ}$) that reproduces 
$\sim$250 km s$^{-1}$ rotation velocity at $\sim$0.65 pc \citep{gre96}.
}
\end{figure}

\begin{figure}
\begin{center}
\includegraphics[angle=0,scale=.2]{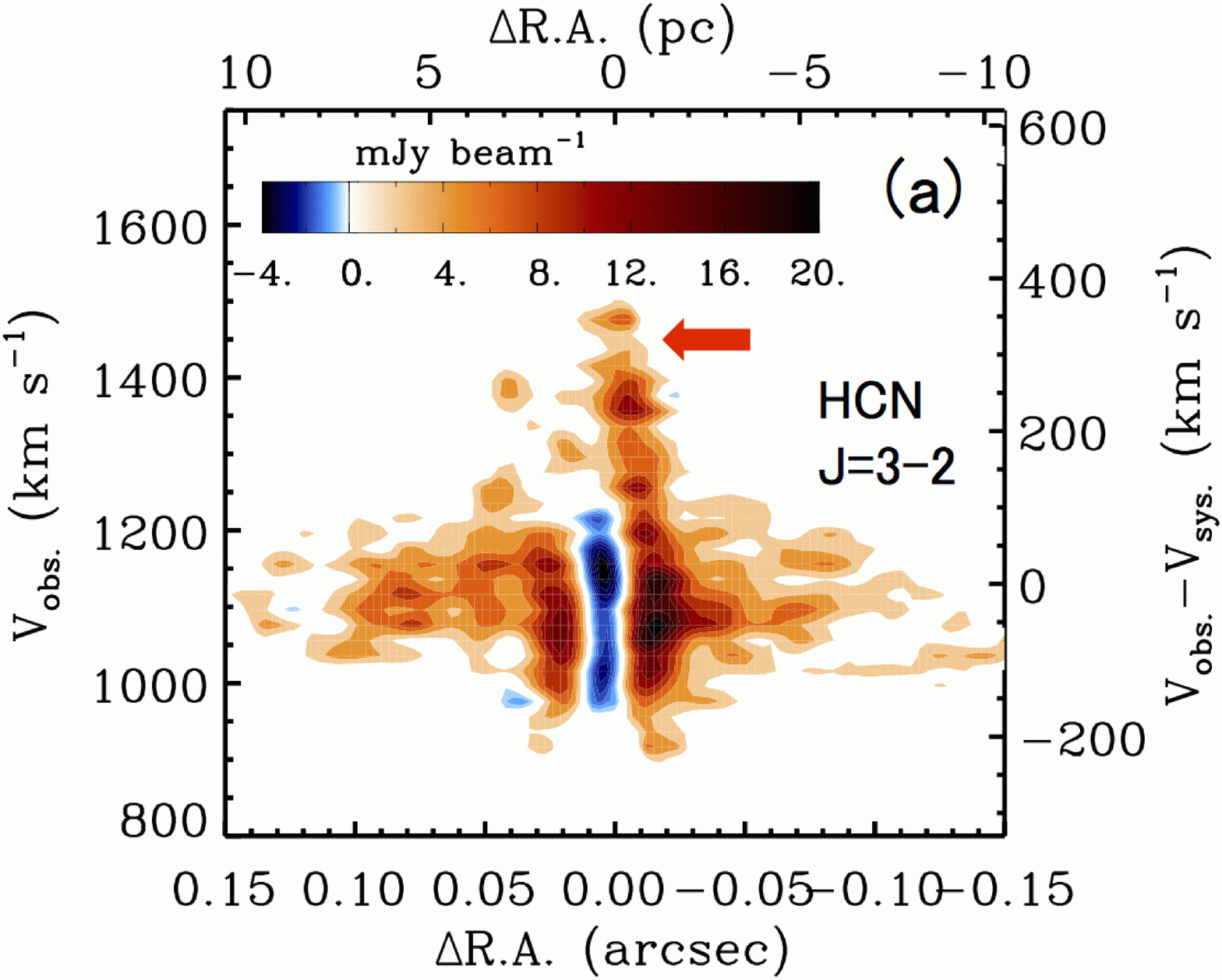} 
\includegraphics[angle=0,scale=.2]{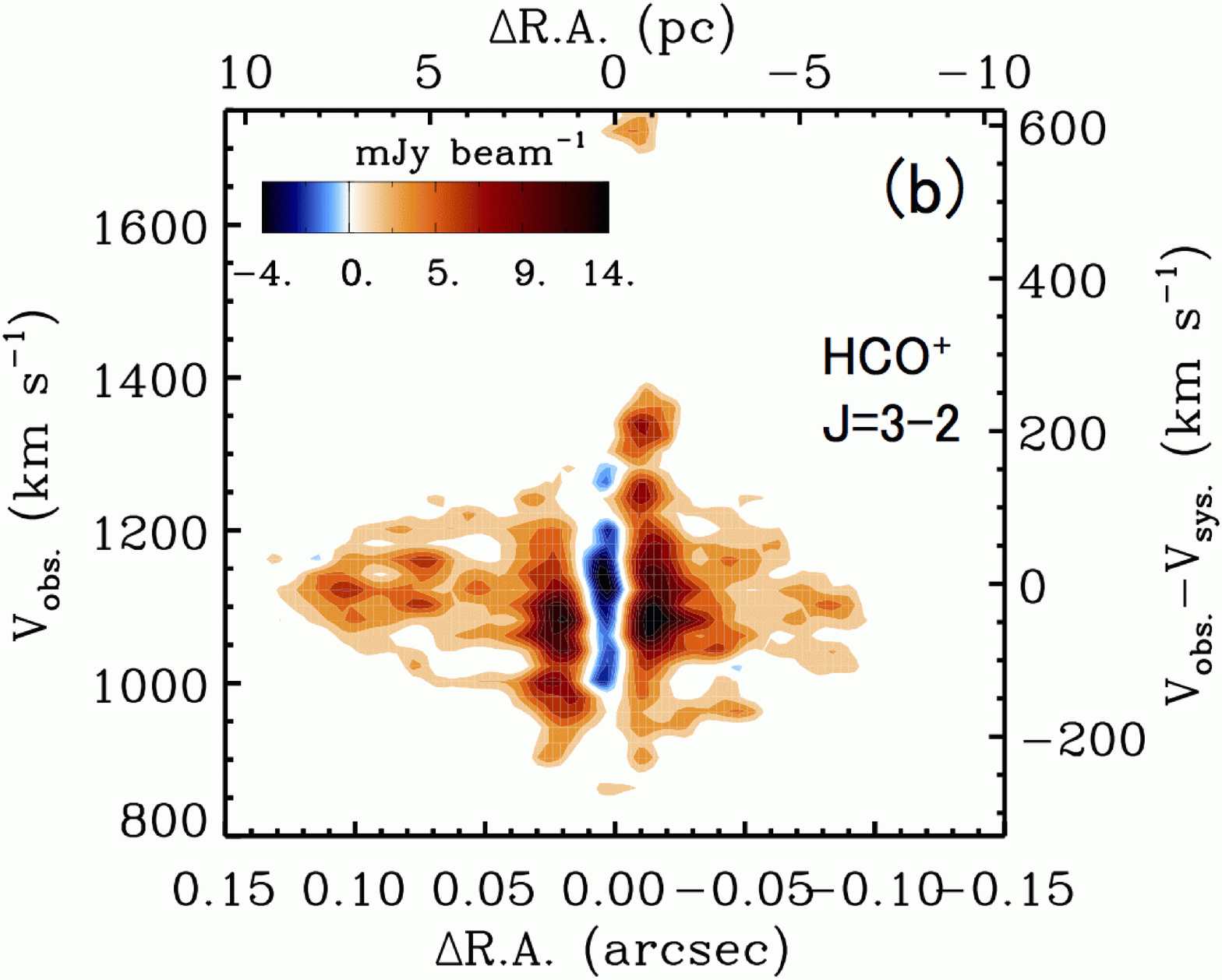} 
\end{center}
\caption{
Position velocity diagram of {\it (a)} HCN J=3--2 and 
{\it (b)} HCO$^{+}$ J=3--2 lines along PA = 115$^{\circ}$ east of north.
The abscissa is offset from the SMBH position in arcsec (bottom) and 
in pc (top).
The ordinate is optical LSR velocity (in km s$^{-1}$; left) and 
velocity relative to the systemic one of V$_{\rm sys}$ = 1130 km s$^{-1}$ 
(in km s$^{-1}$; right).
Southeast is to the left and northwest is to the right.
The thick red leftward arrow in {\it (a)} indicates the redshifted 
high velocity (V $\sim$ 1300--1500 km s$^{-1}$ 
or V$_{\rm sys}$ $+$ [170--370] km s$^{-1}$) HCN J=3--2 emission 
component at the innermost northwestern torus discussed in $\S$4.3.
In {\it (b)}, signal at the very top (V $\gtrsim$ 1700 km s$^{-1}$
or V$_{\rm sys}$ $+$ [$\gtrsim$570] km s$^{-1}$) is 
not from the HCO$^{+}$ J=3--2 line, but from the redshifted HCN-VIB 
J=3--2 line emitted at the innermost northwestern torus (Figure 6b).
}
\end{figure}

\begin{figure}
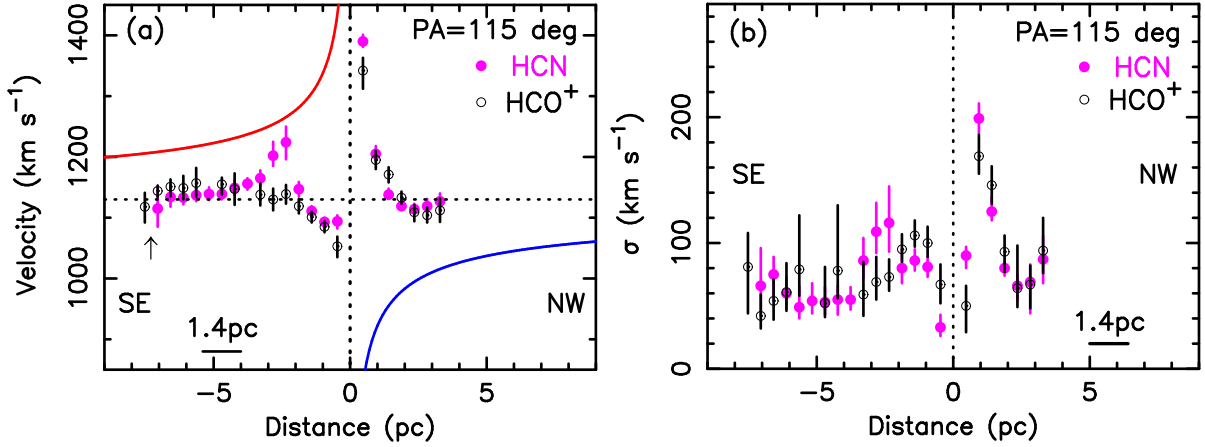

\begin{center}
\includegraphics[angle=-90,scale=.35]{f14a.eps} 
\includegraphics[angle=-90,scale=.35]{f14b.eps} 
\end{center}
\caption{
{\it (a)}:
Velocity of inner to outer torus, along PA $\sim$ 115$^{\circ}$, 
based on Gaussian fit of beam-sized spectrum extracted at each position.
The abscissa is projected distance from the central SMBH (in pc) and 
the ordinate is peak velocity, V$_{\rm c}$ (in km s$^{-1}$), of the 
Gaussian function of A$\times$exp($-$[V$-$V$_{\rm c}$]$^{2}$/2$\sigma^{2}$), 
where A is a constant, V is velocity (in km s$^{-1}$), and $\sigma$ 
is emission line width (in km s$^{-1}$). 
The thin dotted vertical and horizontal straight line indicates 
SMBH position and systemic velocity (V$_{\rm sys}$ = 1130 km s$^{-1}$), 
respectively.
Magenta filled circle: HCN J=3--2. 
Black open circle: HCO$^{+}$ J=3--2.
Southeast is to the left and northwest is to the right.
Peak velocity uncertainty based on the Gaussian fit is 
$\lesssim$20 km s$^{-1}$ in almost all points, except for a few 
points with faint molecular line emission where the uncertainty is 
$\sim$30 km s$^{-1}$.
Data points with uncertainty $\gtrsim$30 km s$^{-1}$ are excluded.
The red and blue solid curved lines are Keplerian velocity for the 
central SMBH mass of 1 $\times$ 10$^{7}$M$_{\odot}$ ({\it i} = 90$^{\circ}$), 
with the eastern (western) part being redshifted (blueshifted).
The thin black upward arrow indicates abrupt velocity change 
from redshifted to blueshifted motion at $\sim$7 pc southeastern side of 
the SMBH (discussed in $\S$4.6).
The redshifted components at 0--1 pc (just right side of the SMBH) 
correspond to the redshifted dense molecular line emission at the 
innermost northwestern torus ($\S$4.3).
The velocity of the redshifted innermost northwestern HCO$^{+}$ J=3--2 
data point (black open circle just right side of the SMBH) can be 
largely affected by the HCN-VIB J=3--2 absorption feature (Figure 6b). 
Blueshifted components are also seen at the innermost ($\lesssim$1 pc) 
southeastern torus (just left side of the SMBH), although its velocity, 
relative to the systemic, is smaller than that in the northwestern side.
The horizontal solid bar at the lower part is our beam size 
($\sim$0$\farcs$02 or $\sim$1.4 pc).
{\it (b)}:
Same as {\it (a)}, but the ordinate is emission line velocity 
width, $\sigma$ (in km s$^{-1}$), of 
A$\times$exp($-$[V$-$V$_{\rm c}$]$^{2}$/2$\sigma^{2}$).
}
\end{figure}

\begin{figure}
\begin{center}
\includegraphics[angle=0,scale=.35]{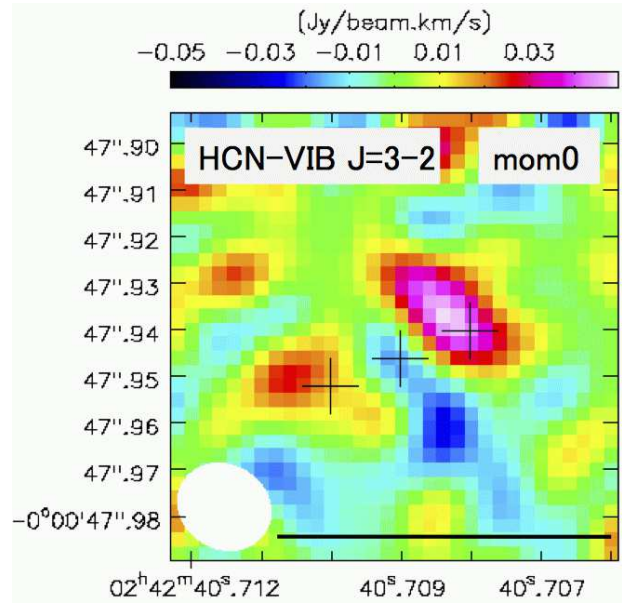} 
\end{center}
\caption{
HCN-VIB J=3--2 line moment 0 map. 
Only signals at V=1170--1430 km s$^{-1}$ 
(V$_{\rm sys}$ $+$ [40--300] km s$^{-1}$)
are integrated, to avoid 
possible contamination from the bright HCO$^{+}$ J=3--2 emission line.
The rms noise is 0.010 (Jy beam$^{-1}$ km s$^{-1}$).
The three crosses are E-peak, C-peak, and W-peak, from left to right.
Signal detection with $\sim$4.8$\sigma$ (white spot) is seen at the 
northwestern side of the C-peak (close to the W-peak).
The horizontal black bar at the lower right corresponds to 5 pc at the 
distance of NGC 1068.     
Beam size is shown as a filled circle in the lower-left region.
}
\end{figure}

\begin{figure}
\begin{center}
\includegraphics[angle=0,scale=.45]{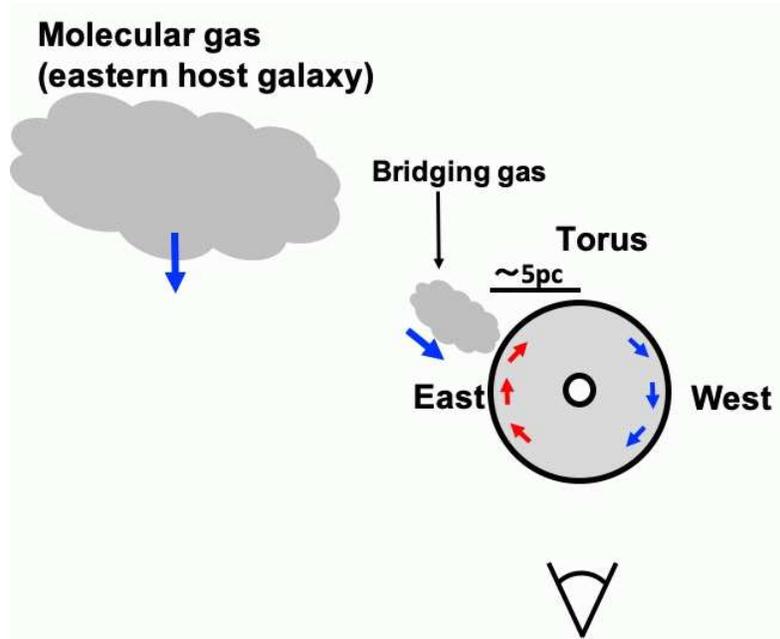}
\end{center}
\caption{
Schematic diagram of molecular gas outside the eastern torus.
East is to the left and west is to the right.
We are seeing from the bottom. 
Blue and red arrow means blueshifted and redshifted motion, with respect to 
the systemic velocity, respectively.
Molecular gas in the eastern host galaxy is at a far side from us 
and is blueshifted.
A blueshifted bridging molecular clump exists just outside the eastern torus.
Torus outer part is redshifted at the eastern side and blueshifted at the 
western side.
}
\end{figure}

\begin{figure}
\begin{center}
\includegraphics[angle=0,scale=.3]{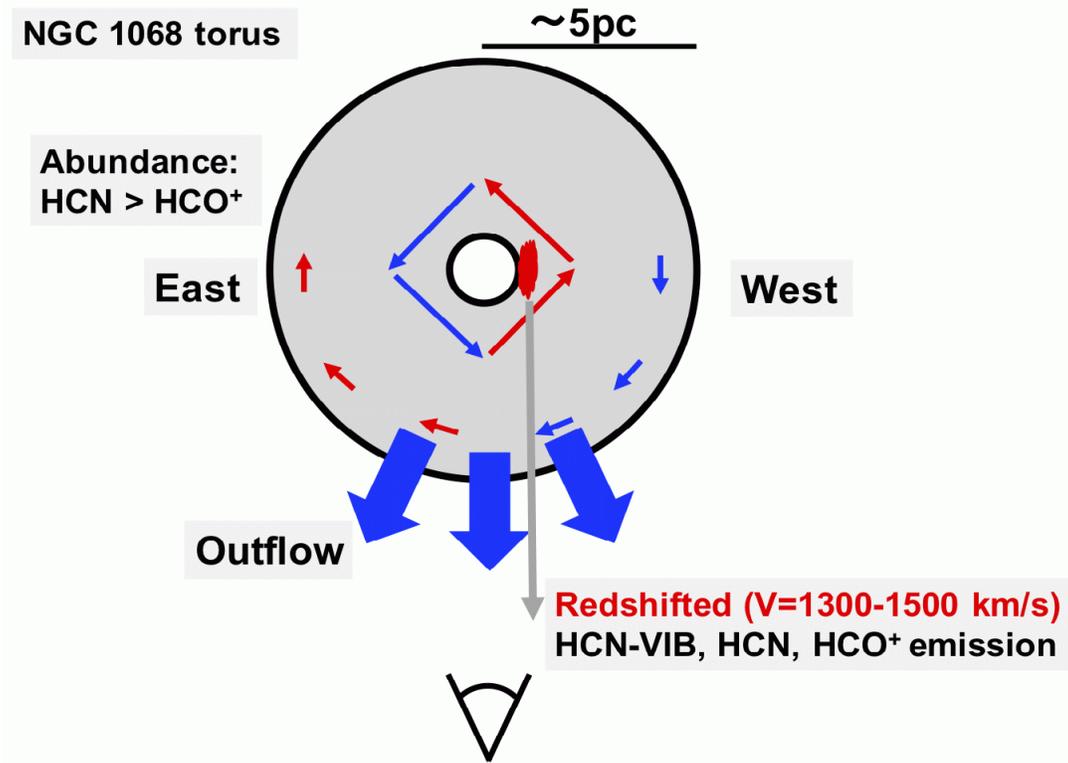} 
\end{center}
\caption{
Schematic diagram of the NGC 1068 torus at $\lesssim$5 pc scale.
East is to the left and west is to the right. 
We are seeing the torus from the bottom.
An innermost almost-Keplerian rotating (thin long arrow) component and 
an outer very slowly counter-rotating (thin short arrow) component are 
present.
Blue and red mean blueshifted and redshifted motions, respectively. 
(The length of the arrow is just for presentation purpose and is not 
scaled to actual velocity.)
Dense molecular outflow (thick blue arrow) toward our direction is 
indicated from the blueshifted absorption wings of HCN J=3--2 and 
HCO$^{+}$ J=3--2 lines in the beam-sized C-peak spectra.
The redshifted (V = 1300--1500 km s$^{-1}$ 
or V$_{\rm sys}$ $+$ [170--370] km s$^{-1}$) HCN-VIB J=3--2, 
HCN J=3--2, 
and HCO$^{+}$ J=3--2 emission lines come from the innermost northwestern 
torus (red filled region).
Higher abundance of HCN than HCO$^{+}$ is suggested throughout the torus.
}
\end{figure}

\begin{figure}
\begin{center}
\vspace*{-0.8cm}
\includegraphics[angle=0,scale=.3]{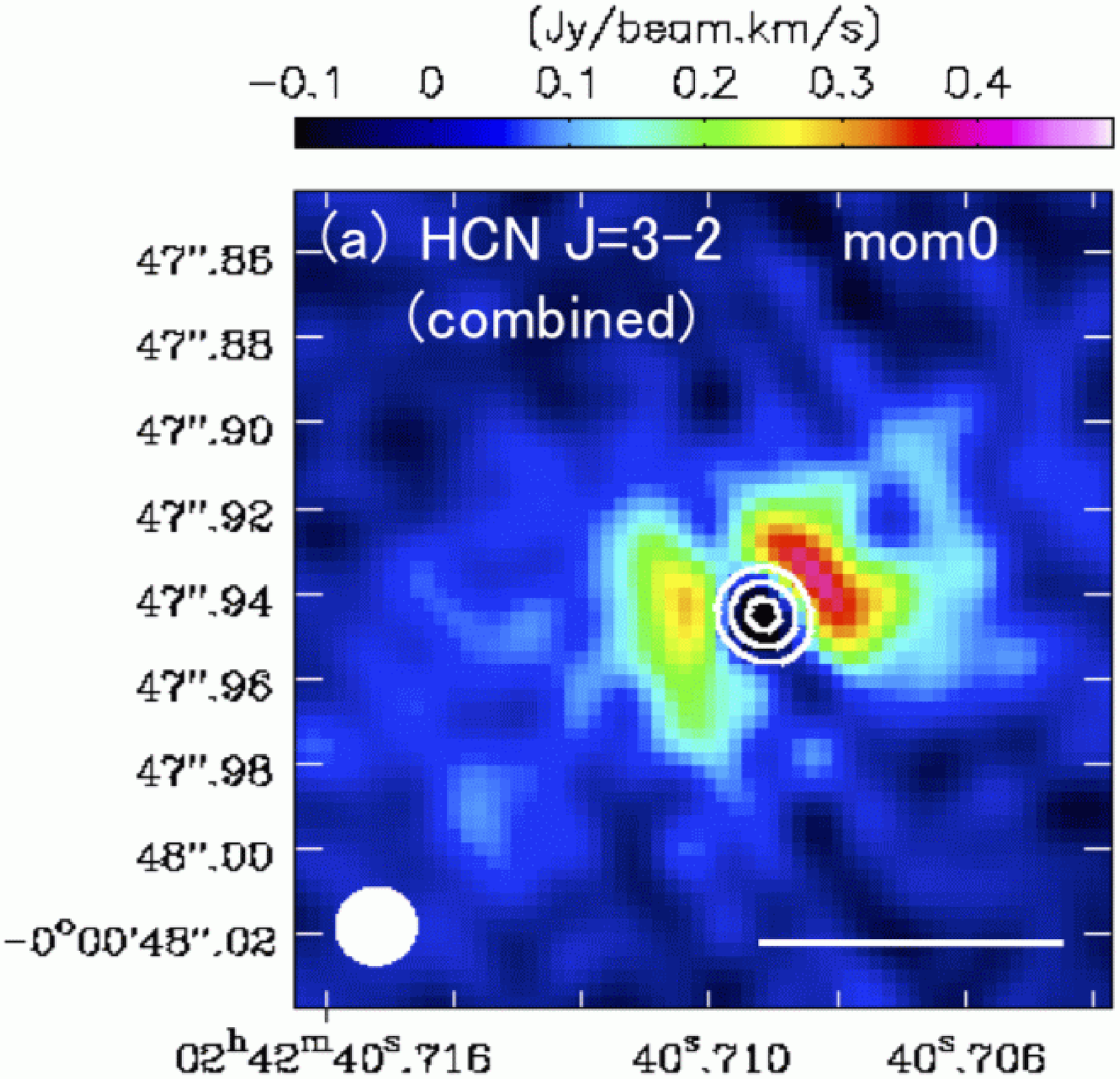}
\includegraphics[angle=0,scale=.3]{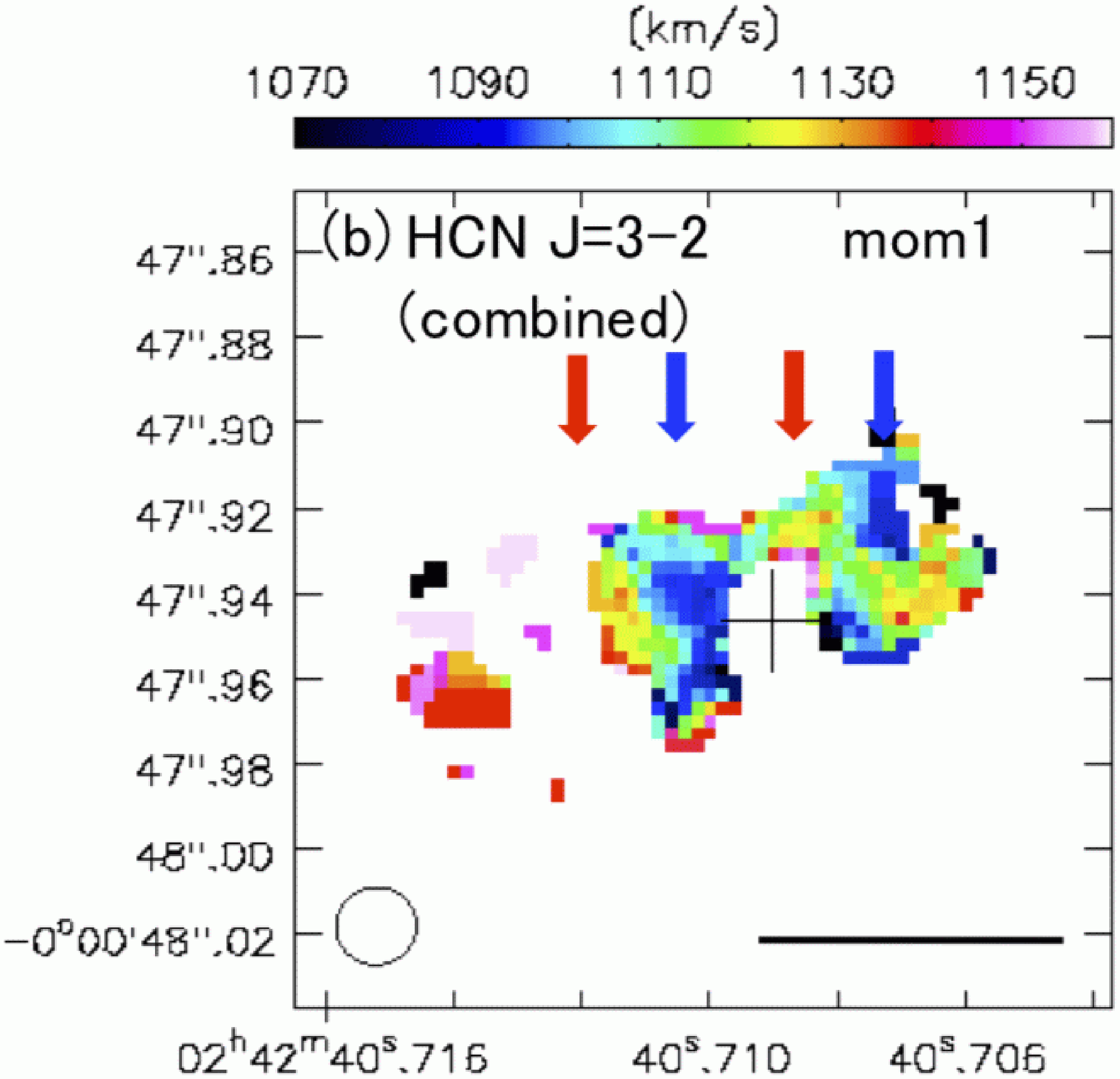} \\
\includegraphics[angle=0,scale=.3]{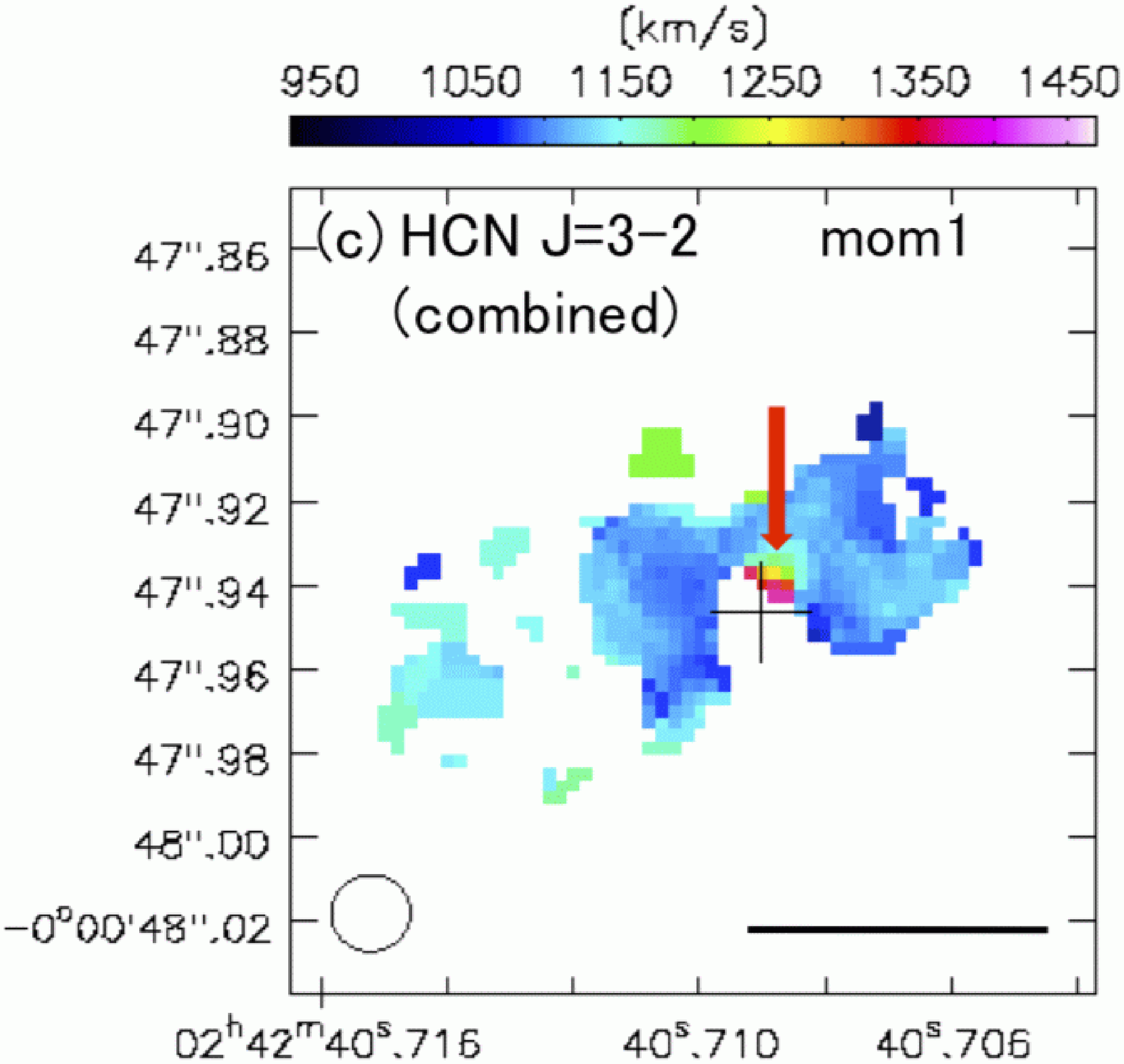}
\includegraphics[angle=0,scale=.3]{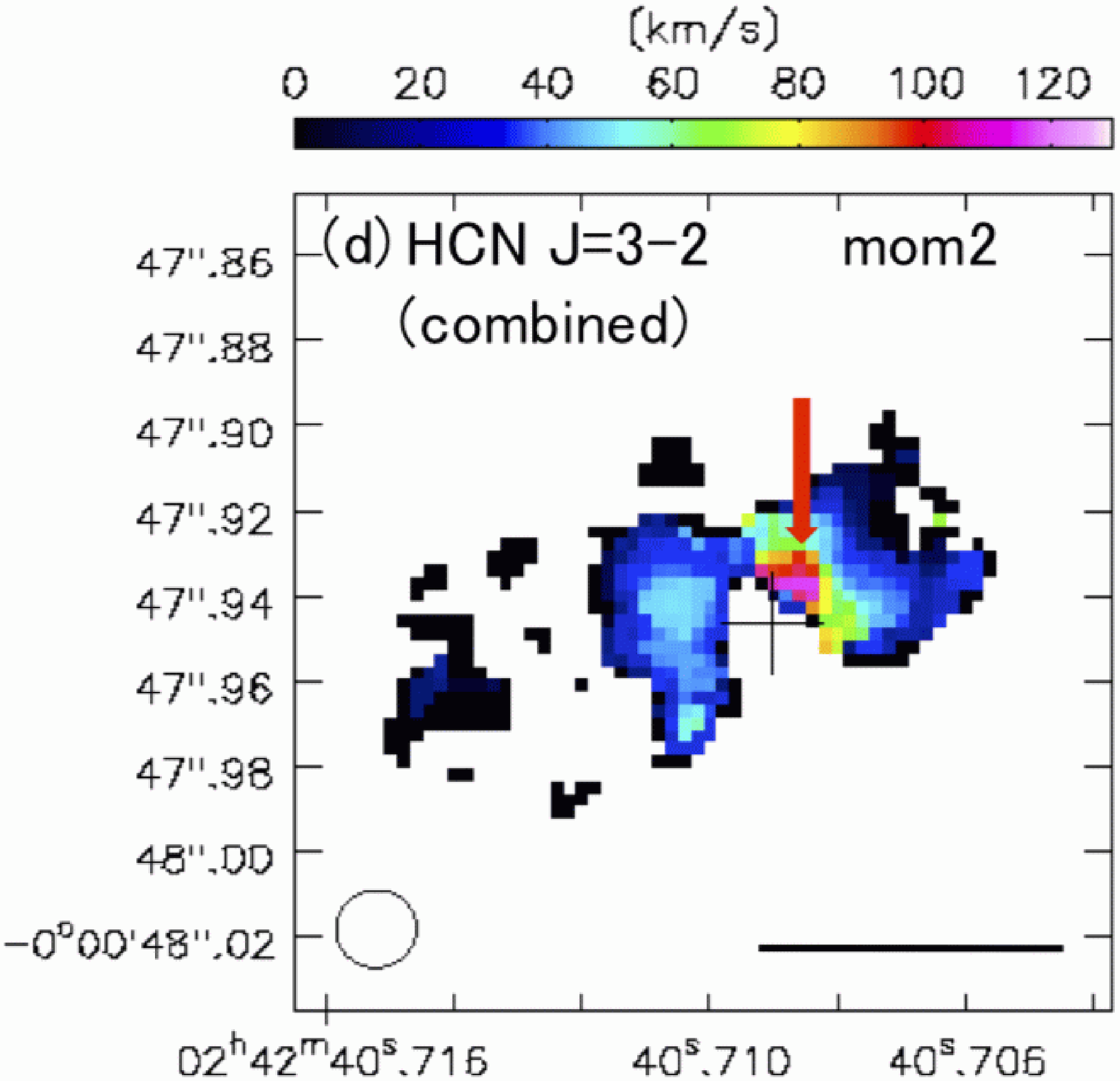} \\
\includegraphics[angle=0,scale=.21]{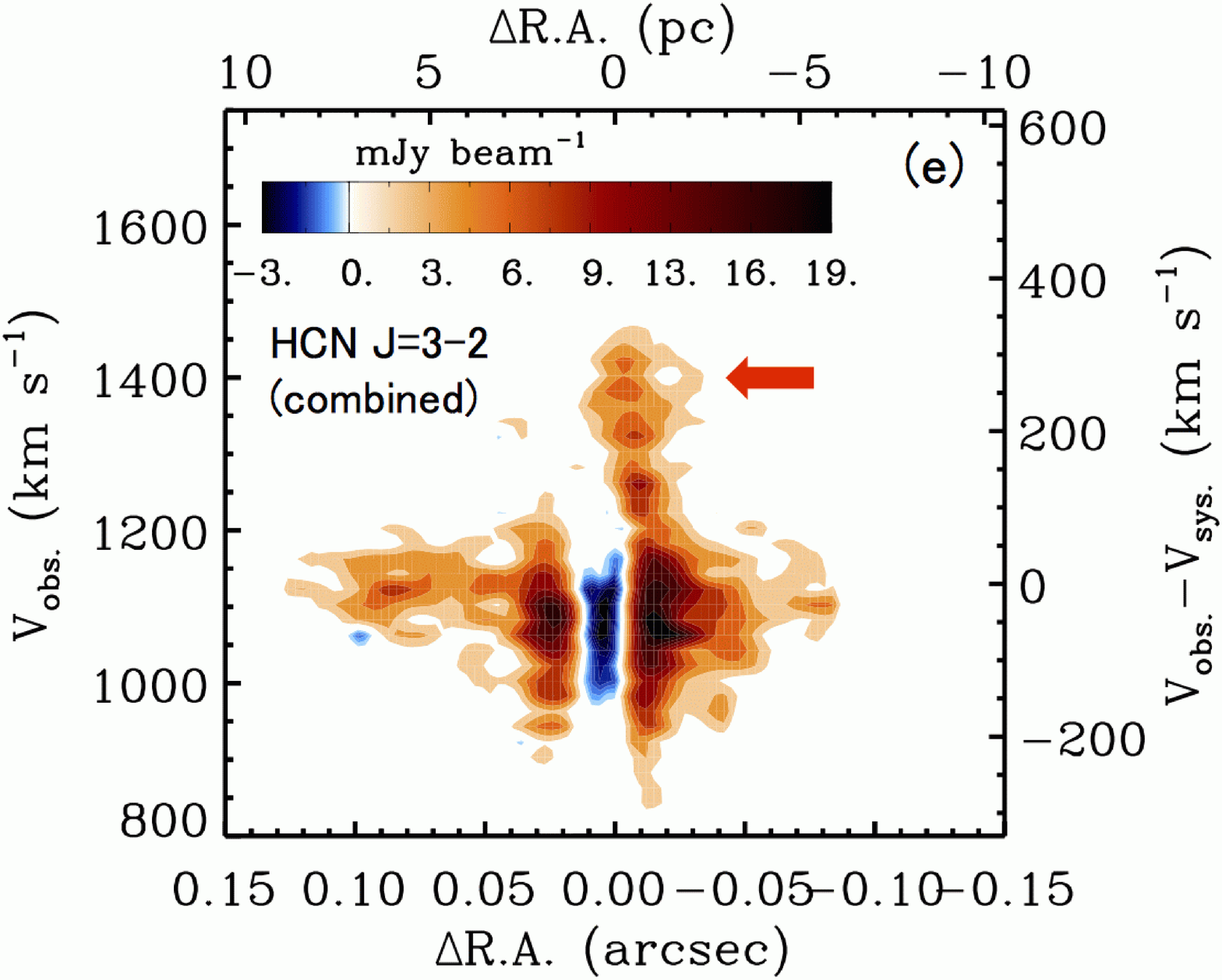} 
\end{center}
\caption{
{\it (a-d)}: Moment 0, 1, and 2 maps of HCN J=3--2 line 
in the combined data 
(our Cycle 6 and Cycle 5 data by \citet{imp19}).
In {\it (a)} moment 0 map, the rms noise level is 0.030 
(Jy beam$^{-1}$ km s$^{-1}$).
For {\it (b,c)} moment 1 maps, figures with two velocity display ranges 
are shown, following Figures 2 and 3 for our Cycle 6 data.
The meanings of white contours, black cross, thick blue and red 
downward arrows, 
horizontal white or black bar, and filled or open circle in the lower-left 
region, are the same as those in Figures 2 and 3. 
An appropriate cutoff ($\sim$5$\sigma$) is applied to moment 1 and 2 
maps in {\it (b--d)} so that they are not dominated by noise. 
{\it (e)}: Position velocity diagram of HCN J=3--2 line 
along PA = 115$^{\circ}$, displayed in the same manner as Figure 13a.
}
\end{figure}

\begin{figure}
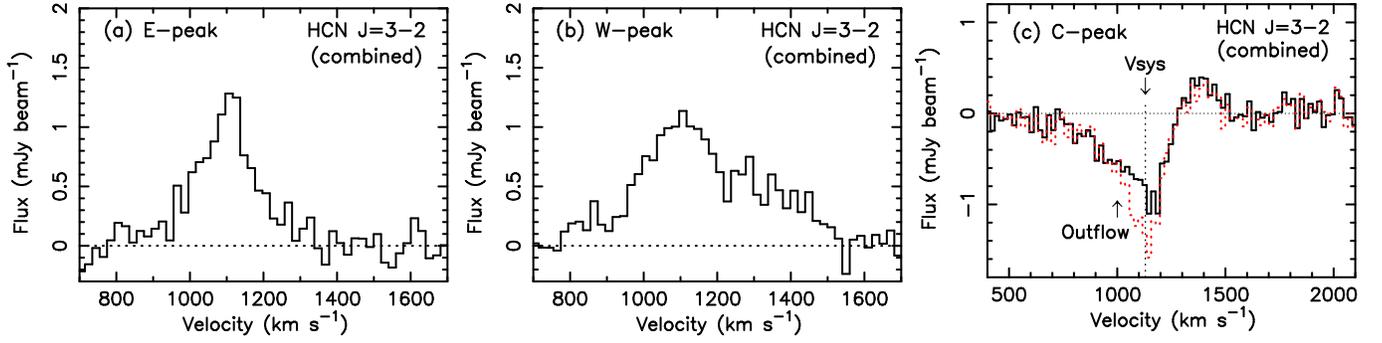

\begin{center}
\includegraphics[angle=-90,scale=.262]{f19a.eps}
\includegraphics[angle=-90,scale=.262]{f19b.eps} 
\includegraphics[angle=-90,scale=.262]{f19c.eps}
\end{center}
\caption{
Spectra of HCN J=3--2 line at the {\it (a)} E-peak, {\it (b)} W-peak, 
and {\it (c)} C-peak, within the beam size 
(0$\farcs$019 $\times$ 0$\farcs$018, PA = $-$79$^{\circ}$), 
in the combined data.
In all plots, the abscissa is optical LSR velocity (in km s$^{-1}$) 
and the ordinate is flux density (in mJy beam$^{-1}$).
The thin horizontal dotted straight line indicates the zero flux level.
In {\it (c)}, the black solid line represents raw data and the red dotted line 
is data corrected for possible emission contamination from 
the innermost torus, the same as in Figure 6 (see $\S$4.7). 
The systemic velocity at V$_{\rm sys}$ = 1130 km s$^{-1}$ and 
a blueshifted broad absorption tail at 750--1100 km s$^{-1}$ 
(V$_{\rm sys}$ $-$ [30--380] km s$^{-1}$) are indicated 
as a vertical dotted line with the note of ``V$_{\rm sys}$'' and 
an upward arrow with the note of ``Outflow'', respectively.
}
\end{figure}

\begin{figure}
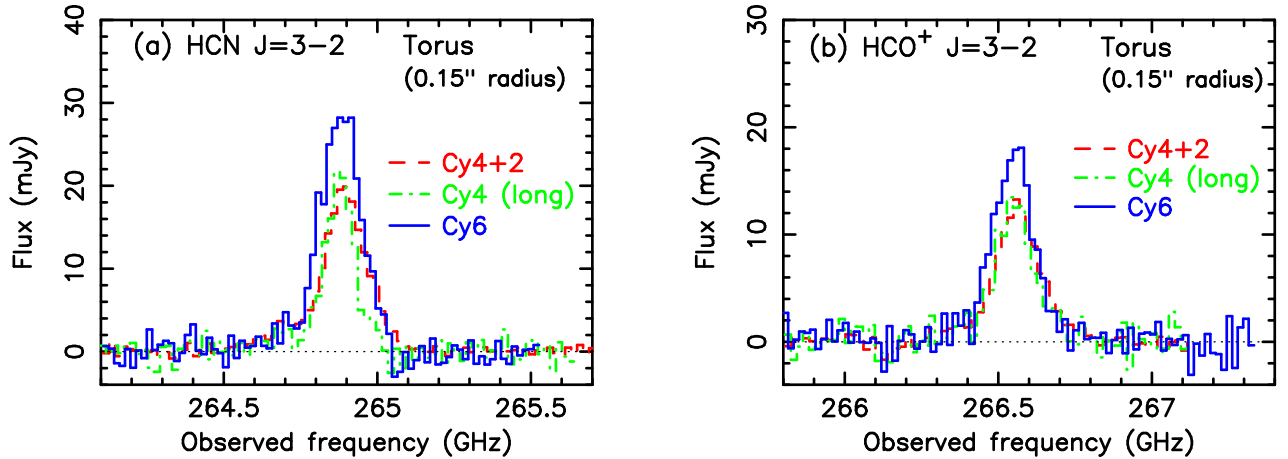

\begin{center}
\includegraphics[angle=-90,scale=.35]{f20a.eps} 
\hspace{1cm}
\includegraphics[angle=-90,scale=.35]{f20b.eps} \\
\end{center}
\vspace{0cm}
\caption{
Comparison of molecular emission line profile at the AGN torus position 
within a 0$\farcs$15 (or $\sim$10.5 pc) radius circular region centered at 
the C-peak (02$^{\rm h}$ 42$^{\rm m}$ 40.709$^{\rm s}$, $-$00$^{\circ}$ 00$'$ 47.946$''$).
{\it (a)}: HCN J=3--2. {\it (b)}: HCO$^{+}$ J=3=2.
The abscissa is observed frequency (in GHz) and the ordinate is 
flux density (in mJy).
The thin horizontal dotted straight line indicates the zero flux level.
Blue solid line: Cycle 6 data. 
Green dash-dotted line: Cycle 4 long baseline data only \citep{ima18a}.
Red dashed line: Cycles 4 and 2 combined data with both long and short 
baselines \citep{ima18a}.
}
\end{figure}

\end{document}